\documentclass[ reprint,superscriptaddress,amsmath,amssymb,aps,pre]{revtex4-2}

\usepackage{amsmath}
\usepackage{amssymb}
\usepackage{graphicx}
\usepackage{hyperref}
\usepackage[arrow, matrix, curve]{xy}
\usepackage{bm}
\usepackage{yhmath}
\usepackage{color}
\newcommand\diff{\mathrm{d}}

\usepackage[usenames,dvipsnames]{xcolor}
\hypersetup{colorlinks=true, linkcolor=BrickRed, urlcolor=blue!50!black, citecolor=blue!50!black}

\usepackage{amsmath,amsthm,amsfonts,amssymb,times,bbm,graphicx,color,epsfig,bm}
\usepackage{url} 
\usepackage{hyperref}

\usepackage[normalem]{ulem}

\makeatletter
\newcommand\hide@visible[1]{%
  \bgroup\fboxsep=.3ex\colorbox{Gray}{begin hide}%
  #1\colorbox{Gray}{end hide}\egroup%
}
\newcommand\hide@hidden[1]{%
  \bgroup\fboxsep=.3ex\colorbox{Gray}{hidden text}%
}
\newcommand\hide@invisible[1]{}
\newcommand\makevisible{\let\hide\hide@visible}
\newcommand\makehidden{\let\hide\hide@hidden}
\newcommand\makeinvisible{\let\hide\hide@invisible}

\makeatother
\makehidden

\begin{document}

\texttt{Published: Physical Review Research 7, 023260 (2025)}
\title{Relating thermodynamic quantities of convex-hard-body fluids to the body's shape }


\author{Thomas Franosch}
\affiliation{Institut f\"ur Theoretische Physik, Universit\"at Innsbruck, Technikerstra{\ss}e, 21A, A-6020 Innsbruck, Austria}
\email{thomas.franosch@uibk.ac.at}
\author{Cristiano De Michele}
\affiliation{Department of Physics, Sapienza  University of Rome, P.le A. Moro 2, 00185 Rome, Italy}
\email{demichel@roma1.infn.it}
\author{Rolf Schilling}
\affiliation{Institut f\"ur Physik, Johannes Gutenberg-Universit\"at Mainz,
Staudinger Weg 9, 55099 Mainz, Germany}
\email{rschill@uni-mainz.de}


\date{\today}

\begin{abstract}
For a fluid of convex  hard  particles, characterized by a length scale  $\sigma_\text{min}$ and an anisotropy parameter $\epsilon$, we develop a formalism allowing one to relate thermodynamic quantities to the body's shape. In a first step its thermodynamics is reduced to that of spherical particles. The latter have a hard core of diameter $\sigma_\text{min }$ and a soft shell with a thickness  $\epsilon \sigma_\text{min}/2$.  Besides their hard core repulsion at $\sigma_\text{min }$ they interact by effective entropic forces which will be calculated. Based on this mapping,  a second step provides a perturbative  method  for the systematic calculation of thermodynamic quantities  with the shape anisotropy $\epsilon$ as smallness parameter. 
 In leading order in $\epsilon $, the equation of state is derived as a functional of the particle's shape. 
To illustrate  these findings, they are applied to a  one- and two-dimensional fluid of ellipses and  compared with results from different analytical   approaches, and our computer simulations. The mapping to spherical particles   also implies that  any phase transition of spherical particles, e.g., the liquid-hexatic transition,  also exists for the nonspherical ones,  and  shifts  linearly with $\epsilon $ for weak shape anisotropy. This is supported by our Monte Carlo simulation. 
\end{abstract}

\maketitle

\section{Introduction}\label{Sec:Introduction}

Many particles in nature, e.g., molecules, are anisotropic. In order to elaborate on their thermodynamic behavior,
one has also to take into account their orientational degrees of freedom (d.o.f.). In this context, fluids of anisotropic colloidal particles have been intensively studied. They are modeled  by  convex hard bodies, mostly  hard ellipsoids. Compared to a fluid of spherical particles new phases occur. The influence of the shape anisotropy on the phase behavior was first studied in a seminal paper by Onsager~\cite{Onsager:AnnalsAcademy_51:1949}. Using  
the standard virial expansion (see, e.g., \cite{Hansen:Theory_of_Simple_Liquids:2013})  
for a three-dimensional (3D) fluid of hard cylinders  and treating the different orientations of a particle as a mixture of particles with different  orientations, Onsager proved that for large enough elongation of the cylinders  and high enough particle density,  a phase exists where, on  average, the cylinders are aligned, although their translational order is still liquid like. Since then, such liquid crystals have been widely investigated (see, e.g., Refs.~\cite{Priestly:Liquid_Crystals:2012,Stephen:RMP_46:1974}).  
Due to  strong experimental progress  in preparing  specially designed hard particles,  the study of their assembly has also gained a lot of interest in material science~\cite{Glotzer:NatureMaterials_6:2007}.
The  shape of hard bodies does not only influence the existence and  the type of phases, but also their thermodynamic properties  in general. In contrast to point particles, their excluded-volume interactions are determined completely by their shape. Consequently, 
one of the important questions is: How does the thermodynamic behavior of hard nonspherical particles depend on their shape?

The additional orientational d.o.f.,  make theoretical investigations of fluids of nonspherical  hard particles  much more involved, particularly for particles of a complex shape. Therefore, computer simulations have predominantly been used to calculate their thermodynamic (see, e.g., Refs.~\cite{Vieillard-Baron:JCP_56:1972,  Frenkel:PRL_52:1984, Odriozola:JCP_136:2012,  Foulaadvand:PRE_88:2013,  Xu:JCP_139:2013,  Bautista-Carbajal:JCP_140:2014, Torres-Diaz:SM_18:2022, Marienhagen:PRE_105:2022}),  as well as glassy properties (see, e.g., Refs.~\cite{DeMichele:PRL_98:2007,  Pfleiderer:EPL_84:2008,  Xu:SM_11:2015, Alhissi:JCP_160:2024}).
But  a variety of analytical approaches also exist. Many of them are extentions of  methods applied to  fluids of spherical particles. Systematic approaches are virial expansions ~\cite{Rowlinson:PROSL_402:1985, Isihara:JCP_18:1950, Hadwiger:Experimentia_7:1951, Kihara:RMP_25:1953, Freasier:MolPhys_32:1976, Nezbeda:ChemPhysLett_41:1976, Rigby:MolPhys_66:1989, Tarjus:MolPhys_73:1991, Rigby:MolPhys_78:1993, You:JCP_123:2005, Masters:JPhys_20:2008, Herold:JCP_147:2017, Marienhagen:PRE_105:2022, Kulossa:PRE_105:2022, Kulossa:MolPhys_0:2023, Kulossa:PRE_111:2025} and perturbation theory~\cite{Gray:Molecular_Fluids:1984,Solana:Perturbation_Theory_Fluids:2013}. A low-order truncation of the virial expansion (e.g., virial coefficients $B_l=0$ for $l \geq 4$) describes the Monte Carlo results for the pressure $p(n,T)$  only satisfactorily for rather low densities $n$. A different approach uses $y=\eta/(1-\eta)$  as an expansion parameter~\cite{Barboy:JCP_71:1979}, where $\eta$ is the packing fraction.  A low-order truncation
of such a  $y$ expansion fits the Monte Carlo  data rather well, even up to higher densities 
(see, e.g., Ref.~\cite{Mulder:MolPhys_55:1985}).  In a perturbative approach, the  pair potential or its Boltzmann factor  is decomposed into an isotropic and anisotropic part. The latter is chosen as a perturbation of the isotropic reference fluid. A particular reference system is a fluid of hard spheres
with diameters $\sigma_0$. For this case the question arises: How to choose an optimal value for $\sigma_0$? A frequently  used criterion is the vanishing of the first-order term in the series expansion of the Helmholtz free energy $F$ (see, e.g.,
Refs.~\cite{Belleman:PRL_21:1968,Nezbeda:JChemSoc_75:1979}). Other criteria were considered as well 
(see, e.g., Ref.~\cite{Percus:AnnNYAcad_221:1954}). We will come back to this point  in Sec.~\ref{SubSec:effective_potential}.  Scaled-particle theory~\cite{Reiss:JCP_31:1959} extended  to nonspherical particles~\cite{Gibbons:MolPhys_17:1969,  Boublik:MolPhys_27:1974, Boublik:MolPhys_42:1981}) and density-functional theory~\cite{Rosenfeld:PRE_50:1994, Rosenfeld:ACS_Symposium:1996, Cinacchi:JPhys_14:2002, Wittmann:JPhys_28:2016} involve approximations which cannot easily be controlled.  Onsager's approach~\cite{Onsager:AnnalsAcademy_51:1949} is a version of such a functional theory.

A rather different approach leading to sphericalization is the following one. As for fluids of spherical particles, several thermodynamic quantities of nonspherical particles such as the equation of state (e.o.s.) can be expressed completely by the orientational-dependent pair potential 
 and its pair-distribution function (see, e.g., Ref.~\cite{Boublik:MolPhys_27:1974}). Approximating the latter for a subclass of pair potentials of the form $v(r_\text{12}/d_c)$ with an  orientational-dependent function $d_c$ by $g^{(2)}(r_\text{12}/d_c)$~\cite{Parsons:PRA_19:1979,Lee:JCP_87:1987}, the translational and orientational d.o.f.\@ decouple. Hard-body interactions belong to this subclass. For two hard bodies,  $d_c$ is their closest distance  depending on the orientations, which will be discussed  in Sec.~\ref{Sec:model}. 
Consequently, thermodynamic quantities of a D-dimensional hard-body fluid become approximated  by those of a fluid of hard spheres of diameter   $[\langle (d_c)^D\rangle_o]^{1/D}$,
obtained by   averaging over the orientational d.o.f.. Finally, using a few low-order virial coefficients from analytical and numerical calculations, several attempts  have led to improved equations of state (see, e.g., Refs.~\cite{Marienhagen:PRE_105:2022, Boublik:MolPhys_42:1981,Song:PRA_41:1990, Vega:MolPhys_92:1997,Solana:MolPhys_113:2015}).
For further details on analytical methods, simulation techniques,  and thermodynamic properties of convex hard body fluids, the reader is referred also to Refs.~\cite{Allen:AdChemPhys_86:1993,Dijkstra:AdvChemPhys_156:2014}.
These investigations have not provided relations between thermodynamic quantities and the body's shape.
 An exception is the  remarkable  result for the second reduced virial coefficient $B^*_2$ for arbitrary  convex hard bodies: It depends only on geometric measures (Minkowski functionals) \cite{Mecke:Minkowski_functionals:2000,Torquato:JStatMech:2022} such as the volume $V_p$, surface $S_p$, and the mean curvature $R_p$ of the convex body~
\cite{Isihara:JCP_18:1950, Hadwiger:Experimentia_7:1951, Kihara:RMP_25:1953, Boublik:MolPhys_29:1975, Herold:JCP_147:2017, Kulossa:PRE_105:2022, Kulossa:MolPhys_0:2023, Kulossa:PRE_111:2025}.
This implies that $B^*_2$ takes the same value for \textit{all} shapes with identical geometric measures. 
Note this result  holds for $B^*_2$  only. 


Therefore, it is  our major goal to investigate how far thermodynamic quantities of  hard nonspherical particles can be related to their shape.This will be achieved by a completely new approach.  By eliminating all  orientational d.o.f.\@  we prove  in a \textit{first} step, that the thermodynamics of a D-dimensional  fluid of convex  hard nonspherical particles can be obtained from a corresponding  fluid of spherical particles. Note, this fluid of spherical particles will be not a reference fluid in the sense discussed above. 
This already has an interesting implication: Any phase transition of a fluid of spherical particles also exists for hard nonspherical convex particles, at least for small shape anisotropies. 
Since 2D fluids of spherical particles  exhibit  two-step melting  with  a liquid-hexatic  and a hexatic-solid  phase transition~\cite{Strandburg:RMP_60:1988}, our mapping   implies that for small shape anisotropy these transitions should   exist for a fluid of   convex hard nonspherical bodies as well. Based on the mapping to spherical particles, in a \textit{second} step, the Helmholtz free energy  will be calculated  perturbatively with the shape anisotropy as perturbation. Combined with a perturbative solution of the contact conditions, this will provide a relation between the e.o.s. and the particle's shape.

The outline of our paper is as follows. In Sec.~\ref{Sec:Model_and_Framework}, we  describe the model fluid and present the theoretical framework. This section also contains the mapping to a fluid of spherical particles and the series expansion of the free energy. A key quantity entering the thermodynamic quantities is  the contact function for two  convex hard bodies. How its  perturbative calculation can be performed by  using  the body's shape as an input will be discussed in Sec.~\ref{Sec:contact-f}.  The dependence of the equation of state on the particle's shape will be obtained  in   Sec.~\ref{Sec:e.o.s.} up to first order in the shape anisotropy, valid for any convex body and any dimension D. This allows us to derive the equation of state for a fluid of
hard ellipses in D=1 and D=2, as well as for  a 3D fluid of hard ellipsoids of revolution.  In addition, a  comparison with results from other analytical methods and from our Monte Carlo simulation will be performed for hard ellipses in  D=1 and D=2.  
 Finally, Sec.~\ref{Sec:conclusions} contains a summary and some concluding remarks. In order not to overload the main text with technical manipulations, details are presented in Appendices.

\section{Model fluid and theoretical framework}\label{Sec:Model_and_Framework}

\subsection{The model}\label{Sec:model}

\begin{figure}
\includegraphics[angle=0,width=0.9\linewidth]{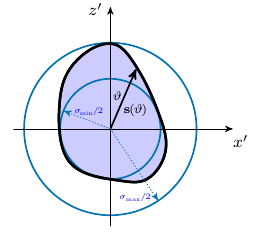}
\caption{Construction for an arbitrary 2D convex hard body,  $\mathcal{B}$, (blue domain) of the smaller ($\mathcal{S}_\text{min}$) and larger circle ($\mathcal{S}_\text{max}$) with radius  $\sigma_{\text{min}}/2$ and $\sigma_{\text{max}}/2$, respectively.  The center of  $\mathcal{B}$ is chosen as the origin of the body-fixed frame and the  $z'$ axis  points to the   tangential contact of  $\mathcal{B}$ and  $\mathcal{S}_\text{max}$. The shape of the hard body is parametrized by $\mathbf{s}(\vartheta)$ where 
$\vartheta$  is the angle between  $\mathbf{s}(\vartheta)$ and the $z'$ axis (see also main text). \label{fig:hard_body}}.
\end{figure}

We investigate a fluid of $N$ identical  convex hard nonspherical particles in a D-dimensional box of volume $V$. Then $n = N/V$ is its number density. To quantify their nonspherical shape, we proceed as follows. The hard body is denoted by $\mathcal{B}$. Two characteristic length scales, $\sigma_\text{min}$ and $\sigma_\text{max}$,  can be introduced.
We denote by $\sigma_\text{min}$ the maximum diameter of a D-dimensional sphere, $\mathcal{S}_\text{min}$, which can be inscribed into $\mathcal{B}$, and by $\sigma_{\text{max}}$ the minimum diameter of a sphere, $\mathcal{S}_\text{max}$, with the center 
coinciding with the center of $\mathcal{S}_\text{min}$ such that $\mathcal{B}$ can be inscribed into $\mathcal{S}_\text{max}$. For an illustration, see the 2D example in Fig.~\ref{fig:hard_body}.  Note, that there can be more than two contact points with $\mathcal{S}_\text{min}$  and more than one with  $\mathcal{S}_\text{max}$.
As the center of the hard body, we choose the common center of  $\mathcal{S}_\text{min}$ and  $\mathcal{S}_\text{max}$,  also taken as the origin of the body-fixed frame. For a generic hard body,  this center  does not coincide with the body's center of mass.
The $z'$ axis of the body-fixed frame can be chosen such that, e.g., one of the tangential contact points with  $\mathcal{S}_\text{max}$ is on the $z'$ axis. In the body-fixed frame, the surface of the convex  hard body can be represented  by spherical coordinates $(s,\vartheta,\varphi)$:
$\mathbf{s}(\vartheta,\varphi) = s(\vartheta,\varphi) \mathbf{e}_s(\vartheta, \varphi)$ with $ \mathbf{e}_s(\vartheta, \varphi)$, the corresponding unit vector. All information on the particle's shape is provided by the \textit{shape function} $s(\vartheta,\varphi)$.  

Choosing $\sigma_{\text{min}}$ as the fundamental length scale and 
$ (\sigma_{\text{max}}/\sigma_{\text{min}} - 1) $   as  anisotropy parameter, we can introduce a  dimensionless shape function $\tilde{s}(\vartheta,\varphi)$, with $0 \leq \tilde{s}(\vartheta,\varphi) \leq 1$, by 
\begin{align}\label{eq:tilde-shape} 
\frac{2}{\sigma_{\text{min}}} s(\vartheta,\varphi) - 1 = (X_0 -1 ) \ \tilde{s}(\vartheta,\varphi) \ ,
\end{align} 
with the aspect ratio $X_0 = \sigma_{\text{max}}/\sigma_{\text{min}}$. 
Hard bodies with zero thickness
or infinite extension in one or more dimensions will not be considered, i.e.,  $\sigma_{\text{min}} > 0$ and $X_0 \leq \text{const} < \infty$. In the following, we will study hard bodies characterized by a family of shape functions: 
 \begin{align}\label{eq:shape} 
s(\vartheta,\varphi;\epsilon) &= \frac{\sigma_\text{min}}{2} \big[1 + \epsilon~\tilde{s}(\vartheta,\varphi) \big] \ . 
\end{align} 
Note that  $\epsilon$ interpolates between the shape of the sphere, $\mathcal{S}_\text{min}$ (for $\epsilon = 0$), and the original hard body, $\mathcal{B}$ (for $\epsilon = \epsilon_\text{max} = X_0 - 1$). 
In Sec.~\ref{Sec:free_energy}, we will elaborate on a perturbative method for the calculation of thermodynamic quantities, where $\epsilon$ plays the role of a smallness parameter.

Some comments are in order. First, there is an alternative route leading to the shape family, Eq.~\eqref{eq:shape}. Instead of using an arbitrary convex body as a   starting point, one could choose a sphere of radius  $\sigma_\text{min}/2$. Deforming this sphere linearly in $\epsilon$ by $ \epsilon~(\sigma_\text{min}/2) \tilde{s}(\vartheta,\varphi)$ with $0 \leq \tilde{s}(\vartheta,\varphi) \leq 1$, generates the family of Eq.~\eqref{eq:shape}. However, the condition of convexity introduces constraints on $\tilde{s}(\vartheta,\varphi)$: The eigenvalues of  the  curvature tensor of the surface obtained from $s(\vartheta,\varphi;\epsilon) \mathbf{e}_s(\vartheta,\varphi) $ have to be non-negative for all $(\vartheta,\varphi)$.
Second, ellipsoids of revolution (or ellipses) are completely determined by their  aspect ratio $X_0$ and 
$\sigma_\text{min}= 2b$, the length of their minor axis. Therefore, their shape function will involve $\tilde{s}(\vartheta,\varphi;\epsilon)$, depending also on $\epsilon= X_0-1$,  in contrast to generic hard bodies. Although the class of ellipsoids of revolution (or ellipses)  is nongeneric, it plays an important role, since the vast majority of analytical studies and computer simulations have been performed for ellipsoids of revolution. There are also   shapes of hard bodies which depend on more parameters than merely $X_0$. An  example is a triaxial  ellipsoid, for which a second aspect ratio $X_1$ exists. Such hard bodies, requiring more than one anisotropy parameter, will not be considered here. Third, whereas the thermodynamic properties will depend sensitively on 
$\epsilon$, their dependence on $\sigma_{\text{min}}$ appears only through the dimensionless density $n^*= n (\sigma_{\text{min}})^D$. Therefore the dependence of $s(\vartheta,\varphi;\epsilon)$ on $\sigma_{\text{min}}$ is suppressed.  Fourth,  it is obvious that the shape function $s(\vartheta,\varphi;\epsilon)$  will depend on the choice of the origin and orientation  of the body-fixed frame. However, thermodynamic quantities, e.g.,  the free energy $F(T,V,N)$  involve only geometrical entities  being independent, on such a choice.  In addition, the reference fluid of the hard spheres $\mathcal{S}_\text{min}$ would gain artificial orientational d.o.f.\@ by changing the reference frame.

For simplicity,  we  restrict ourselves  to 3D (2D) bodies with a rotational (reflection) symmetry axis. However, our analytical approach can also be applied to  arbitrary convex  hard bodies.  Due to the symmetry, the body's center will lie on the symmetry axis and the hard body is parametrized by the family of Eq.~ \eqref{eq:shape}  of shape functions  
which depends only on the angle $\vartheta$ between that  axis  and the unit vector $\mathbf{e}_s(\vartheta, \varphi)$. No additional symmetry, such as a head-tail  symmetry or smoothness of $s(\vartheta;\epsilon)$ is assumed so far.  
 
\begin{figure}
\includegraphics[angle=0,width=0.9\linewidth]{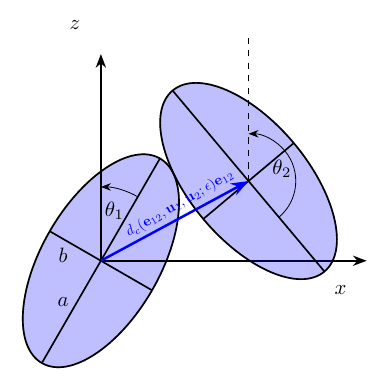}
\caption{Two hard ellipses ($a$ and $b$ are the lengths of the major and minor semiaxis) at contact in the $x$-$z$ plane with orientations $\mathbf{u}_i=(\sin \theta_i,\cos \theta_i)^T, i=1,2$ and center-to-center separation vector 
 $ d_c(\mathbf{e}_{12},\mathbf{u}_1,\mathbf{u}_2;\epsilon) \mathbf{e}_{12}$.  
\label{fig:2_hard_ellipses}}.
\end{figure}

Assuming that (i)  the $z'$ axis of the body-fixed frame is along the  body's symmetry axis  and (ii)  the  $z'$ axis   in the space-fixed frame points in  the direction of the  unit vector $\mathbf{u}$, then the  surface  of the 3D hard bodies is given by the set
\begin{align} \label{eq:surface}
S(\mathbf{u};\epsilon)  = \{ \mathbf{s} \in \mathbb{R}^3 |\  \mathbf{s} & = \mathbf{s}(\vartheta,\varphi,\mathbf{u};\epsilon): = s(\vartheta;\epsilon)  R(\mathbf{u}) \mathbf{e}_s(\vartheta, \varphi),  \nonumber\\
&  0 \leq \vartheta \leq \pi \ , \ 0 \leq \varphi < 2\pi \}  \ ,
\end{align}
where $R(\mathbf{u})   \in \text{SO}(3,\mathbb{R})$  rotates the body-fixed $z'$ axis into the direction of 
$\mathbf{u}$. 

The presence of a rotational symmetry axis allows one to introduce two characteristic length scales,
$\sigma_{\parallel}$, the length of the hard body parallel to the symmetry axis, and  $\sigma_{\perp}$, the maximum length perpendicular to that axis. It is obvious that $\sigma_{\parallel} = \sigma_{\text{max}}$ and 
 $\sigma_{\perp} = \sigma_{\text{min}}$ or vice versa.  Then $\tilde{X}_0 = \sigma_{\parallel}/\sigma_{\perp}$ is an alternative definition of an aspect ratio.  For $X_0 =  \sigma_{\text{max}}/ \sigma_{\text{max}}$ as defined above, it is always 
$X_0 \geq 1$, whereas $\tilde{X}_0 =X_0 > 1$ for prolate convex bodies and $\tilde{X}_0 = 1/X_0 < 1$ for oblate ones. Note that, although we will use $X_0$ in the following, the results presented in this work  hold for prolate and oblate particles. It is the shape function $s(\vartheta;\epsilon)$ in Eq.~\eqref{eq:surface} which is different for prolate and oblate hard particles (see, e.g., Eqs.~ \eqref{eq:s1-prolate}  and \eqref{eq:s1-oblate} for ellipsoids of revolution).

Consider now two bodies  with centers  separated by $\mathbf{d} = d \mathbf{e}_{12}$ ($|\mathbf{e}_{12}  |=1$) and orientations $\mathbf{u}_i$, $i = 1,2$ (see Fig.~\ref{fig:2_hard_ellipses} for two ellipses). Although their interaction energy is rather simple assuming values zero or infinity, the calculation of a partition function requires the knowledge of  a suitable \emph{overlap function} $\psi(d\mathbf{e}_{12},\mathbf{u}_1,\mathbf{u}_2;\epsilon)$. 
If the two convex bodies are in a tangential contact, then   $\psi(d\mathbf{e}_{12},\mathbf{u}_1,\mathbf{u}_2;\epsilon) = 0$.  If both bodies overlap, the overlap function is negative, and if they do not have a common point  it is positive.  An explicit analytical expression for such a function is only known for ellipses \cite{Vieillard-Baron:JCP_56:1972}.  For arbitrary ellipsoids, an iterative scheme for its computation is presented in 
 Ref.~\cite{Perram:JCompPhys_58:1985}. Note that, using the overlap function from Ref.~\cite{Vieillard-Baron:JCP_56:1972}, an additional condition must be satisfied  for ellipses if they do not have a  point in common (see Appendix~\ref{Sec:Appendix_A}).

The key  quantity in our approach is  the \emph{contact function} $d_\text{c}= d_\text{c}(\mathbf{e}_{12}, \mathbf{u}_1, \mathbf{u}_2;\epsilon)$. It is the distance  when both  bodies with orientations  $\mathbf{u}_1, \mathbf{u}_2$, and separation vector $d\,  \mathbf{e}$ are exteriorly tangential (see Fig.~\ref{fig:2_hard_ellipses} for two ellipses). It is determined by the condition
\begin{align} \label{eq:overlap-0}
\psi(d\,\mathbf{e}_{12},\mathbf{u}_1,\mathbf{u}_2;\epsilon) =0 \ .
\end{align}
Due to the convexity, its solution,  $d_\text{c}(\mathbf{e}_{12},\mathbf{u}_1,\mathbf{u}_2;\epsilon)$, is unique. As will be seen  below,   the calculation of the effective multibody potential in the next subsection, and particularly the calculation of the $k$ th order terms  of the perturbation series for the free energy in Sec.~\ref{Sec:free_energy}
 requires the knowledge of $d_\text{c}(\mathbf{e}_{12},\mathbf{u}_1,\mathbf{u}_2;\epsilon)$.
Since the overlap function is only known for hard ellipses, Eq.~\eqref{eq:overlap-0} cannot be used, in general, to determine  $d_\text{c}(\mathbf{e}_{12},\mathbf{u}_1,\mathbf{u}_2;\epsilon)$. However, it can be calculated  directly from the tangential contact conditions, involving  $\mathbf{s}(\vartheta,\varphi,\mathbf{u};\epsilon)$    and the normal vector at the point of contact of both hard bodies (see  Sec.~\ref{Sec:contact-f} for more details). Since  these entities are completely determined  by the shape function $s(\vartheta;\epsilon)$, this  also holds  for $d_\text{c}(\mathbf{e}_{12},\mathbf{u}_1,\mathbf{u}_2;\epsilon)$. Although  the contact conditions are linear in $d_\text{c}(\mathbf{e}_{12},\mathbf{u}_1,\mathbf{u}_2;\epsilon)$, they are  
nonlinear in $s(\vartheta;\epsilon)$.  Therefore, the contact function $d_\text{c}$ will  not  be linear in $\epsilon$, despite the linearity of the  shape function $s$ in Eq.~\eqref{eq:shape}. Nevertheless, since 
$\sigma_\text{min} \leq d_\text{c}(\mathbf{e}_{12},\mathbf{u}_1,\mathbf{u}_2;\epsilon)  \leq  (1+\epsilon) \sigma_\text{min}$ for the shape family, Eq.~\eqref{eq:shape}, it can be represented  as  
\begin{align}\label{eq:contact} 
d_\text{c}(\mathbf{e}_{12},\mathbf{u}_1,\mathbf{u}_2;\epsilon) = \sigma_\text{min} \big[1 + \epsilon \ \tilde{d}_\text{c}(\mathbf{e}_{12},\mathbf{u}_1,\mathbf{u}_2;\epsilon) \big]\ ,
\end{align} 
with $0  \leq \tilde{d}_\text{c}(\mathbf{e}_{12},\mathbf{u}_1,\mathbf{u}_2;\epsilon)  \leq  1$, in close analogy  to the shape function in Eq.~\eqref{eq:shape}. In contrast to the latter, $\tilde{d}_\text{c}$ will depend on $\epsilon$, even for generic hard bodies, as pointed out above. In  Sec.~\ref{Sec:contact-f},  we will demonstrate how the key quantity  $d_\text{c}(\mathbf{e}_{12},\mathbf{u}_1,\mathbf{u}_2;\epsilon)$ can be calculated for a given shape function $s(\vartheta;\epsilon)$.

The main message from the discussion so far is that for all orientations $(\mathbf{u}_1,\mathbf{u}_2)$, the pair potential is infinite for $r < \sigma_\text{min}$ and vanishes for $r \geq \sigma_\text{max}$. This suggests choosing (i) the fluid of hard spheres of diameter $\sigma_\text{min}$ and center positions $\mathbf{r}_i$, $i = 1, \dots, N$  as a \textit{reference} fluid and (ii) considering  the spherical shell $\sigma_\text{min} \leq |\mathbf{r}_{ij} | \leq \sigma_\text{max}$, $\mathbf{r}_{ij} := \mathbf{r}_i -\mathbf{r}_j$, as a
perturbation, in the sense of a measure. To perform a perturbative approach for the free energy, we proceed in two steps. First, for 
$\sigma_\text{min} \leq |\mathbf{r}_{ij}| \leq \sigma_\text{max}$ we eliminate the orientational d.o.f.\@ which reduces the original fluid of hard nonspherical  particles to a fluid of spheres with a hard core of diameter $\sigma_\text{min}$ interacting via  an effective isotropic potential $V_\text{eff}(\mathbf{r}_1,\ldots, \mathbf{r}_N)$. In a second step, the  free energy is calculated for such a  fluid of spherical particles.

\subsection{Mapping to a fluid of spherical particles}\label{SubSec:effective_potential}

In the present subsection, we will demonstrate how the orientational d.o.f.\@ of a fluid of convex  hard bodies can be eliminated. We notice that this elimination and, accordingly, the resulting mapping to spherical particles  does not require small anisotropy.
Elimination of d.o.f.\@ has played an important role in various fields of physics, e.g., in the renormalization-group theory and in soft-condensed-matter theory. It leads to effective interactions. A famous example is the Asakura-Oosawa model~\cite{Asakura:JCP_22:1954,Asakura:JPS_33:1958}, where the elimination of the smaller particles of a binary, colloidal mixture generates attractive interactions (depletion forces) between the larger ones.

As described in the previous subsection we consider a fluid of N convex hard bodies (of revolution) with centers at $\mathbf{r}_i$ and orientations $\mathbf{u}_i$, $i = 1, \dots, N$ interacting via the potential
\begin{align}\label{eq:potential}
V(\mathbf{r}_1 \mathbf{u}_1,\ldots, \mathbf{r}_N \mathbf{u}_N;\epsilon) &= \sum_{i<j} 
v_{\text{HP}}(\mathbf{r}_{ij}, \mathbf{u}_i,\mathbf{u}_j;\epsilon) .
\end{align}
Here, $v_{\text{HP}}(\mathbf{r}_{ij}, \mathbf{u}_i,\mathbf{u}_j;\epsilon)$ is infinite if the center-to-center distance $r_{ij}= |\mathbf{r}_{ij}|$ between two hard bodies is smaller than the value of the contact function $d_c(\mathbf{e}_{ij},\mathbf{u}_i,\mathbf{u}_j;\epsilon)$, and zero otherwise, i.e., $\exp[-\beta v_{\text{HP}} (r_{ij}\mathbf{e}_{ij}, \mathbf{u}_i, \mathbf{u}_j)]   = \Theta (r_{i j} -  d_c(\mathbf{e}_{i j} , \mathbf{u}_i, \mathbf{u}_j ))$.   In the following the $\epsilon$ dependence of the various quantities  will be made explicit only if necessary for clarity.

To reduce the thermodynamics of this fluid to that of spherical particles, we eliminate the orientational d.o.f.\@. Accordingly, the averaging of  the Boltzmann factor over the orientational d.o.f.\@ leads to a coarse-grained excess free energy defined by
$\mathcal{F}(\mathbf{r}_1,\ldots, \mathbf{r}_N)$, 
\begin{align}\label{eq:effpot1}
\exp[-\beta \mathcal{F}(\mathbf{r}_1,\ldots, \mathbf{r}_N) ] 
&:= \langle \exp[- \beta V(\mathbf{r}_1 \mathbf{u}_1,\ldots, \mathbf{r}_N \mathbf{u}_N) ] \rangle_o
\end{align}
with $\beta = 1/k_B T $ the inverse temperature and 
\begin{align}\label{eq:orient-average}
\langle (\cdots) \rangle_o &= \left[\int \prod_{i} \frac{\diff \mathbf{u}_i}{\Omega_D}\right] (\cdots) \ ,
\end{align}
the orientational average over $\Omega_D$, the surface area of a D-dimensional unit sphere.

The basic idea of our  approach is to choose the fluid of hard spheres of diameter $\sigma_\text{min}$  as a reference fluid (unperturbed system) and to take the anisotropic shape quantified by the reduced shape function  $\tilde{s}(\vartheta)$ [cf. Eq.~\eqref{eq:tilde-shape}]  as  the ``perturbation''. Then
 the potential of the reference fluid  is provided by
\begin{align} \label{eq:potential_0} 
V_0(\mathbf{r}_1,\ldots, \mathbf{r}_N) = \sum_{i<j} v_{\text{HS}}(r_{ij}; \sigma_\text{min}) ,
\end{align}
where $ \exp[-\beta v_{\text{HS}}(r_{ij}; \sigma_\text{min})] = \Theta(r_{ij} - \sigma_\text{min})$. We define the cluster function 
\begin{align} \label{eq:cluster-1} 
f(i,j) & = \Theta \Big(r_{ij} - d_c(\mathbf{e}_{ij},\mathbf{u}_{i},\mathbf{u}_{j})\Big)   -  \Theta\Big(r_{ij} - \sigma_\text{min}\Big) .
\end{align}
Using Eqs.~\eqref{eq:potential}, \eqref{eq:potential_0}, \eqref{eq:cluster-1}, and an identity in Ref.~\cite{Franosch:PRL_109:2012, *Franosch:PRL_110:2013, *Franosch:PRL_128:2022} allows us to express the Boltzmann factor  $\exp[ - \beta V(\mathbf{r}_1 \mathbf{u}_1,\ldots, \mathbf{r}_N \mathbf{u}_N) ] =  \prod_{i<j} \Theta \big(r_{ij} - d_c(\mathbf{e}_{ij},\mathbf{u}_{i},\mathbf{u}_{j})\big)$  as follows:
\begin{align}\label{eq:effpot2}
\lefteqn{\prod_{i<j} \Theta \Big(r_{ij} - d_c(\mathbf{e}_{ij},\mathbf{u}_{i},\mathbf{u}_{j})\Big)  = } \nonumber  \\ 
&= \prod_{i<j} \Theta(r_{ij} - \sigma_\text{min})  \prod_{i<j} [ 1 + f(ij)] .
\end{align}
Then we obtain from Eqs.~\eqref{eq:effpot1} and \eqref{eq:effpot2} for the effective interaction potential $V_\text{eff}(\mathbf{r}_1,\ldots, \mathbf{r}_N): = \mathcal{F}(\mathbf{r}_1,\ldots, \mathbf{r}_N) - V_0(\mathbf{r}_1,\ldots, \mathbf{r}_N)$:
\begin{align}\label{eq:effpot3}
V_\text{eff}(\mathbf{r}_1,\ldots, \mathbf{r}_N) = - k_B T \ln \big\langle\prod_{i<j} [ 1 + f(ij) ] \big\rangle_o.
\end{align}

\begin{figure}
\includegraphics[angle=0,width=\linewidth]{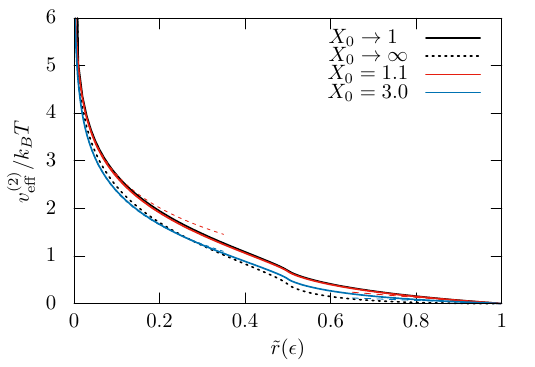}
\caption{Numerical results for  $v^{(2)}_{\text{eff}}(r)$ and $X_0 = 1.1$ (solid red line)  and $3.0$ (solid blue line)  for a 2D fluid of ellipses with semiaxes of length 
$a, b$  and aspect ratio $X_0= a/b = 1+\epsilon$ as a function of the scaled distance
$\tilde{r}(\epsilon) = (r/2b - 1)/\epsilon$. Also shown are the  semianalytical results, Eqs.~\eqref{eq:area-1-1} and \eqref{eq:area-1-2},  for $X_0 \to 1$ (upper black solid line) and Eqs.~\eqref{eq:area_1-infinite}-\eqref{eq:area_3-infinite}  for $X_0 \to \infty$ (lower black dotted line) as well as the analytical asymptotic results, Eqs.~\eqref{eq:area-3} and \eqref{eq:area-2},  for   $r \to 2b^+$  and  $r \to 2a^-$, respectively (corresponding to dashed lines). 
\label{fig:soft_2_body_potential}}
\end{figure}

This is an appropriate place to come back to the choice of  an optimal diameter $\sigma_0$ for  a hard-sphere reference fluid, as discussed in Sec.~\ref{Sec:Introduction}. It is obvious that one should use $\sigma_\text{min} \equiv \text{min}_{\mathbf{u}_{i},\mathbf{u}_{j}}  d_c(\mathbf{e}_{ij},\mathbf{u}_{i},\mathbf{u}_{j}) \leq \sigma_0 \leq \text{max}_{\mathbf{u}_{i},\mathbf{u}_{j}}  d_c(\mathbf{e}_{ij},\mathbf{u}_{i},\mathbf{u}_{j}) \equiv \epsilon \sigma_\text{min} $. Replacing 
in Eqs.~\eqref{eq:potential_0}, \eqref{eq:cluster-1} and \eqref{eq:effpot2}  $\sigma_\text{min}$ by  $\sigma_0$, the identity in Eq.~\eqref{eq:effpot2} does not hold anymore, if $\sigma_0 > \sigma_\text{min}$. For instance, choosing $\sigma_\text{min}  < r_{i_0j_0}  < \sigma_0$ for a fixed pair $(i_0,j_0)$  and $ r_{ij} \geq \epsilon \sigma_\text{min}$ for all other pairs,   there exists a finite domain for the orientational d.o.f.\@, ($\mathbf{u}_{i_0},\mathbf{u}_{j_0}$), such that  the l.h.s. of  Eq.~\eqref{eq:effpot2} equals unity. However, its r.h.s.\@ vanishes because of the factor  $\Theta(r_{i_0j_0} - \sigma_0) = 0$. The identity in Eq.~\eqref{eq:effpot2} is the crucial starting point for a mapping of the hard-body fluid  to a fluid of spherical particles. This demonstrates the particular role of   $ \sigma_\text{min}$.

Due to the elimination of the orientational d.o.f.\@, the effective potential  $V_\text{eff}(\mathbf{r}_1,\ldots, \mathbf{r}_N)$ is isotropic.  Furthermore, $f(ij)$ is only nonzero in the shell 
$\sigma_\text{min} \leq  r_{ij} \leq (1+\epsilon) \sigma_\text{min}$. Therefore, the effective potential 
is nonvanishing only for interparticle distances of the spherical particles  between $\sigma_\text{min}$
and  $(1+\epsilon) \sigma_\text{min}$. 
If two particles are separated by a distance within this shell, we say that there is \emph{bond} between them, 
and a collection of particles with $k$ bonds  where each particle has at least one bond 
is called a $k$-\emph{cluster}.
If each particle has a bond with at most another particle, i.e., there are only single particles or  two-clusters,
the effective potential reduces to $V_\text{eff}(\mathbf{r}_1, \ldots, \mathbf{r}_N) = \sum_{i<j} v^{(2)}_{\text{eff}}({r}_{ij})$ with the effective two-body potential
\begin{align}\label{eq:eff_2_pot}
v^{(2)}_\text{eff}(r) = - k_B T \ln \big[1+\langle f(r\mathbf{e}_{12},\mathbf{u}_1,\mathbf{u}_2) \rangle_o \big] ,
\end{align}
which depends only on the  distance $r$.
Similarly, the higher effective potentials, $v^{(k)}_\text{eff}(r)$ for $k > 2$, can be derived by recursion from Eq.~\eqref{eq:effpot3}, applied to a cluster of $k$ spherical particles and the knowledge of $v^{(m)}_\text{eff}(r)$, \ $2 \leq m \leq k-1$.

The qualitative behavior of $v^{(\text{2})}_\text{eff}(r)$ follows without its explicit calculation. For fixed $r$, $1+\langle f(r\mathbf{e}_{12},\mathbf{u}_1,\mathbf{u}_2) \rangle_o$ is the ratio of the volume in orientational space of $(\mathbf{u}_1,\mathbf{u}_2)$ in which two hard bodies do not overlap to its total volume $(\Omega_D)^2$. For  $r \to  \sigma_\text{min}^{+}$,  orientational d.o.f.\@ become more and more constrained and the ratio goes to zero, implying a logarithmic divergence  of $v^{(2)}_\text{eff}({r}_{ij})$. 
In the opposite limit, $r \to (\epsilon \sigma_\text{min})^{-}$, the orientational d.o.f.\@ become more and more free
such that  the ratio converges to unity, implying  $v^{(2)}_\text{eff}({r}_{ij}) \to 0$.
 Hence, the effective two-body potential interpolates between the hard-core regime $r < \sigma_\text{min}$ and the force-free one $r > \epsilon \sigma_\text{min}$. The qualitative behavior of $v^{(\text{k})}_\text{eff}(\mathbf{r}_1,\dots,\mathbf{r}_{\text{k}})$ for $k > 2$ can be discussed similarly.

Figure~\ref{fig:soft_2_body_potential}  displays the two-body potential for a 2D fluid of ellipses with semiaxes of length $a$ and $b$ and aspect ratio  $X_0 = (a/b) = 1 + \epsilon \geq 1$. We use the scaled distance $\tilde{r}(\epsilon) := (r/2b - 1)/\epsilon$, which  varies between $0$ and $1$ for $ \sigma_\text{min}=2b \leq r \leq \sigma_\text{max}=2a$. The effective two-body potential $v^{(\text{2})}_\text{eff}(r)$  cannot be calculated analytically  for arbitrary aspect ratios. For $X_0=1.1$ and $3.0$, the orientational average in Eq.~\eqref{eq:eff_2_pot}  was performed numerically by   a standard
hit-and-miss Monte Carlo method:  One hard ellipse with a fixed orientation
is placed with its center at the origin, then another one is inserted $N^{(\text{MC})}_\text{t}$ times with its center randomly placed over a circle of radius $r$ and with a random orientation. $r$ is changed linearly in discrete steps. We count  the number of times,  $N_\text{nov}$, where the two hard ellipses do not overlap. Since for given $r$, $f(r\mathbf{e}_{12},\mathbf{u}_1,\mathbf{u}_2)$ equals $0$, if both ellipses do not overlap and $-1$, otherwise, we obtain for the  orientational average $1+\langle f(r\mathbf{e}_{12},\mathbf{u}_1,\mathbf{u}_2) \rangle_o  \approx   N^{(\text{MC})}_\text{nov}/N^{(\text{MC})}_t$. Using this in 
expression~\eqref{eq:eff_2_pot} leads for $N^{(\text{MC})}_\text{t} = 10^9 - 10^{10}$
 to $v^{(2)}_\text{eff}(r)$  shown as  solid red and blue lines for  $X_0=1.1$ and $3.0$, respectively. The corresponding dashed lines represent the asymptotically exact  results, Eqs.~\eqref{eq:area-3} and \eqref{eq:area-2}, for $\tilde{r}(\epsilon) \to 0$ and 
$\tilde{r}(\epsilon) \to 1$, respectively. Comparing the semianalytical results, Eqs.~\eqref{eq:area-1-1} and \eqref{eq:area-1-2}, for 
$X_0 \to 1$  (upper black solid line)  with the results of Eqs.~\eqref{eq:area_1-infinite}-\eqref{eq:area_3-infinite}   for  $X_0 \to \infty$  (lower black dotted line) shows that $v^{(\text{2})}_\text{eff}(r)$ does not depend sensitively on the aspect ratio $X_0$.
There is  another characteristic feature of  $v^{(\text{2})}_\text{eff}(r)$ which can be observed in Fig.~\ref{fig:soft_2_body_potential} for $X_0=1.1$ and $3.0$. In  Appendix~\ref{Sec:Appendix_A} [Eq.~\eqref{eq:def_v2eff-1}], we have shown that the logarithmic term in Eq.~\eqref{eq:eff_2_pot} can be 
  rewritten as $\ln[|\mathcal{D}(r;\epsilon)|/\pi^2]$, where $|\mathcal{D}(r;\epsilon)|$ is the area of the  domain in the two-dimensional orientational space  of both ellipses  in which for given distance $r$ the ellipses do not overlap (cf.  Fig.~\ref{fig:domain}). In  Appendix~\ref{Sec:Appendix_A}, we have also proven that the derivative of 
$|\mathcal{D}(r;\epsilon)|$ with respect to the scaled distance $\tilde{r}(\epsilon) = (r/2b - 1)/\epsilon$ diverges 
at  $\tilde{r}(\epsilon) = 1/2$, corresponding to $r=a+b$ (see  [Eq.~\eqref{eq:area-1-4}]). Because $|\mathcal{D}(r;\epsilon)|$ is finite at $\tilde{r}=1/2$, the derivative of  $\ln[\mathcal{D}(r;\epsilon)|]$ and therefore the derivative of the effective two-particle potential  diverges at $\tilde{r}=1/2$ as well. 
This divergence at  $\tilde{r}(\epsilon) = 1/2$ appears  in Fig.~\ref{fig:soft_2_body_potential}  only  as a kink, due  to the numerical inaccuracy and discreteness of the variable $\tilde{r}$.

\begin{figure}
\includegraphics[angle=0,width=\linewidth]{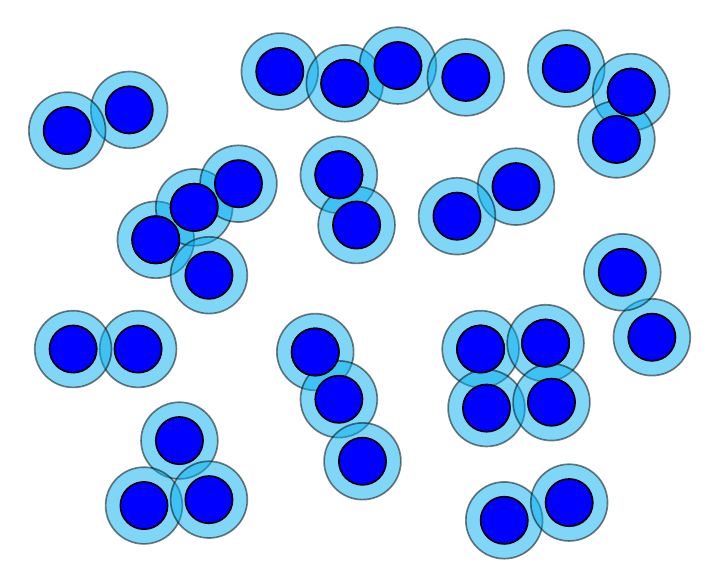}
\caption{Illustration of a cluster configuration of a  two-dimensional fluid of spherical particles. It consists 
of several $k$ clusters with $k=2,3$ and $4$. For $k > 2$, various types exist  with a  different number of 
bonds (number of overlaps).  
Their hard core of diameter $\sigma_\text{min}$ is shown in dark blue and the light blue circle with a diameter
$\sigma_\text{max}$ marks the range above which the excluded volume interactions vanish. Clusters occur if spherical shells (shown in light blue) overlap. Their interaction energy is given be the effective  potential $V_\text{eff}(\mathbf{r}_1, \ldots, \mathbf{r}_N)$.  }
\label{fig:cluster_expansion}
\end{figure}

The reduction  to a fluid of spherical particles with a hard core of diameter $\sigma_\text{min}$ and a soft shell of thickness $\epsilon \sigma_\text{min}/2$ has two implications. First, since the domain of the nonzero effective potential shrinks to zero for $\epsilon \to 0$, i.e., the ``perturbation'' of the reference fluid of spherical hard spheres of diameter $\sigma_\text{min}$ goes to zero, we can already predict that a 
phase transition point of a fluid of  convex nonspherical hard bodies will change linearly with $\epsilon = (\sigma_\text{max}/\sigma_\text{min} -1)$, provided the shape anisotropy is small  
and  the transition point is analytic in $\epsilon$. For instance,  the isotropic-hexatic, the  hexatic-solid phase transition of a 2D fluid of hard ellipses and the isotropic-plastic crystal phase transition of ellipses in 2D and ellipsoids of revolution in 3D should shift linearly with $\epsilon$.
Second, the calculation of thermodynamic quantities for the original fluid of convex nonspherical  hard bodies is  reduced to the calculation of (i) the free energy of the reference fluid and (ii) contributions of  the cluster configurations (two-clusters, three-clusters, etc.) as illustrated in Fig.~\ref{fig:cluster_expansion}. The energy of such a cluster configuration follows from the effective potential, Eq.~\eqref{eq:effpot3}. 
Summing over all such cluster configurations yields the configurational part of the canonical partition function of the original fluid. This second step will be performed in the following subsection. 
Finally, we stress that  the present perturbative approach to eliminate the orientational d.o.f.\@ can also be  extended  to liquids of rigid molecules with a smooth pair potential.

\subsection{Free energy}\label{Sec:free_energy}

The Helmholtz  free energy 
\begin{align}\label{eq:free-energy}
F(T,V,N;\epsilon) = F_{\text{id}}(T,V,N) + F_{\text{ex}}(T,V,N;\epsilon) 
\end{align}
of  a  D-dimensional fluid of $N$ hard nonspherical particles in a volume $V$ with number density $n=N/V$  consists of the ideal gas contribution $ F_{\text{id}}(T,V,N) = Nk_BT \ln\left[n (\lambda_t)^D \mu_o \right]$ and the excess free energy
\begin{align}\label{eq:excess_free1} 
F_{\text{ex}}(T,V,N;\epsilon) = - k_BT \ln Z_{\text{ex}}(T,V,N;\epsilon)  .
\end{align}
Here, $\lambda_t = \sqrt{2 \pi \hbar^2/ m k_B T}$ is the thermal wavelength of the translational d.o.f.\@ and $\mu_o$ is the corresponding contribution from  the  orientational d.o.f.\@, involving the moments of inertia of the hard body. The excess free energy, Eq.~\eqref{eq:excess_free1}, follows from the excess canonical partition function
\begin{align}\label{eq:ex-partition1}
\lefteqn{Z_{\text{ex}}=  Z_{\text{ex}}(T,V,N;\epsilon)= 
} \nonumber \\
&= \int   \Big[\prod_{i=1}^N\frac{\diff \mathbf{u}_i}{\Omega_D} 
\frac{\diff \mathbf{r} _i}{V} \Big] \exp[- \beta V( \mathbf{r}_1\mathbf{u}_1, \dots,\mathbf{r}_N\mathbf{u}_N)] \ .
\end{align}
Now we will perform a cluster expansion $F_\text{ex}(T,V,N;\epsilon)= \sum_{k=0}^{\infty} F_k(T,V,N;\epsilon)$ of the excess free energy 
with  $\epsilon$ as the smallness parameter. Due to the mapping of the hard-body fluid to a fluid of spherical particles, the terms $F_k(T,V,N;\epsilon)$ ($k \geq 0$) acquire a clear physical interpretation. The  zeroth order term, $F_0(T,V,N)$, is the excess free energy of the reference fluid consisting of hard spheres of diameter  $\sigma_\text{min}$, and $F_k(T,V,N;\epsilon)$ for  $k \geq 1$ is the contribution of all cluster configurations  with $k$ bonds, i.e., $k$ overlapping soft shells (cf. Fig.~\ref{fig:cluster_expansion}). 
Note that this physical picture no longer  involves  the orientational d.o.f.\@. It demonstrates that   it is the thickness of the soft shell of the spherical particles which becomes  the perturbation of the reference fluid and not an anisotropic potential as in, e.g., Refs.~\cite{Gray:Molecular_Fluids:1984,Solana:Perturbation_Theory_Fluids:2013,Belleman:PRL_21:1968}. 
Making use of Eq.~\eqref{eq:effpot2} and suppressing the
$\epsilon$ dependence,  Eq.~\eqref{eq:ex-partition1} assumes the form
\begin{align}\label{eq:ex-partition2}
Z_{\text{ex}} 
&= Z^{(0)}_{\text{ex}} \ \langle \langle \prod_{i< j} [1+ f(ij) ] \rangle_o \rangle_t ,
\end{align}
with the excess partition function 
\begin{align}
Z^{(0)}_{\text{ex}} = \int  \Big[\prod_{i=1}^N  
\frac{\diff \mathbf{r} _i}{V} \Big] \exp[- \beta V_0(\mathbf{r}_1,\ldots,\mathbf{r}_N)] 
\end{align}  
of the reference fluid and
the average with respect to the canonical ensemble of the reference fluid:
\begin{align}\label{eq:ex-partition3}
\langle (\cdots) \rangle_t 
 &=  \frac{1}{Z^{(0)}_{\text{ex}}}    \int \Big[\prod_{i=1}^N 
\frac{\diff \mathbf{r} _i}{V} \Big] \exp[- \beta V_0(\mathbf{r}_1,\dots,\mathbf{r}_N) \ (\cdots) \ .
\end{align}
Substituting $Z_{\text{ex}}$ from Eq.~\eqref{eq:ex-partition2} into Eq.~\eqref{eq:excess_free1}, the cluster expansion of $F_\text{ex}(T,V,N)$ reads 
\begin{align}\label{eq:free_energy_expan}
F_\text{ex}(T,V,N) = \sum_{k=0}^\infty F_k(T,V,N) , 
\end{align}
with
\begin{align}\label{eq:free_energy_expan_0}
F_0(T,V,N) = -k_BT \ln Z^{(0)}_{\text{ex}}(T,V,N), 
\end{align}
the excess free energy of the reference fluid  and the cluster contributions $ F_k(T,V,N)$, $k \geq 1$  of  order 
$O(\epsilon^k)$. Note, $\epsilon$ is the anisotropy parameter introduced in Eq.~\eqref{eq:shape}. 
For ellipses and ellipsoids of revolution,  $\epsilon = X_0 -1$ with $X_0$ the aspect ratio.
Explicitly, $F_k$ for $k=1$ and $2$ are provided by
\begin{align}\label{eq:free_energy_expan_1} 
 F_1(T,V,N) = - k_BT \sum_{i < j} \langle \langle f(ij) \rangle_o \rangle_t ,
\end{align} 
\begin{align}\label{eq:free_energy_expan_2} 
\lefteqn{ F_2(T,V,N) \equiv  F^{3}_2(T,V,N) +  F^{2-2}_2(T,V,N) =}   \nonumber \\
=& - 3k_BT \Big\{\sum_{i < j< k} \big[\langle \langle f(ij) f(jk) \rangle_o \rangle_t - \langle \langle f(ij) \rangle_o \rangle_t \langle \langle f(jk) \rangle_o \rangle_t\big]  \nonumber \\
& + \sum_{i < j< k<l} \big[\langle \langle f(ij) f(kl) \rangle_o \rangle_t - \langle \langle f(ij) \rangle_o \rangle_t \langle \langle f(kl) \rangle_o \rangle_t\big]\Big\} \ .
\end{align} 
Also, terms such as $\sum_{i < j} \big(\langle \langle f(ij) \rangle_o \rangle_t\big)^2$ and $\sum_{i < j< k} \langle \langle f(ij)\rangle_o \rangle_t\langle \langle f(jk) f(ji) \rangle_o \rangle_t$ occur. They are of order $N^0$ and do not contribute in the thermodynamic limit. 

The contributions in Eqs.~\eqref{eq:free_energy_expan_1} and \eqref{eq:free_energy_expan_2} possess physical interpretations. The first correction, $F_1$, is the contribution of independent two-clusters. The first line of Eq.~\eqref{eq:free_energy_expan_2}, denoted  by 
$F^3_2$, arises from three-clusters with two bonds (see Fig.~\ref{fig:cluster_expansion}) and its second line is the contribution , $F^{2-2}_2$, of pairs of two-clusters. These pairs are correlated due to the hard-sphere interaction in the reference fluid.

Note that although the form of the results, Eqs.~\eqref{eq:free_energy_expan_1} and \eqref{eq:free_energy_expan_2}, resemble the standard virial expansion, our approch is qualitatively different. First, the averages 
$\langle (\ldots) \rangle_t$ over the translational d.o.f.\@ are performed for the reference fluid which is correlated. Second, in the standard virial series the first and second lines in Eq.~\eqref{eq:free_energy_expan_2}
do not exist  since they  correspond to a reducible cluster.

Making use of Eqs.~\eqref{eq:contact} and \eqref{eq:cluster-1},    one can express  averages of products of the cluster functions by  products of $\tilde{d}_\text{c}(\mathbf{e},\mathbf{u}_1,\mathbf{u}_2;\epsilon)$ averaged over the orientational d.o.f.\@  and $n$-particle distribution functions $g^{(n)}(\mathbf{r}_1,\dots,\mathbf{r}_n)$ . Using the results from 
Appendix~\ref{Sec:Appendix_B}, introducing the dimensionless density $n^*=n\sigma_{\text{min}}^D$ and 
the average over the \textit{bond} orientations $(\mathbf{e}_{ij})$:
\begin{align}\label{eq:bond-av} 
\overline{(\cdots)} = \int_{\Omega_D} \prod_{i<j} \frac{\diff \mathbf{e}_{ij}}{\Omega_D}  \ (\cdots)    ,
\end{align} 
we find in leading order in $\epsilon$ the contribution of a two-cluster 
\begin{align}\label{eq:free_energy_expan_1a} 
\lefteqn{ F_1(T,V,N) =} \nonumber \\
& = \frac{1}{2} N  k_BT  \Big[ \Omega_D  n^* g^{(2)}(\sigma_\text{min}) \overline{\langle \tilde{d}_c(\mathbf{e}_{12},\mathbf{u}_{1},\mathbf{u}_{2};0) \rangle_o  } \Big]  \epsilon \ + O(\epsilon^2),
\end{align} 
and that of a pair of two-clusters 
\begin{align}\label{eq:free_energy_expan_2a} 
\lefteqn{ F^{2-2}_2(T,V,N) = } \nonumber \\
=& - \frac{1}{8} N k_BT  \left[\Omega_D  n^* g^{(2)}(\sigma_\text{min})  \overline{\langle \tilde{d}_c(\mathbf{e}_{12},\mathbf{u}_{1},\mathbf{u}_{2};0) \rangle_o}  \right]^2   \nonumber\\  & \times   \  \left[\Omega_D  n  \, \overline{\xi(\mathbf{e}_{12},\mathbf{e}_{34},\mathbf{e}_{13})^D}\right] \epsilon^2   + O(\epsilon^3) \ ,
\end{align} 
where $\xi(\mathbf{e}_{12},\mathbf{e}_{34},\mathbf{e}_{13})$ is the correlation length of a 
correlation function  $ g^{(4)}(\sigma_\text{min},\sigma_\text{min},r_{13};\mathbf{e}_{12},\mathbf{e}_{34},\mathbf{e}_{13})$ [cf. Eq.~\eqref{eq:corr-length}]. 
Hence, $\xi(\mathbf{e}_{12},\mathbf{e}_{34},\mathbf{e}_{13})$ is a measure of the  correlation between the \textit{bond} orientations $\mathbf{e}_{12}$ and $\mathbf{e}_{34}$ of a pair of two-clusters
with bond-length $\sigma_\text{min}$ separated by $\mathbf{r}_{13} =r_{13}\mathbf{e}_{13}$.  Note that $ g^{(4)}(\sigma_\text{min},\sigma_\text{min},r_{13};\mathbf{e}_{12},\mathbf{e}_{34},\mathbf{e}_{13})$ resembles the correlation function of the local hexatic order parameter $\Psi(\mathbf{r_i}), i = 1,3$ at $\mathbf{r}_{1}$ and $\mathbf{r}_{3}$~\cite{Strandburg:RMP_60:1988}. 
For the contribution of a three-cluster with two bonds, we find 
\begin{widetext}
\begin{align}\label{eq:free_energy_expan_2b} 
 F^{3}_2(T,V,N) =& - \frac{1}{2}N  k_BT \big(\Omega_D  n^*\big)^2   \Big\{\overline{ g^{(3)}(\sigma_\text{min},\sigma_\text{min}; \mathbf{e}_{12} \cdot \mathbf{e}_{23})
\langle \tilde{d}_c(\mathbf{e}_{12},\mathbf{u}_{1},\mathbf{u}_{2};0)
\tilde{d}_c(\mathbf{e}_{23},\mathbf{u}_{2},\mathbf{u}_{3};0) \rangle_o  }  \nonumber\\
& \ - \left[  g^{(2)}(\sigma_\text{min}) \overline{\langle \tilde{d}_c(\mathbf{e}_{12},\mathbf{u}_{1},\mathbf{u}_{2};0) \rangle_o }\right]^2 \Big\} \ \epsilon^2 \ 
+O(\epsilon^3) \ .
\end{align} 
\end{widetext}

Let us comment on the general structure of $F_k(T,V,N;\epsilon)$, where the dependence
on  $\epsilon$ is made explicit.  It is given by a power series: 
$\sum^{\infty}_{l=0} F^{(l)}_k(T,V,N) \ \epsilon^{k+l}$.
Its leading order term, $F^{(0)}_k(T,V,N)$,  consists of products  involving   $\langle \tilde{d}_c(\mathbf{e}_{12},\mathbf{u}_{1},\mathbf{u}_{2};0) \tilde{d}_c(\mathbf{e}_{23},\mathbf{u}_{2},\mathbf{u}_{3};0)\ldots) \rangle_o$ and the   $m$-particle distribution functions 
$g^{(m)}(\mathbf{r}_1,\mathbf{r}_2,\mathbf{r}_3,\ldots) $ ($k \leq m \leq 2k$). These products are   averaged over the bond orientations $(\mathbf{e}_{12},\mathbf{e}_{23},\ldots)$. The higher-order corrections, $F^{(l)}_k(T,V,N)$, for  $l > 0$ involve derivatives 
$\partial^{\nu} \tilde{d}_c(\mathbf{e}_{ij},\mathbf{u}_{i},\mathbf{u}_{j};\epsilon)/\partial \epsilon^{\nu} |_{\epsilon =0}$  and  $\partial^{\mu}g^{(m)}(\mathbf{r}_1,\mathbf{r}_2,\dots)/ \partial (n^*)^{\mu}$. An explicit example is given in Appendix~\ref{Sec:Appendix_B} for $k$=1 and $l=1$. 
Using the series exansion 
 $ \Theta \big[(r_{ij} - \sigma_\text{min}) - \Delta\big] = \Theta(r_{ij} - \sigma_\text{min}) - \delta(r_{ij} - \sigma_\text{min})\Delta  +  \delta'(r_{ij} - \sigma_\text{min}) \Delta^2/2  +  O(\Delta^3)$
(see also Ref.~\cite{Belleman:PRL_21:1968})  with
 $\Delta =(\epsilon  \sigma_\text{min})\tilde{d}_c(\mathbf{e}_{ij},\mathbf{u}_{i},\mathbf{u}_{j}; \epsilon)$ as a perturbation, one also obtains $F_k(T,V,N;\epsilon)$.

To summarize this section, the calculation of the free energy of the fluid of hard nonspherical particles has been reduced to the calculation of (i) the $m$-particle distribution function $g^{(m)}$  of the reference fluid of hard spheres (or disks  in case of a 2D fluid) of diameter $\sigma_\text{min}$ and (ii)  orientational averages of  products of the  dimensionless contact function $\tilde{d}_c(\mathbf{e}_{ij},\mathbf{u}_i,\mathbf{u}_j;\epsilon)$. 
The dependence on the  density $n^*$ comes only from $g^{(m)}$ and its derivatives with respect to $n^*$. 
The contact function is a purely geometrical quantity, completely  independent on the thermodynamic variables $(T,V,N)$. 
Even if the overlap functions was known, the analytical calculation of  $\tilde{d}_c(\mathbf{e}_{ij},\mathbf{u}_i,\mathbf{u}_j;\epsilon)$ would  not be feasable. On the other hand, the two-body quantity $\tilde{d}_c(\mathbf{e}_{ij},\mathbf{u}_i,\mathbf{u}_j;\epsilon)$ is  determined by the shape function  $s(\vartheta;\epsilon)$, a  one-body entity. In the next section,  we will present   a perturbative method  allowing us to express the contact function by the shape function. Using these results,  the perturbative expansion of the free energy with
$\epsilon\tilde{d}_c(\mathbf{e}_{ij},\mathbf{u}_i,\mathbf{u}_j;\epsilon)$  as a ``perturbation'' turns into  an expansion 
with respect to the shape anisotropy $\epsilon\tilde{s}(\vartheta)$ [Eq.~\eqref{eq:shape}].

\section{Contact function: shape dependence}\label{Sec:contact-f}

In this sectionm it is necessary to make the dependence of the various quantities on the shape anisotropy $\epsilon$ explicit.

As  shown in  Sec.~\ref{Sec:free_energy}, the calculation  of the free energy, and therefore
of thermodynamic quantities, requires  knowledge of the reduced contact function $\tilde{d}_c(\mathbf{e}_{12},\mathbf{u}_{1},\mathbf{u}_{2};\epsilon)$ [Eq.~\eqref{eq:contact}].
 The calculation of this function by solving  Eq.~\eqref{eq:overlap-0} for two  convex hard nonspherical bodies is a highly intricate mathematical  problem which has attracted a lot of attention over decades (see, e.g., Refs.~\cite{Vieillard-Baron:JCP_56:1972, Perram:JCompPhys_58:1985, Zheng:PRE_75:2007, Zheng:PRE_79:2009}). Although  an analytical closed expression for an overlap function $\psi(d\mathbf{e}_{12},\mathbf{u}_1,\mathbf{u}_2;\epsilon) $ exists for a pair of hard ellipses, an analytical  solution of
 $\psi(d\mathbf{e}_{12},\mathbf{u}_1,\mathbf{u}_2;\epsilon) = 0$ is possible, in principle. The square of $\tilde{d}_c(\mathbf{e}_{12},\mathbf{u}_{1},\mathbf{u}_{2};\epsilon)$ is one of the four  roots of a quartic equation. Since these roots are more involved, it  may  not be easy to figure out which of these roots is the physically correct one.More important, the overlap function
for ellipses is the only analytically known example. Therefore, it is highly desirable to  elaborate on a quite different approach than in the past. This becomes possible using the contact conditions for two convex hard bodies with a \textit{smooth} surface which \textit{directly} involve the shape function $s(\vartheta,\varphi;\epsilon)$  .  Although this can be done for arbitrary  hard bodies, we will restrict ourselves again to bodies with at least one rotational axis such that the shape function itself does not depend on the azimuth angle  $\varphi$.

The derivation of the contact conditions uses elementary tools of differential geometry of curved surfaces embedded in  3D space  or curved lines in 2D space. This first step is straightforward and will be described for bodies in 3D. The solution of the contact conditions yields the contact point parametrized by the angles 
$\big(\vartheta(\mathbf{e}_{12},\mathbf{u}_{1},\mathbf{u}_{2};\epsilon) ,\varphi(\mathbf{e}_{12},\mathbf{u}_{1},\mathbf{u}_{2};\epsilon)\big)$,  and the contact function  $d_c(\mathbf{e}_{12},\mathbf{u}_{1},\mathbf{u}_{2};\epsilon)$ for given orientations  $\mathbf{u}_{1}$, $\mathbf{u}_{2}$ and direction $\mathbf{e}_{12}$ of the center-to-center vector.
 Of course, the equations for the  contact conditions cannot be solved, in general. But, in a second step, we will develop a perturbative method which will allow us to solve these equations iteratively,
 even for the generic family of Eq.~\eqref{eq:shape} of  convex hard bodies.  For the second step (resembling the standard perturbation theory in quantum mechanics),   we will describe  the essential conceptual steps only. Details can be found in  Appendix~\ref{Sec:Appendix_C}.

\subsection{Contact conditions}\label{Sec:contact-conditions}

 As stressed in Sec.~\ref{Sec:model}, the contact conditions for convex hard bodies of revolution are completely determined by the 
 shape function $s(\vartheta;\epsilon)$ characterizing the set $S(\mathbf{u};\epsilon)$ [Eq.~\eqref{eq:surface}].  Abbreviating  $\omega = (\vartheta,\varphi)$, a point on the surface  of the body with orientation $\mathbf{u}$ is provided by  
 \begin{align}\label{eq:vec-smain}  
 \mathbf{s}(\omega,\mathbf{u};\epsilon)  = s(\vartheta;\epsilon) R(\mathbf{u})\mathbf{e}_s(\omega),
 \end{align} 
with $\mathbf{e}_s(\omega) = 
 ( \sin \vartheta  \cos\varphi, \sin\vartheta \sin\varphi, \cos\vartheta)^T$. The unit vector  $\mathbf{e}_s(\omega)$
together with  $\mathbf{e}_{\vartheta}(\omega)= \partial \mathbf{e}_s(\omega)/ \partial  \vartheta$ and  $\mathbf{e}_{\varphi} (\omega) = (\partial \mathbf{e}_s(\omega)/ \partial  \varphi)/ \sin\vartheta$ build a local  orthonormal basis. 

If two such convex hard bodies with orientations $\mathbf{u}_i$, $i = 1,2$ and a center-to-center separation  vector $\mathbf{d}_{12}= d \, \mathbf{e}_{12}$ are in an exterior  tangential contact,  their two centers and the contact point  form a triangle. Its side vectors  are provided by  $ \mathbf{s}(\omega_i,\mathbf{u}_i;\epsilon)$, $i = 1,2$, 
and  $d_c(\mathbf{e}_{12},\mathbf{u}_1,\mathbf{u}_2;\epsilon) \mathbf{e}_{12}$.
Therefore, the first contact condition reads
\begin{align}\label{eq:contact-condition-main-1} 
\mathbf{s}(\omega_1,\mathbf{u}_1;\epsilon) & = d_c( \mathbf{e}_{12},\mathbf{u}_1,\mathbf{u}_2;\epsilon) \ \mathbf{e}_{12} +  \mathbf{s}(\omega _2,\mathbf{u}_2;\epsilon)  .
\end{align}  
Due to the  tangential contact, the tangential planes of both hard bodies at the contact point are identical, however, with normal vectors $\mathbf{n}(\omega _i,\mathbf{u}_i;\epsilon) $ pointing in  opposite directions. Accordingly, the second contact condition becomes 
\begin{align}\label{eq:contact-condition-main-2} 
\mathbf{n}(\omega _1,\mathbf{u}_1;\epsilon) & = - \  \mathbf{n}(\omega _2,\mathbf{u}_2;\epsilon)  \ .
\end{align}    
 The normal vector
 $\mathbf{n}(\omega _i,\mathbf{u}_i;\epsilon) = \mathbf{t}_{\vartheta}(\omega _i,\mathbf{u}_i;\epsilon) \times \mathbf{t}_{\varphi}(\omega _i,\mathbf{u}_i;\epsilon)$  
 at the surface point $\mathbf{s}(\omega_i,\mathbf{u}_i;\epsilon)$ is the cross product of  
the orthonormal  tangential vectors
$\mathbf{t}_{\vartheta}(\omega _i,\mathbf{u}_i;\epsilon)= (\partial  \mathbf{s}/ \partial \vartheta)/| \partial  \mathbf{s}/ \partial \vartheta|(\omega _i,\mathbf{u}_i;\epsilon)$ and 
$\mathbf{t}_{\varphi}(\omega _i,\mathbf{u}_i;\epsilon) = (\partial  \mathbf{s}/ \partial \varphi )/| \partial  \mathbf{s}/ \partial \varphi|(\omega _i,\mathbf{u}_i;\epsilon)$.
As shown in Appendix  \ref{Sec:Appendix_C}, it is determined by 
 \begin{align}\label{eq:normal-vector-main} 
\mathbf{n}(\omega _i,\mathbf{u}_i;\epsilon) & = R(\mathbf{u}_i) \frac{ - \partial s/\partial \vartheta_i(\vartheta_i;\epsilon)\mathbf{e}_{\vartheta}(\omega_i) + s(\vartheta_i;\epsilon)\mathbf{e}_{s}(\omega_i) }{\sqrt{(\partial s/\partial \vartheta_i (\vartheta_i;\epsilon))^2 + ( s(\vartheta_i;\epsilon))^2}} \ .
\end{align}   
Given the shape function $s(\vartheta;\epsilon)$ and  $(\mathbf{e}_{12},\mathbf{u}_1,\mathbf{u}_2)$,  the solution of Eqs.~\eqref{eq:contact-condition-main-1} and  \eqref{eq:contact-condition-main-2} 
together with Eqs.~\eqref{eq:vec-smain} and \eqref{eq:normal-vector-main} yields the polar angles 
$\omega_i(\mathbf{e}_{12},\mathbf{u}_1,\mathbf{u}_2;\epsilon)$ (corresponding to the contact point)  and the contact function $d_c(\mathbf{e}_{12},\mathbf{u}_1,\mathbf{u}_2;\epsilon)$ as a functional of $s(\vartheta;\epsilon)$. Due to the overall rotational invariance of the system, these three functions will only depend on the scalars 
$\mathbf{e}_{12}\cdot \mathbf{u}_1$, $\mathbf{e}_{12}\cdot \mathbf{u}_2$ and  $\mathbf{u}_1\cdot \mathbf{u}_2$.

\subsection{Perturbative solution}\label{Sec:perturb}

As already stressed above, Eqs.~\eqref{eq:contact-condition-main-1} and \eqref{eq:contact-condition-main-2}  cannot  be solved analytically, in general.  For spherical bodies of diameter $\sigma_\text{min}$, their solution is obvious. With respect to the origin of the first  sphere,  the contact point is at    $(\sigma_\text{min}/2) \mathbf{e}_{12}$ such that $d_c(\mathbf{e}_{12},\mathbf{u}_1,\mathbf{u}_2;\epsilon=0) = \sigma_\text{min}$. The latter is 
independent on $(\mathbf{e}_{12},\mathbf{u}_1,\mathbf{u}_2)$.  Deforming the spherical shape smoothly, as described by the family of shapes in Eq.~\eqref{eq:shape},   the contact point and the contact distance $d_c$ will change smoothly. Therefore, we elaborate on a perturbative approach using the strength of deformation, $\epsilon$, as a smallness parameter.  Note the smoothness concerns only the dependence of $s$ on  $\epsilon$ but not yet its dependence on $\vartheta$. This motivates a Taylor expansion of  $\omega_i(\mathbf{e}_{12},\mathbf{u}_1,\mathbf{u}_2;\epsilon)$ and $d_c(\mathbf{e}_{12},\mathbf{u}_1,\mathbf{u}_2;\epsilon)$ with respect to $\epsilon$. Suppressing  the dependence  on $(\mathbf{e}_{12},\mathbf{u}_1,\mathbf{u}_2)$,  we use 
\begin{align}\label{eq:omega-series-main} 
 \omega_i( \epsilon)  = \sum^{\infty}_{\nu=0}  \omega^{(\nu)}_i   \  \epsilon^{\nu}       ,  
 \qquad  \omega^{(\nu)}_i=( \vartheta^{(\nu)}_i, \varphi^{(\nu)}_i)          , 
\end{align}  
and 
\begin{align}\label{eq:d-series-main} 
d_c(\epsilon) = \sigma_\text{min} \big[1 + \sum^{\infty}_{\nu=1}  \tilde{d}_{\nu} \epsilon^{\nu}\big]   ;
\end{align}   
compare Eq.~\eqref{eq:contact}.
Including ellipsoids of revolution and ellipses with  aspect ratio
$X_0= 1 + \epsilon$, we also introduce the Taylor expansion of $s(\vartheta;\epsilon)$,
\begin{align}\label{eq:s-series-main} 
 s(\vartheta;\epsilon) = \frac{\sigma_\text{min}}{2} \big[1 + \sum^{\infty}_{\nu=1}  \tilde{s}_{\nu} (\vartheta)  \  \epsilon^{\nu} \big]    ;
\end{align} 
 compare Eq.~\eqref{eq:shape}.  
 For the generic family of Eq.~\eqref{eq:shape} of shapes, this reduces to  $\tilde{s}_{\nu} (\vartheta) \equiv \tilde{s}(\vartheta) \delta_{\nu,1}$ as discussed in Sec.~\ref{Sec:model}. Note for $\epsilon=0$, Eqs.~\eqref{eq:s-series-main}  and ~\eqref{eq:d-series-main}  reduce to  the body's radius $\sigma_\text{min}/2$ and contact function $d_c(\epsilon=0) = \sigma_\text{min}$ of hard spheres of diameter   $\sigma_\text{min}$, respectively.

To perform the perturbative analysis, we have to replace $ \omega_i$ in Eqs.~\eqref{eq:contact-condition-main-1} and \eqref{eq:contact-condition-main-2}
by $ \omega_i(\epsilon) $.
Then, using series Eqs.~\eqref{eq:omega-series-main}, \eqref{eq:d-series-main},  and  \eqref{eq:s-series-main}),   both sides of Eqs.~\eqref{eq:contact-condition-main-1}   and  \eqref{eq:contact-condition-main-2} are expanded  with respect to $\epsilon$. Comparing the coefficients of terms of order $\epsilon^{\mu}$ on both sides leads to relations between  
$\big( \omega^{(0)}_i, \omega^{(1)}_i, \omega^{(2)}_i,\ldots\big)$,  $\big(\tilde{d}_{1},\tilde{d}_{2},\tilde{d}_{3},\ldots \big)$,  $\big(\tilde{s}_{1},\tilde{s}_{2},\tilde{s}_{3},\ldots \big)$  and the derivatives of $ \tilde{s}_{\mu}(\vartheta)$ for $\mu \geq 1$. For generic bodies of revolution, only $\tilde{s}(\vartheta)$ and its derivatives
 occur.
This procedure has been  elaborated in Appendix \ref{Sec:Appendix_C} up to the first. order in $\epsilon$. 
The zeroth order yields $\omega^{(0)}_i(\mathbf{e}_{12},\mathbf{u}_1,\mathbf{u}_2)$ and the first order determines, besides $\omega^{(1)}_i(\mathbf{e}_{12},\mathbf{u}_1,\mathbf{u}_2)$, the linear order contribution $\tilde{d}_{1}(\mathbf{e}_{12},\mathbf{u}_{1},\mathbf{u}_{2})$ of the  contact function. Only the latter enters  the thermodynamic quantities.
It is provided by 
\begin{align}\label{eq:d-expan-0} 
\tilde{d}_{1}(\mathbf{e}_{12},\mathbf{u}_{1},\mathbf{u}_{2}) =&  \frac{1}{2}  \big[\tilde{s}_1\big( \arccos(\mathbf{e}_{12} \cdot \mathbf{u}_1) \big)   \nonumber\\ 
& \ + \tilde{s}_1\big(\arccos( - \  \mathbf{e}_{12} \cdot \mathbf{u}_2)\big) \big]   .
\end{align}  

For an ellipse, $\tilde{s}_{\nu}(\vartheta)$
can be calculated  by substituting  $x(\vartheta;\epsilon) = s(\vartheta;\epsilon) \sin\vartheta$ and
 $z(\vartheta;\epsilon) = s(\vartheta;\epsilon) \cos\vartheta$ into the quadratic equation $(x/b)^2 + (z/a)^2 = 1$ and using Eq.~\eqref{eq:s-series-main} with $\sigma_\text{min} = 2b$ (the length of the minor axis) and $X_0= a/b= 1 + \epsilon$.
For ellipsoids of revolution, one proceeds similarly.  We find for ellipses and prolate ellipsoids 
\begin{align}\label{eq:s1-prolate} 
\tilde{s}_{1}(\vartheta)   = \cos^2\vartheta  
\end{align} 
and
\begin{align}\label{eq:s1-oblate} 
\tilde{s}_{1}(\vartheta)   = 1 -\cos^2 \vartheta  
\end{align} 
for oblate ellipsoids. Then, Eq.~\eqref{eq:d-expan-0} yields, e.g., for ellipses and prolate ellipsoids  with Eq.~\eqref{eq:s1-prolate}:
\begin{align}\label{eq:d-expan-0-ellip} 
\tilde{d}_{1}(\mathbf{e}_{12},\mathbf{u}_{1},\mathbf{u}_{2}) = \frac{1}{2} 
\big[ (\mathbf{e}_{12} \cdot \mathbf{u}_{1})^2 \ + \ ( \mathbf{e}_{12} \cdot \mathbf{u}_{2})^2 \big]  .
\end{align} 
Solving  $\psi(d\mathbf{e}_{12},\mathbf{u}_1,\mathbf{u}_2;\epsilon)  = 0$ with the overlap function for ellipses [Eqs.~\eqref{eq:overlap}, \eqref{eq:g-alpha}, \eqref{eq:G}], one finds 
$d_c(\mathbf{e}_{12},\mathbf{u}_{1},\mathbf{u}_{2};\epsilon) = 2b[1+\epsilon \tilde{d}_{1}(\mathbf{e}_{12},\mathbf{u}_{1},\mathbf{u}_{2}) + O(\epsilon^2)]$  with $\tilde{d}_{1}(\mathbf{e}_{12},\mathbf{u}_{1},\mathbf{u}_{2})$ identical to the result  of Eq.~\eqref{eq:d-expan-0-ellip}, derived from Eq.~\eqref{eq:d-expan-0}  for general shapes.

The general result of Eq.~\eqref{eq:d-expan-0}  demonstrates that the contact function for \textit{generic} hard bodies of revolution in leading order in the anisotropy parameter $\epsilon$ is completely determined by the dimensionless shape function $\tilde{s}(\vartheta)$ since  $\tilde{s}_1(\vartheta) \equiv \tilde{s}(\vartheta)$.
The coefficients $\tilde{d}_{\nu}$, $\nu > 1$ will also  involve  derivatives  of $\tilde{s}(\vartheta)$. Since the derivative's order increases with $\nu$,   an increasing degree of smoothness of the shape function is required.
The dependence of the first-order  coefficient [Eq.~\eqref{eq:d-expan-0}]  on the orientations $(\mathbf{u}_{1},\mathbf{u}_{2})$ is additive. The coefficients for $\nu > 1$, however,   will also depend on  $(\mathbf{u}_{1} \cdot \mathbf{u}_{2})$, i.e., on the ``interaction'' between both orientations. Although our focus is not on the isotropic-nematic phase transition, this may imply that the study of nematic order requires taking into account higher-order terms in the series expansions, Eqs.~\eqref{eq:omega-series-main}, ~\eqref{eq:d-series-main}, and also \eqref{eq:s-series-main}, for the case of ellipses or ellipsoids of revolution.

\section{Equation of state}\label{Sec:e.o.s.}
To illustrate the application of the cluster expansion presented in Sec.~\ref{Sec:Model_and_Framework}, we derive the equation of state (e.o.s.) to leading order in the shape anisotropy $\epsilon$. We will make the dependence on $\epsilon$ and on the dimensionless density $n^*$ explicit.

\subsection{Equation of state:  General shapes, ellipses, and ellipsoids}\label{Sec:p-general}

The result in Eq.~\eqref{eq:d-expan-0} allows us to calculate the leading-order contribution of $F_1$. 
First, we elaborate on the case of  a generic hard body of revolution.
Since then $ \tilde{s}_1 =  \tilde{s}$ and  $\tilde{d}_{c}(\mathbf{e}_{12},\mathbf{u}_{1},\mathbf{u}_{2};0)   \equiv \tilde{d}_{1}(\mathbf{e}_{12},\mathbf{u}_{1},\mathbf{u}_{2})$, we obtain from Eq.~\eqref{eq:d-expan-0} 
 and 
$\langle
 \tilde{s}(\arccos(\mathbf{e}_{12}  \cdot \mathbf{u}_{1}))  \rangle_o = \langle  \tilde{s}(\arccos(- \mathbf{e}_{12}  \cdot \mathbf{u}_{2}))  \rangle_o$    
$ \overline{\langle \tilde{d}_c(\mathbf{e}_{12},\mathbf{u}_{1},\mathbf{u}_{2};0) \rangle_o } = \langle  \tilde{s}(\arccos(\mathbf{e}_{12}  \cdot \mathbf{u}_{1})) 
\rangle_o $.  
Substituting into Eq.~(\ref{eq:free_energy_expan_1a}) leads to
\begin{align}\label{eq:free_energy_expan_1b} 
F_1(T,V,N;\epsilon)  =&  N  k_BT  \frac{\Omega_D}{2}   \ n^*  g^{(2)}(\sigma_\text{min};n^*)    \nonumber\\
& \times \langle \tilde{s}\big(\arccos(\mathbf{e}_{12} \cdot \mathbf{u}_1)\big)  \rangle_o \   \epsilon     +  
O(\epsilon^2)    .
\end{align}  
We remind the reader here and below  that $n^*=n (\sigma_\text{min})^D$ is the dimensionless density of the reference fluid.
Using Eq.~\eqref{eq:free_energy_expan_0}, the excess pressure $p_0 = - \partial F_0/ \partial V|_{T,N}$ of the D-dimensional reference fluid is provided by~\cite{Hansen:Theory_of_Simple_Liquids:2013}
\begin{align}\label{eq:p-ex} 
p_0(T,n) =  n k_BT  \frac{\Omega_D}{2D}   n^*  g^{(2)}(\sigma_\text{min};n^*)    .
\end{align}  
Using these results, we obtain for the e.o.s. in the  form of the compressibility factor $Z(T,n^*;\epsilon) =  p(T,n;\epsilon)/n k_B T$:
\begin{align}\label{eq:e.o.s.-gen} 
 Z(n^*;\epsilon) =&  1 +  \frac{\Omega_D}{2D} n^* \Big\{  g^{(2)}(\sigma_\text{min};n^*)  \ +  D \langle \tilde{s}\big(\vartheta(\mathbf{e}_{12} \cdot \mathbf{u}_1)\big) \rangle_o  \nonumber\\
& \times  \frac{\diff}{\diff n^*} \big[  n^* g^{(2)}(\sigma_\text{min};n^*) \big]  \  \epsilon \ + O(\epsilon^2) \Big\}  .
\end{align}  
The compressibility factor, $Z(n^*;\epsilon)$, should not be confused with the partition function.
Equation~\eqref{eq:p-ex} reveals that the term involving the derivative with respect to $n^*$ is related to the isothermal compressibility  of the reference system. 

Hence, we succeeded in determining the e.o.s. of a D-dimensional fluid of hard nonspherical particles  of revolution as a \textit{functional} of its shape function,  up to the first order in the shape anisotropy $\epsilon$.

We now turn to ellipsoids of revolution and ellipses. Equations~\eqref{eq:s1-prolate} and  \eqref{eq:s1-oblate}  imply 
$\langle \tilde{s}_{1}\big(\arccos(\mathbf{e}_{12} \cdot \mathbf{u}_1)\big) \rangle_o =   a_{\kappa}/3$  with  $a_\text{prolate}=1$  and $a_\text{oblate}=2$. Then we obtain  from  Eq.~\eqref{eq:e.o.s.-gen} and $D=3$ up to order
$\epsilon$:
\begin{align}\label{eq:e.o.s.-ellipsoid} 
 Z^{\text{ellipsoid}}(n^*;\epsilon)  
=&   1 \ +   \frac{2\pi}{3} n^* \  \Big\{     g^{(2)}(\sigma_\text{min};n^*)  \nonumber \\
 &+    a_{\kappa} \frac{\diff}{\diff n^*} \big[  n^* g^{(2)}(\sigma_\text{min};n^*) \big]  \  \epsilon   \nonumber\\
 &       + O(\epsilon^2)  \Big\}.
\end{align}  
For a 2D fluid of ellipses, it follows with  $\langle \tilde{s}_{1}\big(\arccos(\mathbf{e}_{12} \cdot \mathbf{u}_1)\big) \rangle_o = 1/2$
\begin{align}\label{eq:e.o.s.-ellipse-2D} 
 Z^{\text{ellipse}}_{2D}(n^*;\epsilon)  =&  1 \ + \  \frac{\pi}{2} n^* \  \Big\{  g^{(2)}(\sigma_\text{min};n^*) \big)   \nonumber\\
&    +    \frac{\diff}{\diff n^*} \big[  n^* g^{(2)}(\sigma_\text{min};n^*) \big] \  \epsilon   \nonumber\\
&    +  O(\epsilon^2)  \Big\}  .
\end{align}  
Approximating $ Z^{\text{ellipse}}_{2D}(n^*;\epsilon) $ by  neglecting $ O(\epsilon^2)$ will be called  the ``first-order approximation'' (1st-O) in the following.
Our method is also directly transferable to  a 1D fluid of ellipses with centers on the $x$ axis and orientational d.o.f.\@ in the $x$-$z$-plane, which was studied using the transfer-matrix method~\cite{Lebowitz:JStat_49:1987}. This model has the great  advantage
that the reference fluid is the Tonks gas of hard rods of length $\sigma_\text{min}$ for which the e.o.s. \cite{Tonks:PhysRev_50:1936} and the pair-distribution function~\cite{Salsburg:JCP_21:1953} are known analytically. The calculation of the corresponding compressibility factor requires some caution, since the solid angle of the translational d.o.f.\@, $\Omega_1 = 2$, differs from the solid angle, $\Omega_2 = 2\pi$,
of the orientational d.o.f.\@.
Taking into account that  $\Omega_D$ in Eqs.~\eqref{eq:free_energy_expan_1b} and \eqref{eq:p-ex} is the solid angle of the translational d.o.f.\@,  it follows  that 
\begin{align}\label{eq:e.o.s.-ellipse-1D} 
 Z^{\text{ellipse}}_\text{1D}(n^*;\epsilon)  =    \frac{1}{1-n^*} + \ &  \frac{1}{2}  \frac{n^*}{(1-n^*)^2}   \  \epsilon  
 +  O(\epsilon^2) \  .
\end{align} 
Neglecting $ O(\epsilon^2)$ will again be called  ``first-order approximation'' (1st-O).
The first term  is the compressibility factor of the Tonks gas. It is the sum of the contribution  $n k_BT$ of the 1D ideal gas and
the excess pressure $p_\text{0}=  n k_BT   \ n^*/(1-n^*)$ of the Tonks gas.  The excess pressure  $p_\text{0}$ follows from Eq.~\eqref{eq:p-ex} using $D=1$, $\Omega_1=2$,  and $g^{(2)}(\sigma_\text{min};n^*) = 1/(1 - n^*)$~\cite{Salsburg:JCP_21:1953}.
The second term follows from Eq.~\eqref{eq:e.o.s.-gen} again using  $\Omega_1=2$,  $g^{(2)}(\sigma_\text{min};n^*) = 1/(1 - n^*)$, and  $\langle \tilde{s}\big(\arccos(\mathbf{e}_{12} \cdot \mathbf{u}_1)\big)  \rangle_o = 1/2$ for ellipses.

\subsection{Equation of state:  Equal-volume sphericalization}\label{Sec:p-general-approx}

In  Sec.~\ref{Sec:free_energy}, we  presented a theoretical framework in which  the excess free energy,  $F_\text{ex}(T,V,N;\epsilon)$,  of the  convex-hard-body fluid can be obtained by a cluster expansion using a fluid of hard spheres of diameter $\sigma_\text{min}$ as the  unperturbed system and the anisotropy parameter $\epsilon$ as the smallness parameter.  For cluster configurations consisting of a single  and a pair of two-clusters, we have calculated their leading-order contributions to the excess free energy. 

 As discussed in the Introduction, ``sphericalization'' is a frequently used approximation for hard-body fluids, i.e., their thermodynamic quantities are approximated by those of a fluid of hard spheres with an effective diameter. In this  subsection, it will be  shown that  the mapping in Sec.~\ref{SubSec:effective_potential}  together with our perturbative approach in Sec.~\ref{Sec:free_energy}   leads for the same number density to a ``sphericalization'' for which the hard sphere's volume is identical to that of the hard bodies. This follows from the observation that
  a subclass of an  infinite number of particular cluster configurations can be summed up. This sum is  contained  in the excess free energy $F^\text{hs}_\text{ex}(T,V,N;\sigma(\epsilon))$ of  a fluid of hard spheres of an effective diameter $\sigma(\epsilon)$ and the same number density $n$.  The equal-volume condition $V_p(\epsilon) \equiv V^\text{hs}_p(\epsilon)= (\Omega_D/D)(\sigma(\epsilon)/2)^D$ determines $\sigma(\epsilon)$. 
 In particular,
$V_p(\epsilon)$ is a functional of the shape function:
\begin{align}\label{eq:volume-1} 
V_p(\epsilon) = & \frac{1}{D} \int_{\Omega_D} \diff \mathbf{e}_s \ (s(\vartheta;\epsilon))^D  =  \frac{\Omega_D}{D}
 \langle (s(\vartheta;\epsilon))^D \rangle_o .
\end{align} 
Here we used the orientational average, Eq.~\eqref{eq:orient-average}, for a single particle.
Substituting  $s(\vartheta;\epsilon)$ from Eq.~\eqref{eq:shape}, we obtain
\begin{align}\label{eq:volume-2} 
V_p(\epsilon) = &V_p(0) \ \big[1 + \ D \langle \tilde{s}(\vartheta) \rangle_o \ \epsilon + O(\epsilon^2)\ \big]  ,
\end{align}
with $V_p(0) = (\Omega_D/D) (\sigma_\text{min}/2)^D$, the volume of the hard spheres of the reference fluid.
Then, $V_p(\epsilon) =  V^\text{hs}_p(\epsilon)$  leads to 
\begin{align}\label{eq:effective_diameter}
\sigma(\epsilon) & =  \sigma_\text{min}  \Big[\frac{V_p(\epsilon)}{V_p(0)} \Big]^{1/D}   \nonumber\\
& =  \sigma_\text{min} \big[1 + \langle \tilde{s}(\vartheta) \rangle_o \ \epsilon + O(\epsilon^2)\ \big]  .
\end{align} 

The excess free energy,  $F^\text{hs}_\text{ex}(T,V,N;\sigma(\epsilon))$, is identical to the excess free energy $F_0(T,V,N(\epsilon))$ of the reference fluid with an effective number density 
$n(\epsilon)  = N(\epsilon)/V =  n V_p(\epsilon)/V_p(0) = n\big[1 + D \langle \tilde{s}(\vartheta) \rangle_o \ \epsilon + O(\epsilon^2) \big]$.
The cluster expansion, Eq.~\eqref{eq:free_energy_expan}, of the excess free energy of the hard-body fluid reads  
$F_\text{ex}(T,V,N;\epsilon)= F_0(T,V,N) + F_1(T,V,N;\epsilon) + O(\epsilon^2)$, where the leading-order term, Eq.~\eqref{eq:free_energy_expan_1a}, of $F_1(T,V,N;\epsilon)$ is the contribution of a single two-cluster.  To make progress,  we show that the sum $F_0(T,V,N) + F_1(T,V,N;\epsilon) $  can be simplified. This is achieved by using (i)  $F_0(T,V,N(\epsilon)) =: N f_0(T,n(\epsilon))$, 
(ii) $ - \partial F_0/ \partial V|_{T,N} =  k_BT n^2 \partial f_0(T,n)/ \partial n|_{T}  =  p_0 =   n k_BT   ( \Omega_D/2D )   n^*  g^{(2)}(\sigma_\text{min};n^*)$, and
 (iii)  $\langle \tilde{d}_c(\mathbf{e}_{12},\mathbf{u}_{1},\mathbf{u}_{2};0) \rangle_o =  \langle \tilde{s}(\arccos(\mathbf{e}_{12} \cdot \mathbf{u}_1)) \big\rangle_o \equiv \langle \tilde{s}(\vartheta) \rangle_o$,  [follows from  Eq.~\eqref{eq:d-expan-0}] and expansion up to $O(\epsilon)$.
 Because $ F_k(T,V,N;\epsilon) =  O(\epsilon^k)$ for all $k$,  we obtain from Eq.~\eqref{eq:free_energy_expan} and perform the steps  above:
\begin{align}\label{eq:excess_free_energy_approx}
 F_\text{ex}(T,V,N;\epsilon) =  F^\text{hs}_\text{ex}(T,V,N;\sigma(\epsilon)) + O(\epsilon^2)  .
\end{align}

Now it is crucial to note that 
also the leading-order term  $O(\epsilon^k)$ of all cluster configurations contributing to $F_k(T,V,N;\epsilon)$, $k \geq 2$ and consisting of a $k$-tuple of two-clusters is contained in the corresponding leading-order term $\mathcal{O}(\epsilon^k)$ of 
$F^\text{hs}_\text{ex}(T,V,N;\sigma(\epsilon))$. This can be ``proven'' as follows. The cluster expansion discussed in Sec.~\ref{Sec:free_energy} can also be applied to the fluid of hard spheres of effective diameter $\sigma(\epsilon)$, again
using  the fluid of hard spheres of diameter $\sigma_\text{min}$ as a reference fluid. In the cluster 
function, $f(ij)$ [cf. Eq.~\eqref{eq:cluster-1}], one only  has to replace the contact distance $d_c(\mathbf{e}_{12},\mathbf{u}_{1},\mathbf{u}_{2};\epsilon )$ for two convex  hard bodies by $\sigma(\epsilon)$, the contact function of two hard spheres of diameter $\sigma(\epsilon)$. Let us denote this cluster function by $f^\text{hs}(ij)$. Of course, it  does not depend on $(\mathbf{e}_{12},\mathbf{u}_{1},\mathbf{u}_{2})$. 
 Comparison of Eqs.~\eqref{eq:d-series-main} and ~\eqref{eq:effective_diameter} shows that  this replacement corresponds in linear order in $\epsilon$ to replacing
 $\tilde{d}_c(\mathbf{e}_{12},\mathbf{u}_{1},\mathbf{u}_{2};0) \equiv \tilde{d}_1(\mathbf{e}_{12},\mathbf{u}_{1},\mathbf{u}_{2})$ by $\langle \tilde{s}(\vartheta) \rangle_o$. 
 
 The orientational d.o.f.\@ of the fluid of convex  hard  bodies contribute in leading order $O(\epsilon^k)$ a factor $[\langle \tilde{d}_c(\mathbf{e}_{12},\mathbf{u}_{1},\mathbf{u}_{2};0) \rangle_o]^k$ to the free energy of cluster configurations of $k$-tuple of two-clusters [see, e.g., Eq.~\eqref{eq:free_energy_expan_2a} for $k=2$ and Appendix~\ref{Sec:Appendix_B}]. For the cluster expansion of $F^\text{hs}_\text{ex}(T,V,N;\sigma(\epsilon))$, these $k$-tuple of two-clusters yield in leading order  the factor $[\langle \tilde{s}(\vartheta) \rangle_o]^k$.   For generic hard bodies, Eq.~\eqref{eq:d-expan-0} 
implies  $\langle \tilde{d}_c(\mathbf{e}_{12},\mathbf{u}_{1},\mathbf{u}_{2};0) \rangle_o =  \langle \tilde{s}\big(\vartheta^{(0)}_1(\mathbf{e}_{12},\mathbf{u}_1,\mathbf{u}_2)\big) \big\rangle_o$.  The latter is identical to  $\langle \tilde{s}(\vartheta) \rangle_o$, i.e., the factor stemming from the orientational d.o.f.\@  is identical for the $f$ and  $f^\text{hs}$ expansions.
This also  holds  for the remaining factor arising from the translational d.o.f.\@. This factor involves  $m$-particle distribution functions of the reference fluid which are identical for both expansions since the reference fluid is the same.
Therefore, the leading-order contribution of the $k$-tuple of two-clusters to   
$F_\text{ex}(T,V,N;\epsilon)$ and  $F^\text{hs}_\text{ex}(T,V,N;\sigma(\epsilon))$ is identical.

The remaining cluster configurations, also contributing to the leading-order term  of $F_k(T,V,N;\epsilon)$,  contain, in addition, three-clusters, four-clusters, etc.  Again, the translational d.o.f.\@ contribute in the $f$ and the $f^\text{hs}$ expansion  the same terms. However,  this is not true for the orientational d.o.f.\@. In the $f$ expansion, the  orientational d.o.f. contribute  a factor $\langle \tilde{d}_c(\mathbf{e}_{12},\mathbf{u}_{1},\mathbf{u}_{2};0) 
 \tilde{d}_c(\mathbf{e}_{23},\mathbf{u}_{2},\mathbf{u}_{3};0) \cdots \rangle_o$  [see, e.g., Eq.~\eqref{eq:free_energy_expan_2b}] whereas the factor  $\langle \tilde{s}(\vartheta;0) \rangle_o \langle \tilde{s}(\vartheta;0) \rangle_o \cdots $ appears in the $f^\text{hs}$ expansion. Both factors are different.
 Equating these factors corresponds to  a ``mean-field''-like approximation:  $\langle \tilde{d}_c(\mathbf{e}_{12},\mathbf{u}_{1},\mathbf{u}_{2};0)  \tilde{d}_c(\mathbf{e}_{23},\mathbf{u}_{2},\mathbf{u}_{3};0) \ldots \rangle_o  \approx    \langle \tilde{d}_c(\mathbf{e}_{12},\mathbf{u}_{1},\mathbf{u}_{2};0) \rangle_o  \langle \tilde{d}_c(\mathbf{e}_{12},\mathbf{u}_{1},\mathbf{u}_{2};0) \rangle_o \ldots$.  This ``mean-field''-like approximation 
 corresponds to replacing the original convex  hard body by a hard sphere with the same volume. 
 Yet, three-clusters, four-clusters, etc. become more important for larger anisotropy parameters. Therefore, for $\epsilon$  small enough, $F^\text{hs}_\text{ex}(T,V,N;\sigma(\epsilon))$  is likely a more accurate  approximation of $F_\text{ex}(T,V,N;\epsilon)$ than   $F_0(T,V,N) + F_1(T,V,N;\epsilon)$ up to the linear order of Eq.~\eqref{eq:free_energy_expan_1b}. 
Equation~\eqref{eq:excess_free_energy_approx} together with the ideal gas contribution yields for the compressibility factor of the   convex-hard-body fluid
\begin{align}\label{eq:n-scaled-2} 
Z(n^*;\epsilon) & = Z^\text{hs}(n;\sigma(\epsilon)) + O(\epsilon^2) ,
\end{align}  
with
\begin{align}\label{eq:n-scaled-3} 
 Z^\text{hs}(n;\sigma(\epsilon)) = 1 + \frac{\Omega_D}{2D} \ n\sigma(\epsilon)^D g^{(2)}\big(\sigma(\epsilon);n\big) \ ,
\end{align} 
the compressibilty factor of a hard-sphere fluid with the  \textit{same} number density, $n$, as of the hard-body fluid and an effective diameter  $\sigma(\epsilon)$.  Due to the discussion above,  $Z^\text{hs}(n;\sigma(\epsilon))$ is probably a better approximation of
$Z(n^*;\epsilon)$ than the systematic expansion up to $O(\epsilon)$ on the r.h.s.\@ of Eq.~\eqref{eq:e.o.s.-gen}, 
if the anisotropy parameter is not too large. In the following, the approximation  $Z(n^*;\epsilon) \approx Z^\text{hs}(n;\sigma(\epsilon))$, with $\sigma(\epsilon)$ from the first  line of Eq.~\eqref{eq:effective_diameter},    will be called  the ``equal-volume sphericalization'' (EVS).

\subsection{Comparison with the virial expansion}\label{Sec:comparison-1}

Since our approach resembles the standard virial expansion, we compare both approaches.
The standard virial expansion for the compressibility factor of a fluid of nonspherical particles described by the family of Eq.~\eqref{eq:shape} of generic shapes (characterized by the anisotropy parameter $\epsilon$) reads
\begin{align}\label{eq:virial-gen} 
Z(n^*;\epsilon) & = 1 + \sum^{\infty}_{l=2} B_l(\epsilon) \ (n^*/\sigma_\text{min}^D)^{l-1} ,
\end{align} 
and for the excess pressure, Eq.~\eqref{eq:p-ex}, of the reference fluid of hard spheres of diameter $\sigma_\text{min}$: 
\begin{align}\label{eq:virial-ex} 
p_0(T,n) & = k_BT n \sum^{\infty}_{l=2} B_l(0) \ n^{l-1} \ .
\end{align} 
Equations~\eqref{eq:p-ex} and \eqref{eq:virial-ex} lead to the virial series of the pair-distribution function (of the reference fluid) at contact
\begin{align}\label{eq:virial-g-contact} 
g^{(2)}(\sigma_\text{min};n^*) & = \frac{2D}{\Omega_D \ (\sigma_\text{min})^D} \sum^{\infty}_{l=2} B_l(0) \ (n^*/\sigma_\text{min}^D)^{l-2} \ .
\end{align} 
Substitution of $g^{(2)}(\sigma_\text{min};n^*)$ from Eq.~(\ref{eq:virial-g-contact}) and its derivative with respect to
$n^*$ into Eq.~(\ref{eq:e.o.s.-gen}) and comparing with Eq.~(\ref{eq:virial-gen}) allows us to express the virial coefficients for an arbitrary  convex-hard-body fluid by the corresponding coefficients for the reference fluid of hard spheres:
\begin{align}\label{eq:comp-virial} 
B_l(\epsilon) =& \Big[1 + D (l-1) \big\langle \tilde{s}\big(\arccos(  \mathbf{e}_{12} \cdot \mathbf{u}_1);0\big) \big\rangle_o \ \epsilon \nonumber\\
& + O(\epsilon^2 ) \Big] \ B_l(0)  
\end{align} 
for all $l$. Note that because we include ellipses and ellipsoids of revolution, $\tilde{s}(\vartheta;\epsilon)$ alos depends  on $\epsilon$ [see corresponding comment below Eq.~\eqref{eq:shape}]. 

The reduced virial coefficients are defined by $B^*_l(\epsilon) = B_l(\epsilon)/[V_p(\epsilon)]^{l-1} $, with $V_p(\epsilon)$ from Eq.~\eqref{eq:volume-2}.
Since $\langle \tilde{s}(\vartheta;0) \rangle_o = \langle \tilde{s}\big(\vartheta(\mathbf{e}_{12} \cdot \mathbf{u}_1);0\big) \big\rangle_o$, Eq.~\eqref{eq:comp-virial} implies that the linear term in $\epsilon$ of $B_l(\epsilon)/[V_p(\epsilon)]^{l-1}$ vanishes. Accordingly, we obtain
\begin{align}\label{eq:reduced-virial} 
B^*_l(\epsilon) = \big[1 + O(\epsilon^2 ) \big] \ B^*_l(0)
\end{align}
for \textit{all} $l$. Note that the term of $O(\epsilon)$ of $B^*_l(\epsilon)$ vanishes for all  convex hard bodies, including shapes which are continuous but not smooth. For $l=2$, this is consistent with the expansion of the general result for $B^*_2(\epsilon)$ for ellipses~\cite{Kulossa:MolPhys_0:2023} and for ellipsoids of revolution~\cite{Freasier:MolPhys_32:1976}. Analytical results for $B^*_l(\epsilon)$ and $l \geq 3$ do not seem to exist. 
The result in Eq.~\eqref{eq:reduced-virial} is obvious.   The reduced virial coefficients  
$B^*_l(\epsilon) = \partial^{l-1} Z(n^*;\epsilon)/ \partial \eta(\epsilon)^{l-1}/ (l-1)!  $ are essentially the derivatives of the compressibility factor  with respect to the packing fraction $\eta(\epsilon) = n (\Omega_D/D) (\sigma(\epsilon)/2)^D$. Then 
Eq.~\eqref{eq:n-scaled-2} implies Eq.~\eqref{eq:reduced-virial}.

\subsection{Comparison with the Monte Carlo results}\label{Sec:comparison-2}

For a  fluid of hard ellipses, we will compare our analytical results, Eqs.~\eqref{eq:e.o.s.-ellipse-2D} , \eqref{eq:e.o.s.-ellipse-1D},   and   \eqref{eq:n-scaled-2}  [neglecting $O(\epsilon)^2$]  for the compressibility factor   with the result  from our Monte Carlo (MC) simulation and scaled particle theory. In the following, $a$ and $b$  denote the length of the major and minor semiaxis, respectively, and $X_0  = 1 + \epsilon  = a/b$ is the aspect ratio of the ellipses with $\epsilon$  the anisotropy parameter. 

The MC simulation for the 1D and 2D fluid of ellipses was performed at constant pressure, $p$, and temperature, $T$, for $N=100$ and $N=2610$ ellipses, respectively.  Periodic boundary conditions were chosen. In total,  $2\times10^8$ MC steps were performed for the 1D fluid and $10^8$ MC steps for the 2D one. The initial equilibration part was discarded  by inspecting the behavior of the density.  As an overlap criterion we used the one proposed by  Perram and Wertheim~\cite{Perram:JCompPhys_58:1985}. The pressure was changed such that the corresponding dimensionless number densities $n^*= (N/L)(2b)$ and  $n^*= (N/A)(2b)^2$ for the 1D fluid of length $L$ and the 2D one of area $A$,  respectively, vary  between $n^*_\text{min} \approx 0.010$ and $n^*_\text{max} \approx 0.8$. For the 2D fluid and $X_0 < 1.5$, $n^*_\text{max}$  is still  below the transition density $n^*_{i-p}(X_0)$ of the phase  transition  from the isotropic liquid to the plastic crystalline phase. A nematic phase occurs only above $n^*_{i-p}(X_0)$, provided  $X_0 \gtrsim 2.5$  \cite{Bautista-Carbajal:JCP_140:2014}. Note, our 2D system is not large enough  to exhibit the two-step melting process with the hexatic phase in between. 

\begin{figure}
\includegraphics[angle=0,width=0.982\linewidth]{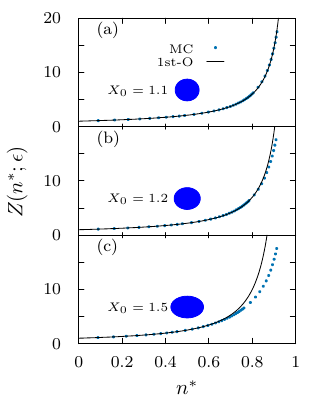}
\caption{Density dependence of the compressibility factor $Z(n^*;\epsilon)$ of the 1D fluid of ellipses for aspect ratios $X_0 = 1+\epsilon = 1.1, 1.2$, and $1.5$. The degree of the shape anisotropy is illustrated by the corresponding blue ellipses.  Shown is the analytical result (1st-O) from   Eq.~\eqref{eq:e.o.s.-ellipse-1D} [neglecting $O(\epsilon^2)$] and  the Monte Carlo result (MC).   \label{fig:1D-ellipse-Z}}
\end{figure}

\begin{figure}
\includegraphics[angle=0,width=\linewidth]{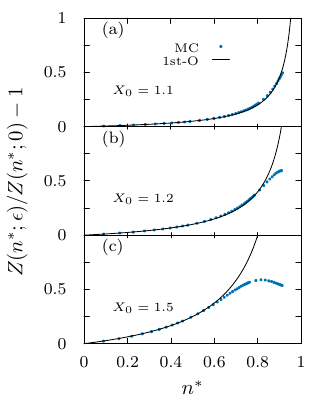}
\caption{The relative compressibility factor $[Z(n^*;\epsilon)/Z(n^*;0) - 1]$ of the 1D fluid of ellipses as a function of $n^*$ for aspect ratio $X_0 = 1+ \epsilon = 1.1, 1.2$ and $1.5$. Shown is the analytical  result  from Eq.~\eqref{eq:e.o.s.-ellipse-1D} (1st-O)  and  the Monte Carlo (MC) result.  \label{fig:1D-ellipse-Z_rel}}
\end{figure}

First, the comparison is performed for the 1D fluid where the centers of the ellipses can only move along the $x$ axis and rotate in the $x-z$ plane. 
For this model, there are no results from  scaled-particle theory. A comparison with  $ Z^\text{hs}(n;\sigma(\epsilon))$ from Eq.~\eqref{eq:n-scaled-3} is also  not possible, since the analytical result  in Eq.~\eqref{eq:n-scaled-2}  requires that the spatial  dimension of the  convex hard body and that of the fluid are identical.    Figure~\ref{fig:1D-ellipse-Z} compares  $Z^\text{ellipse}_\text{1D}(n^*;\epsilon)$ [Eq.~\eqref{eq:e.o.s.-ellipse-1D}]  with the corresponding MC result  for aspect ratios $X_0 =1 + \epsilon = 1.1, 1.2$, and $1.5$. To observe the direct influence of the orientational d.o.f.\@, we have plotted in Fig.~\ref{fig:1D-ellipse-Z_rel} the relative compressibility factor, 
$[Z(n^*;\epsilon)/Z(n^*;0) - 1]$, which measures the deviation of $Z(n^*;\epsilon)$ from the compressibilty factor $Z(n^*;0)$ of the reference fluid of hard rods of length 
$\sigma_\text{min}=2b$. Figure~\ref{fig:1D-ellipse-Z}  demonstrates  good agreement. Particularly for $\epsilon =  0.1$, the agreement is very good, even up to $n^*= 0.85$, which is already close to $n^*_\text{cp}=1$, the maximum density for closest packing. Of course, for the smallest value $\epsilon = 0.1$, the compressibility factor  $Z(n^*;0.1)$ does not differ much from  $Z(n^*;0)$. Accordingly, it is more conclusive  to compare the relative compressibility factors in  Fig.~\ref{fig:1D-ellipse-Z_rel}. We still observe  very good  agreement for  $\epsilon = 0.1$  up to $n^* = 0.85$. Even for $\epsilon = 0.5$, both results agree  rather satisfactorily   up to $n^* \approx 0.6$.

\begin{figure}
\includegraphics[angle=0,width=0.965\linewidth]{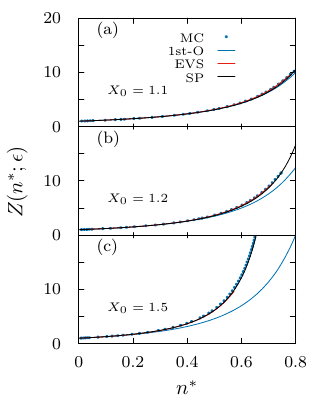}
\caption{Density dependence of the compressibility factor $Z(n^*;\epsilon)$ of the 2D fluid of ellipses for aspect ratios $X_0 = 1 + \epsilon = 1.1, 1.2$, and $1.5$. Shown are our analytical  results (1st-O), (EVS) [Eqs.~\eqref{eq:e.o.s.-ellipse-2D}, \eqref{eq:n-scaled-2}] and from scaled-particle theory (SP) [Eq.~\eqref{eq:Z-scp}]  as well as the Monte Carlo (MC) result . \label{fig:2D-ellipse-Z}}.
\end{figure}

\begin{figure}
\includegraphics[angle=0,width=\linewidth]{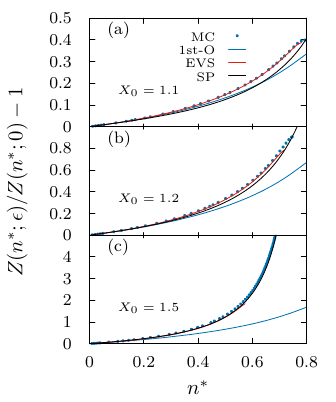}
\caption{Density dependence of the relative compressibility factor $[Z(n^*;\epsilon)/Z(n^*;0) - 1]$ of the 2D fluid of ellipses for aspect ratios $X_0 =  1 + \epsilon = 1.1, 1.2$, and $1.5$.  Shown are our analytical  results (1st-O), (EVS) [Eqs.~\eqref{eq:e.o.s.-ellipse-2D} and \eqref{eq:n-scaled-2}] and from scaled-particle theory (SP) [Eq.~\eqref{eq:Z-scp}]  as well as the Monte Carlo (MC) result . \label{fig:2D-ellipse-Z_rel}}
\end{figure}

Now we  turn to the 2D fluid of ellipses. The comparison of the MC result with  $Z^\text{ellipse}_\text{2D}(n^*;\epsilon)$ [Eq.~\eqref{eq:e.o.s.-ellipse-2D}] requires  the knowledge of the pair-distribution function of the reference fluid at contact, $ g^{(2)}(\sigma_\text{min};n^*)$, and the  derivative of $n^* g^{(2)}(\sigma_\text{min};n^*)$ with respect to $n^*$.  
The former follows directly from the pressure [cf. Eq.~\eqref{eq:p-ex}] and the latter from the compressibility.
In contrast to the 1D fluid, both quantities are  not known analytically. 
These quantities could be obtained by the MC simulation. Since  the simulation  was performed at constant pressure, $p$, the corresponding  density $n^*(p;X_0)$ for $X_0 > 1$ will differ from the density $n^*(p;X_0=1)$ of the reference fluid. To avoid this problem, we have chosen the e.o.s. derived in Ref.~\cite{Kolafa:MolPhys_104:2006} for a 2D fluid of hard disks and $n^* \leq n^*_\text{max}=0.880$. This e.o.s. is very precise. For instance,  at the maximum density $n^*= 0.860$ of our MC simulation of the reference fluid, its compressibility factor differs from the simulational result by $0.25\%$. Using the compressibility factor 
$Z^\text{disks}_\text{SP}(n^*) = 1/(1 - \pi n^*/4 )^2$~\cite{Helfand:JCP_34:1961} for the hard-disk fluid from 
scaled-particle theory, it deviates  from the MC result by 1.65 \%. The e.o.s. of   Ref.~\cite{Kolafa:MolPhys_104:2006} allows us to explicitly calculate  the first-order result from Eq.~\eqref{eq:e.o.s.-ellipse-2D} and of the equal volume sphericalization  with $Z^\text{hs}(n;\sigma(\epsilon))$ from Eq.~\eqref{eq:n-scaled-3}. 

Besides the  equal-volume sphericalization (EVS),  $Z(n^*;\epsilon) \approx Z^\text{hs}(n;\sigma(\epsilon))$ [Eq.~\eqref{eq:n-scaled-2}], the comparison also encompasses the result from scaled-particle theory for a 2D fluid of convex hard bodies~\cite{Boublik:MolPhys_29:1975}:
\begin{align} \label{eq:Z-scp}
Z_\text{SP}(\eta) & = \frac{1}{1-\eta}  + \gamma \frac{\eta}{(1-\eta)^2 } \ .
\end{align}
Here, $\eta =N A_p/A= n A_p$ is the packing fraction and 
\begin{align} \label{eq:b2-star}
\gamma& = \ \frac{(S_p)^2}{4\pi A_p} 
\end{align}
the shape parameter.  These two parameters depend on $A_p$ and $S_p$, the area and the perimeter, respectively, of an arbitrary two-dimensional  convex hard body. For ellipses,  $A_p(X_0) = \pi b^2 X_0$ and $S_p(X_0) = 4b E(e(X_0)) X_0$. The eccentricity is defined by     $e(X_0)= \sqrt{1 - X_0^{-2}}$, and $E(e)$ is the complete elliptical integral of the second kind~\cite{bworld}. For hard disks, it is $\gamma =1$ and  $\eta = \pi n^*/4 $. Then, Eq.~\eqref{eq:Z-scp} reduces to $Z^\text{disks}_\text{SP}(n^*)$ from above.

Figures~\ref{fig:2D-ellipse-Z} and~\ref{fig:2D-ellipse-Z_rel}, respectively,  present the density dependence of   the compressibility factor, $Z(n^*;\epsilon)$, and  of the relative compressibility factor, $[Z(n^*;\epsilon)/Z(n^*;0) - 1]$,  for the various analytical results and the results from the MC simulation. The relative  compressibility factor  is a direct measure of the shape anisotropy. It is zero for all densities if the shape anisotropy vanishes. We note that the graph (SP) in 
Fig.~\ref{fig:2D-ellipse-Z_rel} for  scaled-particle theory  displays $Z_\text{SP}(\pi n^*X_0/4)/Z(n^*,0)$ and not $Z_\text{SP}(\pi n^*X_0/4)/Z_\text{SP}(\pi n^*/4)$. Since EVS involves the compressibility factor of a fluid of hard disks of diameter $\sigma(\epsilon) =  2b \sqrt{1+\epsilon}$ [follows from the first line of Eq.~\eqref{eq:effective_diameter}] and density $n$,  $n^*$ is limited to  $n^* \leq 0.860/(1+\epsilon)$. Consequently, the maximum range for the EVS results in Figs~\ref{fig:2D-ellipse-Z} and~\ref{fig:2D-ellipse-Z_rel} become reduced by the factor $1/X_0$.   To illustrate the influence of the body's anisotropy, we plot these quantities  versus $n^*$ and not versus the packing fraction 
 (area fraction), $\eta(X_0)= \pi n^*X_0/4$, which itself depends on  $X_0= 1+\epsilon$. 
Both figures  demonstrate    good agreement  between our first-order result  (1st-O) [Eq.~\eqref{eq:e.o.s.-ellipse-2D}], and the result (MC) from the MC simulation.  However, this agreement  becomes more and more restricted to a smaller range of $n^*$ if  $\epsilon$ increases. But,  both figures also show that the equal-volume sphericalization (EVS) represents a significant improvement, compared to the first-order result (1st-O).  Figure~\ref{fig:2D-ellipse-Z}   reveals that the MC points are well described by both the SP result from scaled-particle theory and  the EVS result. However,  the more sensitive quantity in Fig.~\ref{fig:2D-ellipse-Z_rel} demonstrates the strength of the equal-volume sphericalization  and its superiority with respect  to scaled-particle theory, at least for smaller shape anisotropies.  

As discussed  in Sec.~\ref{Sec:Introduction}, further approximate versions for  the e.o.s. exist in the literature.  For instance, the e.o.s. in Ref.~\cite{Solana:MolPhys_113:2015} is an improvement of $Z_\text{SP}(\eta)$. Although these various e.o.s.  describe
the compressibility factor rather well up to higher densities even  for  larger aspect ratios, most of them (including scaled-particle theory) have some shortcomings, in contrast to EVS.  First, expanding the corresponding compressibility factor with respect to $\epsilon$ reveals that the zeroth- and first-order term differ from the exact result, Eq.~\eqref{eq:e.o.s.-ellipse-2D}. Consequently, for fixed density $n^*$, those e.o.s.\@  do not yield precise results if the anisotropy becomes weaker and weaker.  This is  why the quality of scaled-particle theory is not satisfactory for small shape anisotropies, as demonstrated by Fig.~\ref{fig:2D-ellipse-Z_rel}. The e.o.s. derived in Refs.~\cite{Belleman:PRL_21:1968,Nezbeda:JChemSoc_75:1979,Parsons:PRA_19:1979,Lee:JCP_87:1987,
Marienhagen:PRE_105:2022}  do not suffer from this deficiency. However, for finite $\epsilon$  and for the \textit{same} number density, the sphericalization of the hard-body fluid is different.  The volume of the spheres of their reference  fluid, either  $(\Omega_D/D) [\langle d_\text{c} \rangle_o]^D$~\cite{Belleman:PRL_21:1968,Nezbeda:JChemSoc_75:1979}
 or  $(\Omega_D/D)\langle(d_\text{c})^D\rangle_o$\cite{Parsons:PRA_19:1979,Lee:JCP_87:1987}, differs from $V_p=(\Omega_D/D)\langle s^D\rangle_o$ [Eq.~\eqref{eq:volume-1} ], the volume of the  hard body. This is true because the orientational average of, e.g., the Dth power of the contact function $d_\text{c}$ is not identical to that average of the D-th power of the shape function $s$. 
The second drawback concerns the divergence of the compressibility factor, e.g., for the 2D fluid of ellipses at  $\eta_\text{cp}(\epsilon) = \pi/2\sqrt{3}$, the maximum value for the packing fraction, which does not depend on the aspect ratio~\cite{Toth:ActaSci_12:1950}. This implies that   $ Z^\text{hs}(n;\sigma(\epsilon))$  diverges at the correct value  $n^*_\text{cp}(\epsilon)= 2/\sqrt{3}(1+\epsilon)$, whereas  the other e.o.s.\@  involving $1/(1-\eta)$  (except for Eq.(11)  of Ref.~\cite{Marienhagen:PRE_105:2022}) diverge at the higher value  $n^*_\text{div}(\epsilon) = 4/\pi(1+\epsilon)$, being unphysical.

\begin{figure}
\includegraphics[angle=0,width=\linewidth]{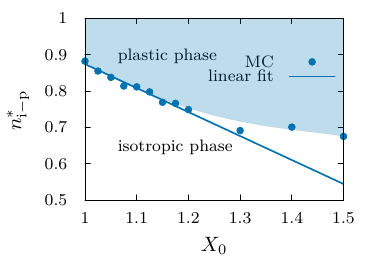}
\caption{Transition density $n^*_{i-p}$ for the  isotropic-plastic transition of the 2D  fluid of ellipses as a function of $X_0$. 
The linear fit  was obtained by a least-squares fit of the first six Monte Carlo (MC) points. The B\'ezier-method was applied to illustrate the domain of the plastic-crystalline phase, including the coexistence regime (light-blue domain). \label{fig:iso-plastic}}.
\end{figure}

Finally, we check the  prediction that all transition points of any  phase transition of a fluid of hard spheres should change linearly with the aspect ratio. This will be performed for   the isotropic-plastic transition of a  2D fluid of ellipses. The corresponding transition for $X_0=1$ is the transition of the fluid of hard disks from  the isotropic liquid to a triangular ``lattice''. In order to observe  the hexatic phase and the two-step melting,  the number of particles in the simulation must be significantly  larger than $N=2610$~\cite{Bernard:PRL_107:2011}. 
To determine  the transition point $n^*_{i-p}(X_0)$ of the transition from the isotropic liquid to the plastic crystal, we used $\Psi_6 = (1/N) \sum^N_{k=1} \langle \exp[6i\phi_{kl}] \rangle$ as an order parameter. $\phi_{kl}$  describes the orientation of the nearest-neighbor bond $(k,l)$ with respect to a fixed axis and  $\langle (\cdots) \rangle$
is the average over all neigbors $l$ of $k$. These neighbors were obtained from a Delaunay triangulation.  The pressure was increased by small steps until the order parameter exhibits a jump. The density corresponding to the occurrence of this 
jump was taken as the transition density  which is displayed in  Fig.~\ref{fig:iso-plastic}. Since this transition is discontinuous (first-order transition),  $n^*_{i-p}(X_0)$ is the freezing line.  The melting line and, accordingly, the coexistence region was also determined.  Since the main 
purpose has only been to check our prediction of a linear dependence of $ n^*_{i-p}(X_0)$ on $X_0$ for small enough anisotropy, and since our MC-data scatter, the melting line and the coexistence region are not shown.
The straight line  in  Fig.~\ref{fig:iso-plastic} supports our prediction of a linear shift of  $ n^*_{i-p}(X_0)$ for small values of $X_0$-1. There is little doubt that the  corresponding densities for the isotropic-hexatic  and hexatic-''solid'' 
transition will shift linearly with $X_0$, as well. We also applied the B\'ezier method  to illustrate the domain of the plastic-crystalline phase from the MC points. This domain is shown in light blue. Its boundary is the freezing line of the transition from the isotropic phase to the plastic-crystalline one.

\section{Summary and Conclusions}\label{Sec:conclusions}

The main goal of the present work was to provide in two steps a relatonship between thermodynamic quantities of a D-dimensional fluid of convex hard bodies and the body's shape. First we 
proved that the thermodynamics of a D-dimensional fluid of   convex hard particles is equivalent to that of a D-dimensional fluid of spherical particles   with a hard core of diameter  $\sigma_\text{min}$ and a soft shell with a thickness, $\epsilon \sigma_\text{min}/2$. Here, $\sigma_\text{min}$ is the unique  diameter of the largest sphere which can be inscribed into the hard body (see  Fig.~\ref{fig:hard_body}) and  $\epsilon \geq 0$ is a measure of the shape anisotropy. Besides their hard-core repulsion at $\sigma_\text{min}$, the spherical particles interact by entropic $k$-body forces, $k \geq 2$ if
their soft shells overlap,  forming clusters as illustrated  in Fig.~\ref{fig:cluster_expansion}.  Note, this equivalence holds for arbitrarily large $\epsilon$. For weak shape anisotropy the
effective potential can be considered  as a perturbation to the reference fluid of the hard spheres of diameter  
$\sigma_\text{min}$.  This has two implications. First, any phase transition of hard-sphere fluids, should also exist  for  fluids of weakly nonspherical hard bodies. In particular, in 2D  the two-step melting scenario~\cite{Strandburg:RMP_60:1988} consisting of the isotropic-hexatic and the hexatic-''solid'' transition should survive.  Second, any phase-transition point  should  shift linearly with 
$\epsilon$.  Our Monte Carlo results for the isotropic-plastic phase transition of a  2D fluid of ellipses are consistent with these predictions (see Fig.~\ref{fig:iso-plastic}). Using   much larger systems (as in Ref.~\cite{Bernard:PRL_107:2011}),  there is little doubt that the two-step melting scenario will be observed with transition points shifting linearly with $\epsilon$.

The equivalence of the hard-body fluid to a fluid of spherical particles may also be useful from a  practical point of view. Computer simulations of fluids involving orientational  degrees of freedom are rather expensive. Therefore, the mapping could be used to instead simulate  the fluid of spherical particles, e.g.,  with the effective two-body interactions (see Fig.~\ref{fig:soft_2_body_potential}).
For a 2D fluid, this would be interesting since it was shown that the isotropic-hexatic phase transition crosses over from a discontinuous (first-order ) to a continuous phase  transition if the pair potential becomes soft enough~\cite{Kapfer:PRL_114:2015}. In this context, one could check whether the spherical particles with a hard core and a soft shell are soft enough or not.

In a second step, the mapping  to cluster configurations of  spherical particles  has allowed us to perform a systematic  expansion  for, e.g.,  the Helmholtz free energy $F(T,V,N;\sigma_\text{min},\epsilon)$. Its zeroth order,  $F_0(T,V,N;\sigma_\text{min})$, is the free energy of  the reference  fluid of hard spheres of diameter  
$\sigma_\text{min}$ and the overlap  of the soft  shells is the perturbation (see Fig.~\ref{fig:cluster_expansion}).  Each term $F_k(T,V,N;\sigma_\text{min},\epsilon)$, $k \geq 1$ of the expansion has a physical interpretation
in terms of cluster configurations.  It  involves the orientational average of $k$-fold products of the reduced contact function  $\tilde{d}_{c}(\mathbf{e}_{12},\mathbf{u}_{1},\mathbf{u}_{2};\epsilon)$ [Eq.~\eqref{eq:contact}].
This function is not accessible analytically, even for ellipses.
Therefore, we have exploited the fact that $\tilde{d}_{c}(\mathbf{e}_{12},\mathbf{u}_{1},\mathbf{u}_{2};\epsilon)$ is completely determined by the reduced shape function $\tilde{s}(\vartheta;\epsilon)$  [cf. Eq.~\eqref{eq:shape}] and its derivatives.  Elaborating on a  perturbative method to solve the  contact conditions [Eqs.~\eqref{eq:contact-condition-main-1} and \eqref{eq:contact-condition-main-2}] for  convex  hard bodies  turns the original perturbation series of  $F(T,V,N;\sigma_\text{min},\epsilon)$ with $\epsilon \tilde{d}_{c}(\mathbf{e}_{12},\mathbf{u}_{1},\mathbf{u}_{2};\epsilon)$ as the perturbation into a series where the shape anisotropy
$\epsilon\tilde{s}(\vartheta;\epsilon)$, has become the perturbation. This result allows one  to express thermodynamic quantities as a functional of the shape function itself. As an illustration, we have calculated in Sec.~\ref{Sec:p-general} the e.o.s. as a functional of $\tilde{s}(\vartheta;\epsilon)$ up to linear order in $\epsilon$. 

The applicability and validity of our theoretical framework has been checked for a 1D and 2D fluid of ellipses  with aspect ratio $X_0 = 1 + \epsilon$ by  comparing 
our  first order results [Eqs.~\eqref{eq:e.o.s.-ellipse-1D} and \eqref{eq:e.o.s.-ellipse-2D}]  for the compressibility factor $Z(n^*;\epsilon)$ and  the relative quantity $[Z(n^*;\epsilon)/Z(n^*;0)-1]$, with  results from our MC simulation. For the 1D fluid,  Figs.~\ref{fig:1D-ellipse-Z} and  ~\ref{fig:1D-ellipse-Z_rel} show  very good agreement with the MC results, even up to high densities $n^*=n\sigma_\text{min} \approx 0.9$ for $\epsilon=0.1$. 
For smallness parameter $\epsilon=0.5$ (which is not much smaller than unity), there is still  good agreement  up to  $n^* \approx 0.65$. For the 2D fluid, Figs.~\ref{fig:2D-ellipse-Z} and  ~\ref{fig:2D-ellipse-Z_rel} demonstrate   similar  good agreement, however,  compared to the 1D fluid,  for a smaller range of the density.

The equal-volume sphericalization (EVS) represents  significant improvement  if the shape anisotropy is not too high (cf.  Fig.~\ref{fig:2D-ellipse-Z_rel}). It  approximates the original hard-body fluid by a fluid of hard spheres with the \textit{same} number density, $n$, and a volume identical to the hard body's volume. The quality of EVS is supported analytically by the observation that a  partial summation of an infinite number of cluster configurations yields part of the  free energy of a fluid of hard spheres with properties described above.  
In contrast to other sphericalizations  (see, e.g., Refs. \cite{Belleman:PRL_21:1968,Nezbeda:JChemSoc_75:1979,Parsons:PRA_19:1979,Lee:JCP_87:1987,
Percus:AnnNYAcad_221:1954,Song:PRA_41:1990,Vega:MolPhys_92:1997,Solana:MolPhys_113:2015}), it is the most natural and simplest one since it involves only the hard body's volume, which is accessible, in contrast to, e.g., the orientational average of the contact function occurring in Refs.~\cite{Belleman:PRL_21:1968,Nezbeda:JChemSoc_75:1979}.  Interestingly, the quality of EVS is also supported by the MD simulation of a fluid of hard ellipsoids: Even for larger aspect ratios, $X_0= 2$ and $3$, the isotropic part of the pair-correlation function at short distances is well approximated at the \textit{same} number density by the pair-correlation function of a fluid of hard spheres having the same volume as an ellipsoid~\cite{Talbot:JCP_92:1990}.

To conclude, our theoretical framework may stimulate further studies of fluids of  convex  hard particles   to determine further influences of the shape anisotropy on the behavior of fluids of nonspherical particles.
It would be interesting  going beyond  first order in $\epsilon$ which will involve  derivatives of the shape function. This would also allow one to study the challenging  inverse problem:  Can one reconstruct the body's shape from the dependence of the e.o.s. on number density, $n$, and anisotropy parameter $\epsilon$?  The cluster expansion could also be applied  to calculate structural quantities such as the pair-distribution function $g^{(2)}(\mathbf{r}_{12},\mathbf{u}_{1},\mathbf{u}_{2})$. Finally, exploiting the cluster description, also  valid for large shape anisotropies, one could  study the isotropic-nematic phase formation. For instance, for prolate ellipsoids in 3D, a three-cluster with strongly overlapping soft shells corresponds  to three aligned ellipsoids, forming a nematic nucleus.

\begin{acknowledgments}
We gratefully acknowledge helpful comments on our manuscript by M. P. Allen  and J. Wagner,  and, particularly, for  drawing our attention to related work.
C.D.M.\@ acknowledges financial support from the European Union - Next Generation EU (Grant No. MUR-PRIN2022
TAMeQUAD CUP:B53D23004500006) and from ICSC -- Centro
Nazionale di Ricerca in High Performance Computing, Big
Data, and Quantum Computing, funded by the European
Union -- NextGeneration EU. R.S.\@  acknowledges financial support by the Deutsche Forschungsgemeinschaft (DFG, German Research Foundation SFB TRR 146, Project No. 233630050.  
This research was funded
in part by the Austrian Science Fund (FWF) No. 10.55776/P35673. For
open access purposes, the author has applied a CC BY public copy-right license to any author accepted manuscript version arising
from this submission.
\end{acknowledgments}

\appendix

\section{Calculation of the effective two-body potential for ellipses}\label{Sec:Appendix_A}
In this Appendix,  we calculate the effective two-body potential $ v^{(2)}_\text{eff}(r;\epsilon)$ for two ellipses with  aspect ratio $X_0=a/b = 1+ \epsilon$  ($\epsilon \geq 0)$ and a center-to-center distance $r$.  Here, $a$ and  
$b$ are  the lengths of the major and minor semiaxes, respectively. As Eq.~\eqref{eq:eff_2_pot} shows, its calculation 
requires the orientational average $\langle f(r\mathbf{e}_{12},\mathbf{u}_1,\mathbf{u}_2) \rangle_o$ of the cluster function.
Because $ v^{(2)}_\text{eff}(r;\epsilon)$  vanishes for $r <  \sigma_\text{min} =2b$  and  $r >  \sigma_\text{max} = 2a$,  we can restrict ourselves to the range $2b \leq r \leq 2a$. In this range, it follows from
Eq.~\eqref{eq:cluster-1} $[1+\langle f(r\mathbf{e}_{12},\mathbf{u}_1,\mathbf{u}_2) \rangle_o] = \langle  \Theta \big(r - d_c(\mathbf{e}_{12},\mathbf{u}_{1},\mathbf{u}_{2};\epsilon)\big) \rangle_o$.  Here, the Heaviside function $\Theta(x)$ equals zero for $x < 0$ and is unity for $x \geq 0$.
Since $\langle  \Theta \big(r - d_c(\mathbf{e}_{12},\mathbf{u}_{1},\mathbf{u}_{2};\epsilon)\big) \rangle_o = \langle  \Theta \big(\psi(r\mathbf{e_{12}},\mathbf{u}_1,\mathbf{u}_2;\epsilon) \big) \rangle_o $, we obtain from Eq.~\eqref{eq:eff_2_pot} 
\begin{align} \label{eq:def_v2eff-1}
 v^{(2)}_\text{eff}(r;\epsilon) &= -k_B T \ln \left[|\mathcal{D}(r;\epsilon)|/\pi^2\right],  \quad \text{for }     2b \leq r \leq 2a,
 \end{align}
 where $|\mathcal{D}(r;\epsilon)|$ is the area of the domain  $\mathcal{D}(r;\epsilon)$  in the space of $(\mathbf{u}_{1},\mathbf{u}_{2})$ with  a boundary following from $\psi(r\mathbf{e_{12}},\mathbf{u}_1,\mathbf{u}_2;\epsilon) =0$. Accordingly, the calculation of $ v^{(2)}_\text{eff}(r;\epsilon)$ requires 
 the overlap function  \cite{Vieillard-Baron:JCP_56:1972}:
\begin{align}\label{eq:overlap}
\lefteqn{ \psi(r\mathbf{e}_{12},\mathbf{u}_1,\mathbf{u}_2;\epsilon) = } \nonumber \\
  =&  4 \big[g_1(r\mathbf{e}_{12},\mathbf{u}_1,\mathbf{u}_2;\epsilon)^2 - 3g_2(r\mathbf{e}_{12},\mathbf{u}_1,\mathbf{u}_2;\epsilon) \big] 
 \nonumber \\
& \times \big[g_2(r\mathbf{e}_{12},\mathbf{u}_1,\mathbf{u}_2;\epsilon)^2 - 3g_1(r\mathbf{e}_{12},\mathbf{u}_1,\mathbf{u}_2;\epsilon) \big]   \nonumber \\
& - \big[9 - g_1(r\mathbf{e}_{12},\mathbf{u}_1,\mathbf{u}_2;\epsilon) g_2(r\mathbf{e}_{12},\mathbf{u}_1,\mathbf{u}_2;\epsilon) \big]^2 \ ,
\end{align}
with
\begin{align}\label{eq:g-alpha} 
g_{\alpha}(r\mathbf{e}_{12},\mathbf{u}_1,\mathbf{u}_2;\epsilon) =& 1 + G(\mathbf{u}_1,\mathbf{u}_2;\epsilon)  \nonumber \\
& -r^2\Big[\Big(\frac{\mathbf{e}_{12} \cdot \mathbf{u}_\alpha}{a}\Big)^2 + \Big(\frac{\mathbf{e}_{12} \cdot \mathbf{u}_\alpha'}{b}\Big)^2 \Big] , \end{align} 
where $\mathbf{u}_\alpha'$ is a unit vector perpendicular to $\mathbf{u}_\alpha$ and 
\begin{align}\label{eq:G} 
G(\mathbf{u}_1,\mathbf{u}_2;\epsilon) & = 2 + \Big(\frac{a}{b} - \frac{b}{a}\Big)^2\Big[1 - (\mathbf{u}_1 \cdot \mathbf{u}_2)^2 \Big] .
\end{align} 
$\psi(r\mathbf{e}_{12} ,\mathbf{u}_1,\mathbf{u}_2;\epsilon)$ is negative  when both ellipsoids overlap; it becomes zero when they are in tangential contact. The ellipses do not overlap if  $\psi(r\mathbf{e}_{12},\mathbf{u}_1,\mathbf{u}_2;\epsilon)$ is positive and, in addition, at least one of the functions $g_{\alpha}(r\mathbf{e}_{12},\mathbf{u}_1,\mathbf{u}_2;\epsilon)$ is negative~\cite{Vieillard-Baron:JCP_56:1972}.

\begin{figure}
\includegraphics[angle=0,width=\linewidth]{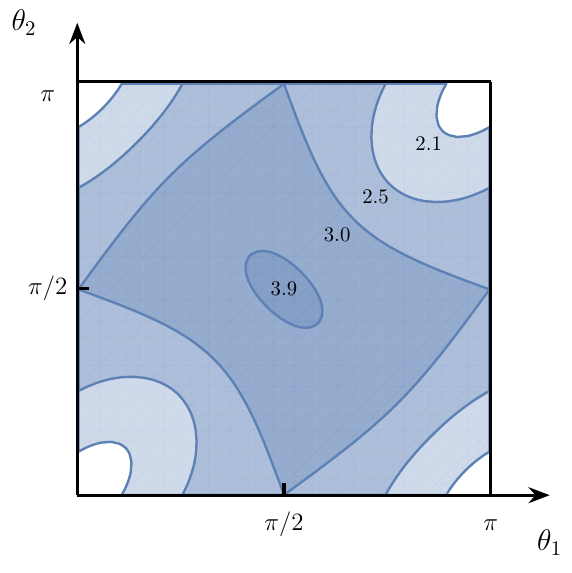}
\caption{Domain $\mathcal{D}(r;\epsilon)$  for two ellipses with aspect ratio $a/b=1+\epsilon=2.0$.  The numbers in the domains are the values for $r$. The white regions at the four corners correspond to $r/b= 2.1> 2$. The region expands to include the light-blue region for $r/b=2.5$. For $r/b= a/b+1= 3.0$, the domain touches the axis. For $r/b=3.9 < 2a/b = 4.0$, the domain consists of the entire square except for the small ellipse-like region  in the center. 
\label{fig:domain}}
\end{figure}

By rotational invariance, we can fix  $\mathbf{e}_{12}=(1,0)^T$   and parametrize $\mathbf{u}_{\alpha}=(\sin \theta_{\alpha},\cos \theta_{\alpha})^T$ in terms of the angles $\theta_\alpha$ between the major axes and  the $z$ axis, such that the variables $(r \mathbf{e}_{12},\mathbf{u}_1,\mathbf{u}_2)$  are replaced by 
$(r,\theta_1,\theta_2)$. Due to the head-tail symmetry, $(\theta_1, \theta_2) \in [0,\pi]^2$. Then 
the area $|\mathcal{D}(r;\epsilon)|$ is provided by
\begin{align} \label{eq:area}
 |\mathcal{D}(r;\epsilon)| &=  \int_{0}^{\pi} d\theta_1 \int_{0}^{\pi} d\theta_2 \ \Theta[\psi(r,\theta_1,\theta_2;\epsilon)]   .
 \end{align}
The qualitative  shape of the   domain   $\mathcal{D}(r;\epsilon)$ is displayed  in Fig.~\ref{fig:domain}  for different regimes of $r$ and $X_0=1+\epsilon= 2.0$. 
Note that $\psi(r,\theta_1,\theta_2;\epsilon)$, and therefore  $\mathcal{D}(r;\epsilon)$ too   is invariant under reflection at both diagonals, i.e., $(\theta_1,\theta_2) \mapsto (\theta_2,\theta_1)$ and
 $(\theta_1,\theta_2) \mapsto (\pi - \theta_2, \pi - \theta_1)$. 
 Using Eq.~\eqref{eq:G},  it is easy to see that $G(\theta_1,\theta_2;\epsilon) =2 +  O(\epsilon^2)$. Neglecting for small shape anisotropy, the term of order $O(\epsilon^2)$, the domain $\mathcal{D}(r;\epsilon)$ gains an additional reflection symmetry 
 $\theta_1 \mapsto \pi  - \theta_1$ and/or  $\theta_2 \mapsto \pi  - \theta_2$. 

 The integrations in  Eq.~\eqref{eq:area} cannot be preformed analytically, in general. But they can be carried out approximately for the full regime  $2b \leq r \leq 2a$  if the shape anisotropy is small,  $\epsilon \ll 1$. 
  Their complete calculation for arbitrary $X_0$  becomes possible  for separations $r\in [2b, 2a]$ close to the boundaries of the interval.  For these three cases, we shall show below that  
\begin{align} \label{eq:g-alpha_A}
  g_{\alpha}(r,\theta_1,\theta_2;\epsilon) = -1 +  \Delta_{\alpha}(r,\theta_1,\theta_2;\epsilon)   ,
 \end{align}  
 such that  the correction becomes small, $|\Delta_{\alpha}(r,\theta_1,\theta_2;\epsilon)| \ll  1$. Substituting this result into Eq.~\eqref{eq:overlap} yields
for the overlap function  $\psi(r,\theta_1,\theta_2;\epsilon) =  -96 \  [ \Delta_{1}(r,\theta_1,\theta_2;\epsilon) +  \Delta_{2}(r,\theta_1,\theta_2;\epsilon)] + O(\Delta_{\alpha}^2) $. Accordingly, to leading order in $\Delta_{\alpha}(r,\theta_1,\theta_2;\epsilon)$ the boundary of the domain  $\mathcal{D}(r;\epsilon)$ follows from 
\begin{align}\label{eq:boundary}
\Delta_{1}(r,\theta_1,\theta_2;\epsilon) +  \Delta_{2}(r,\theta_1,\theta_2;\epsilon) = 0   .
 \end{align} 
 What remains is the calculation of $\Delta_{\alpha}(r,\theta_1,\theta_2;\epsilon)$ for the various limiting cases.

\subsection{Case I: Small shape anisotropy} 
First, we consider  $\epsilon:= X_0 - 1 \ll 1$ and   $2b \leq r \leq 2a$.
Introducing the reduced distance $\tilde{r}(\epsilon)=  (r/2b - 1)/\epsilon \in [0,1]$ and taking into account  $G(\theta_1,\theta_2;\epsilon) - 2 =O(\epsilon^2)$, we obtain from Eqs.~\eqref{eq:g-alpha}, \eqref{eq:G}, and  \eqref{eq:g-alpha_A} that $\Delta_{\alpha}(r,\theta_1,\theta_2;\epsilon) = -8 \epsilon \big[\tilde{r}(\epsilon) - \sin^2 \theta_{\alpha}\big] +  O(\epsilon^2)$, i.e., in leading order in 
$\epsilon$ the condition, Eq.~\eqref{eq:boundary},   for the boundary of $\mathcal{D}(r;\epsilon)$ becomes
\begin{align} \label{eq:boundary-1}
\tilde{r}(\epsilon) - \frac{1}{2} [\sin^2 \theta_1 + \sin^2 \theta_2] = 0,
 \end{align} 
 which has the additional reflection symmetry as discussed above. Therefore,  Eq.~\eqref{eq:area} reduces to 
\begin{align} \label{eq:area_1}
 |\mathcal{D}(r;\epsilon)| &= 4 \  \int_{0}^{\pi/2} d\theta_1 \int_{0}^{\pi/2} d\theta_2 \ \Theta[\psi(r,\theta_1,\theta_2;\epsilon)]       .
 \end{align} 
 Equation~\eqref{eq:boundary-1} implies that the calculation of the integrals in Eq.~\eqref{eq:area_1} require us to distinguish between $0 \leq \tilde{r}(\epsilon) \leq 1/2$  and $1/2 \leq \tilde{r}(\epsilon) \leq 1$,   which corresponds to $b \leq r \leq (a+b)$ and $(a+b) \leq r \leq a$, respectively (see also  Fig.~\ref{fig:domain}).

\subsubsection{Left interval: $0 \leq \tilde{r}(\epsilon) \leq 1/2$}

In this case, the condition in Eq.~\eqref{eq:boundary-1} implies 
$0 \leq \theta_1 \leq \arcsin[\sqrt{2\tilde{r}(\epsilon)}]$     and 
$0 \leq \theta_2 \leq\arcsin[\sqrt{2\tilde{r}(\epsilon) - \sin^2 \theta_1}]$.  This leads to
\begin{align} \label{eq:area-1-1}
 |\mathcal{D}(r;\epsilon)| &= 4 \int_{0}^{\arcsin[\sqrt{2\tilde{r}(\epsilon)}]} d\theta_1   \arcsin\big[\sqrt{2\tilde{r}(\epsilon) - \sin^2 \theta_1}\big]   .
\end{align}

\subsubsection{Right  interval:    $1/2 \leq \tilde{r}(\epsilon) \leq 1$}
From  Eq.~\eqref{eq:boundary-1}, it follows that 
$0 \leq \theta_2 \leq \arcsin[\sqrt{2\tilde{r}(\epsilon)-\sin^2 \theta_1}]$,  if   
$\arcsin[\sqrt{2\tilde{r}(\epsilon)- 1}] \leq  \theta_1 \leq \pi/2$,  
and   $0 \leq \theta_2 \leq \pi/2$,  if $0 \leq \theta_1 \leq \arcsin[\sqrt{2\tilde{r}(\epsilon)- 1}] $. 
Then it follows:
\begin{align} \label{eq:area-1-2}
 |\mathcal{D}(r;\epsilon)| =&  2\pi  \arcsin\big[\sqrt{2\tilde{r}(\epsilon)- 1}\big]   \nonumber\\ 
 & + 4   \int_{\arcsin[\sqrt{2\tilde{r}(\epsilon)-1}]}^{\pi/2} d\theta_1 \arcsin\big[\sqrt{2\tilde{r}(\epsilon)-\sin^2 \theta_1}\big] .
\end{align} 
Note that the dependence of $|\mathcal{D}(r;\epsilon)|$  on the aspect ratio, i.e., on $\epsilon$, arises only through the 
reduced distance $\tilde{r}(\epsilon)$, a feature which holds for  $ v^{(2)}_\text{eff}(r;\epsilon)$ as well.   

The special value $\tilde{r}(\epsilon)=1/2$ corresponds to  $r=(a+b)$ which is the minimal center-to-center distance of the ellipses  if their major semi-axes  are orthogonal to each other. For this case   $|\mathcal{D}(r;\epsilon)|$ can be calculated analytically. From either of the Eqs.~\eqref{eq:area-1-1}, \eqref{eq:area-1-2}, it follows
\begin{align} \label{eq:area-1-3}
 |\mathcal{D}(r=a+b;\epsilon)| = \frac{\pi^2}{2}.
\end{align} 
This result is obvious because for $\tilde{r}(\epsilon)=1/2$, Eq.~\eqref{eq:boundary-1} implies  
$\theta_2 = \pi/2  \pm \theta_1 $  for  $0 \leq \theta_1 \leq   \pi/2$  and 
$\theta_2 = \pi/2 \pm (\pi - \theta_1)$ for  $\pi/2 \leq \theta_1 \leq \pi$.  Accordingly,  $\mathcal{D}(r;\epsilon)$ is a square with edge length $\pi/\sqrt{2}$ and its  area equals  $(\pi^2/2)$. Note that increasing the shape anisotropy more and more, deforms this square to  boundary $3$ in  Fig.~\ref{fig:domain}.
Furthermore, it follows  from Eqs.~\eqref{eq:area-1-1} and \eqref{eq:area-1-2} that $ |\mathcal{D}(r;\epsilon)|$ is singular at $r=a+b$:
\begin{align} \label{eq:area-1-4}
\lim_{\tilde{r} \to (1/2)^\pm} \big(d |\mathcal{D}(r;\epsilon)|/d\tilde{r}\big)(r) = +\infty                   \  ,
\end{align} 
where $\tilde{r} \to (1/2)^\pm$ corresponds to $r \to (a+b)^\pm$. 
Note this result has been proven for $X_0 \to 1$ only. Whether this divergence at $r=a+b$ exists for all $X_0$ is not clear. However, the property that the excluded domain [the complement of $\mathcal{D}(r;\epsilon)$ within the square] touches for  $r= a+b$ and all $X_0$ the sides of the square (see Fig.~\ref{fig:domain}) implies
that  $ |\mathcal{D}(r;\epsilon)|$ is singular at $r= a+b$. 
The numerical results  in Fig.~\ref{fig:soft_2_body_potential}  support that $v^{(2)}_\text{eff}(r;\epsilon)$ and therefore $|\mathcal{D}(r;\epsilon)|$ too  is singular at $\tilde{r} = 1/2$, corresponding to $r=a+b$.  However, whether both  slopes diverge for all $\epsilon$ is not clear.

Substituting the results, Eqs.~\eqref{eq:area-1-1} and \eqref{eq:area-1-2}  for $ |\mathcal{D}(r;\epsilon)|$ into Eq.~\eqref{eq:def_v2eff-1} and calculating the integral in Eqs.~\eqref{eq:area-1-1}  and ~\eqref{eq:area-1-2} numerically yields the effective two-body potential for $X_0 \to 1$ shown in Fig.~\ref{fig:soft_2_body_potential}.

 \subsection{Case II: Close to furthest distance} 
  
 Independent of the aspect ratio $X_0$, the range  of $(\theta_1, \theta_2)$ where the ellipses overlap becomes more and more restricted  to a neighborhood $(\pi/2, \pi/2)$ if $r$ approaches $2a$ from below (see Fig.~\ref{fig:domain}).  Substituting $\theta_{1} = \pi/2 - \eta_{1}, \theta_2 = \pi/2 -\eta_2 $ into   $g_{\alpha}(r,\theta_1,\theta_2;\epsilon)$  [Eq.~\eqref{eq:g-alpha}] and  expanding up to the leading order in $\eta_1$  and  $\eta_2$, the function $g_{\alpha}(r,\theta_1,\theta_2;\epsilon)$ assumes the form of Eq.~\eqref{eq:g-alpha_A} and   in leading order in $\delta_a  :=  1 - r/2a$
we find $\Delta_{\alpha}(r,\eta_1,\eta_2;\epsilon)= [(X_0^2-1)/X_0]^2(\eta_1 - \eta_2)^2 -  4(X_0^2-1) \eta_{\alpha}^2 + 8 \delta_a $. Then the condition in Eq.~\eqref{eq:boundary} becomes
\begin{align} \label{eq:boundary-2}
A(\delta_a) (\eta_1^2 + \eta_2^2) + 2 B(\delta_a) \eta_1 \eta_2   = 1     ,
 \end{align} 
 with coefficients $A(\delta_a)= (X_0^4 -1)/8X_0^2 \delta_a $ and $B(\delta_a)=  (X_0^2-1)^2/8X_0^2\delta_a$. This is an equation for an ellipse tilted by an angle $\pi/4$ with semiaxes 
 $1/\sqrt{A(\delta_a)\pm B(\delta_a)}$. Its  area is given by $\mathcal{A}(\delta_a) =\pi/\sqrt{A(\delta_a)^2 - B(\delta_a)^2}$. Since   $ |\mathcal{D}(r;\epsilon)| = \pi^2 - \mathcal{A}(\delta_a)$,  we find to leading order in $\delta_a$:
\begin{align} \label{eq:area-2} 
 |\mathcal{D}(r;\epsilon)| &= \pi^2 - 4\pi  \frac{X_0}{X_0^2-1}\  \delta_a        .
\end{align}
Note that there is no  divergence of  $|\mathcal{D}(r;\epsilon)|$ at $X_0=1$ since from $r  \leq  2a$ it follows 
$\delta_a  \leq (X_0-1)$ and, consequently, small but fixed $\delta_a $ forces  $X_0 -1$ to remain finite.

 \subsection{Case III: Close to closest distance}

 The case where the two ellipses approach their closest distance is similar to the previous one.
For an arbitrary but finite  aspect ratio, $X_0$, the range of the angles $(\theta_1, \theta_2)$ becomes more and more restricted to small neighborhoods of $(0,0)$, $(0,\pi)$, $(\pi,0)$, and $(\pi,\pi)$; see Fig.~\ref{fig:domain}. The periodic continuation of $\mathcal{D}(r;\epsilon)$ in the direction of  $\theta_1 $ and $\theta_2 $  allows us to determine   $\Delta_{\alpha}(r,\theta_1,\theta_2;\epsilon)$ by expanding  $g_{\alpha}(r,\theta_1,\theta_2;\epsilon)$ [Eq.~\eqref{eq:g-alpha}]  up to leading order in $\theta_1, \theta_2 $.  This expansion is of the form of Eq.~\eqref{eq:g-alpha_A} and   in leading order in
$\delta_b  := r/2b -1$ we find
$\Delta_{\alpha}(r,\theta_1,\theta_2;\epsilon)= [(X_0^2-1)/X_0]^2(\theta_1 - \theta_2)^2 + 4[(X_0^2-1)/X_0^2]\theta_{\alpha}^2 - 8 \delta_b$.  Substituting into the condition of Eq.~\eqref{eq:boundary} yields again  Eq.~\eqref{eq:boundary-2} for an ellipse with  $A(\delta_b)$  and $-B(\delta_b)$.
Therefore,  $|\mathcal{D}(r;\epsilon)|$ is given by its area
$\mathcal{A}(\delta_b) =\pi/\sqrt{A(\delta_b)^2 - B(\delta_b)^2}$. Then it follows
in leading order in $\delta_b$:
\begin{align} \label{eq:area-3}
 |\mathcal{D}(r;\epsilon)| &=  4\pi \frac{X_0}{X_0^2 -1} \delta_b   .
\end{align}
Substituting $ |\mathcal{D}(r;\epsilon)|$  from Eqs.~\eqref{eq:area-2} and \eqref{eq:area-3} into Eq.~\eqref{eq:def_v2eff-1}
and expressing $\delta_a$ and $\delta_b$ by  $\tilde{r}(\epsilon)$,
one obtains the asymptotic results shown in Fig.~\ref{fig:soft_2_body_potential}.
 In order that the r.h.s.\@ of Eq.~\eqref{eq:area-3} for large $X_0$ is small,  the following must apply  $\delta_b \ll 1/X_0$.
Note that  the limits  $\delta_b \to 0$, i.e.,  $r \to 2b$ and $X_0 \to \infty$  do not commute. Therefore, the limit $X_0 \to \infty$ for fixed
 $\delta_b$  has to be discussed separately,  which will be done in the next subsection.

\begin{figure}[htp]
\includegraphics[angle=0,width=0.9\linewidth]{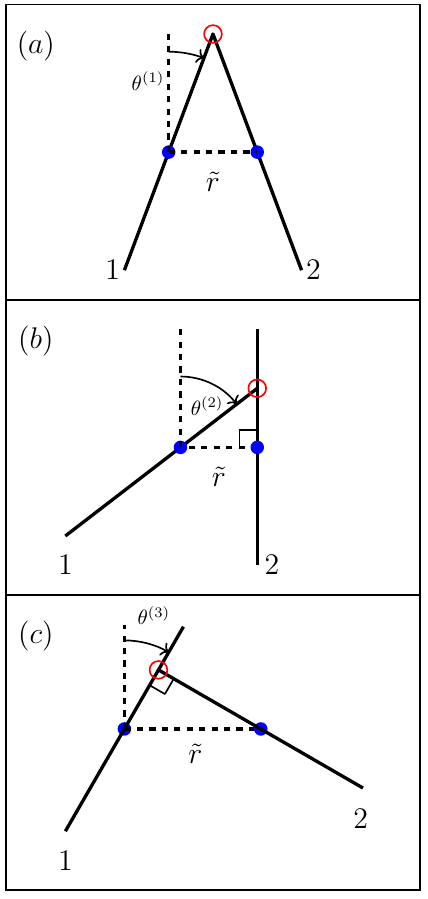}
\caption{Illustration of the characteristic angles $\theta^{(i)}, i=1,2,3$ of needle 1 in the three distinguished  contact configurations (a)-(c) of  needles 1 and 2 of unit length.  The red open circles mark the point of contact and  $\tilde{r}$ is the distance between the needle's centers (blue dots). Configurations (b) and (c) involve a right angle.}
\label{fig:two-needles}
\end{figure}

\subsection{Case IV:  Large shape anisotropy}

In contrast to the three cases studied above,  the dependence of the overlap function on $\theta_1$ and $\theta_2$ does not separate for $X_0 \to \infty$. Nevertheless, the condition $\psi(r\mathbf{e_{12}},\mathbf{u}_1,\mathbf{u}_2;\epsilon) =0$  simplifies significantly. Keeping the length $a$ of the major semiaxis  finite,  choosing   $r= 2a \tilde{r}(\epsilon)$, where $0 \leq \tilde{r}(\epsilon) \leq 1$,  and taking the limit $b \to 0$   we obtain from Eqs.~\eqref{eq:g-alpha} and \eqref{eq:G}: 
$g_{\alpha}(r\mathbf{e}_{12},\mathbf{u}_1,\mathbf{u}_2;\epsilon)= (X_0)^2 [\sin^2 (\theta_1-\theta_2) - 4\tilde{r}(\epsilon)^2 \cos^2 \theta_{\alpha}]  + O(1)$ for $X_0 \to \infty$. Substituting into Eq.~\eqref{eq:overlap}  it follows that 
\begin{align} \label{eq:overlap-infinite}
& \psi(r\mathbf{e_{12}},\mathbf{u}_1,\mathbf{u}_2;\epsilon)  = (X_0)^8\big\{[\sin^2(\theta_1-\theta_2) - 4\tilde{r}(\epsilon)^2 \cos^2 \theta_1] \ \nonumber\\  
& \ \times [\sin^2(\theta_1-\theta_2)- 4\tilde{r}(\epsilon)^2  \cos^2 \theta_2] + O((1/X_0)^2)\big\}  \ .
\end{align}
Therefore, for $X_0 \to \infty$ the boundary of $\mathcal{D}(r)$ follows from 
\begin{align} \label{eq:boundary-infinite-1}
|\sin(\theta_1-\theta_2)| &= 2\tilde{r}(\epsilon) \ |\cos \theta_1|    \nonumber\\  
|\sin(\theta_1-\theta_2)| &= 2\tilde{r}(\epsilon)\  |\cos \theta_2|  
\end{align} 
for $0 \leq \theta_i \leq \pi$, $i=1, 2$. Due to the invariance of   $\mathcal{D}(r)$ under   $(\theta_1,\theta_2) \to (\theta_2,\theta_1)$  and  $(\theta_1,\theta_2) \to (\pi-\theta_2,\pi-\theta_1)$, it suffices to discuss 
$0 \leq  \theta_1 \leq \pi/2$ only.

Fixing $a$ and taking the limit $b \to 0$, the calculation of $\mathcal{D}(r;\epsilon)$  reduces   to that of  $\mathcal{D}(\tilde{r})$, which  is the accessible domain in the orientational space 
$(\mathbf{u}_1,\mathbf{u}_2)$ of two infinitely thin hard rods (needles) of unit length and center-to-center distance $\tilde{r}$. It is straightforward to determine the boundaries of $\mathcal{D}(\tilde{r})$. This shows, in particular, that the first  line of  Eq.~\eqref{eq:boundary-infinite-1} is the contact condition for the case where the lower end of needle 2  touches   needle 1. Similarly, its second line is the contact condition if  the upper end of needle 1 touches needle 2.   Note that  $0 \leq  \theta_1 \leq \pi/2$ and   $0 \leq  \theta_2 \leq \pi$ imply that only these two types of contacts can occur.

The discussion of the two-needle system shows the existence of three characteristic angles of needle 1 (see Fig.~\ref{fig:two-needles}):
\begin{align} \label{eq:angles}
\theta^{(1)}(\tilde{r}) &=  \arcsin(\tilde{r}), \nonumber\\  
\theta^{(2)}(\tilde{r}) &=  \arcsin(2\tilde{r}), \nonumber\\  
\theta^{(3)}(\tilde{r}) &=  \arccos(1/2\tilde{r})   \  .
\end{align} 
These characteristic angles  imply the existence of three different regimes for $\tilde{r}$: (i) $0 \leq \tilde{r} \leq 1/2$, (ii)  $1/2 \leq \tilde{r} \leq 
1/\sqrt{2}$, and (iii)   $1/\sqrt{2} \leq \tilde{r} \leq 1$. Note $\theta^{(1)}(\tilde{r})$ is defined for all $ \tilde{r}$,
 $\theta^{(2)}(\tilde{r})$  for $ \tilde{r} \leq 1/2$, and   $\theta^{(3)}(\tilde{r})$  for $\tilde{r} \geq 1/2$, only.
In  regimes (i)-(iii) we determine the subdomains contributing to   $\mathcal{D}(r)$ which involve 
\begin{align} \label{eq:boundary-infinite-2}
\theta^{<}_2(\theta_1) & =\theta_1 + \arcsin(2\tilde{r} \cos \theta_1), \nonumber\\  
\theta^{>}_2(\theta_1) & =\theta_1 + \pi - \arcsin(2\tilde{r} \cos \theta_1)     ,
\end{align} 
where $\arcsin(x)$  is the branch in the interval $[-\pi/2,\pi/2]$. Note $\theta^{<}_2(\theta_1)$  and $\theta^{>}_2(\theta_1)$  are solutions of the first line of Eq.~\eqref{eq:boundary-infinite-1}.

\subsubsection{Left interval:  $0 \leq \tilde{r} \leq 1/2$} 

In the lower square $0 \leq \theta_1 \leq \pi/2$, $0 \leq \theta_2  \leq \pi/2$,  there is the  subdomain
$\theta_1 \leq \theta_2 \leq \theta^{<}_2(\theta_1)$ for  $0 \leq \theta_1 \leq \pi/2$  and its mirror image at the diagonal $\theta_2=\theta_1$. In the upper square $0 \leq \theta_1 \leq \pi/2$, $\pi/2  \leq \theta_2  \leq \pi$,
we have  the subdomain  $\theta^{>}_2(\theta_1) \leq \theta_2 \leq \pi -\theta_1$ for  $0 \leq  \theta_1 \leq \theta^{(1)}(\tilde{r})$ and its mirror image at the diagonal $\theta_2= \pi - \theta_1$. Use of Eq.~\eqref{eq:boundary-infinite-2}   and accounting for  the identical contribution following for $\pi/2 \leq \theta_1 \leq \pi$, we find

\begin{align} \label{eq:area_1-infinite}
 |\mathcal{D}(r;\epsilon)| =& 4  \int_{0}^{\pi/2} d\theta_1 \arcsin(2\tilde{r} \cos \theta_1)    \nonumber\\
 & + 4 \int_{0}^{\theta^{(1)}(\tilde{r})} d\theta_1 \arcsin(2\tilde{r} \cos \theta_1)  -4 (\theta^{(1)}(\tilde{r}))^2     .
 \end{align}

\subsubsection{Intermediate interval: $1/2 \leq \tilde{r} \leq 1/\sqrt{2}$}
The complete  lower square $0 \leq \theta_1 \leq \pi/2$, $0 \leq \theta_2  \leq \pi/2$   is a  subdomain of $\mathcal{D}(r)$. In the upper square $0 \leq \theta_1 \leq \pi/2$, $\pi/2  \leq \theta_2  \leq \pi$,
we have  the subdomain  consisting of $\theta^{<}_2(\theta_1) \leq \theta_2 \leq \theta^{>}_2(\theta_1)$ for  $\theta^{(3)}(\tilde{r}) \leq  \theta_1 \leq \theta^{(1)}(\tilde{r})$ and  $\theta^{<}_2(\theta_1) \leq \theta_2 \leq \pi - \theta_1$ for  $\theta^{(1)}(\tilde{r}) \leq  \theta_1 \leq \pi/2$,
as well as of its miror image  at the diagonal $\theta_2= \pi - \theta_1$.   Then we obtain

\begin{align} \label{eq:area_2-infinite}
 |\mathcal{D}(r;\epsilon)| = &\pi^2 + 8 \int_{\theta^{(3)}(\tilde{r})}^{\theta^{(1)}(\tilde{r})} d\theta_1 \arcsin(2\tilde{r} \cos \theta_1)    \nonumber\\
 & +  4 \int_{\theta^{(1)}(\tilde{r})}^{\pi/2} d\theta_1 \arcsin(2\tilde{r} \cos \theta_1)  - \big(\pi- 2\theta^{(1)}(\tilde{r})\big)^2     .
 \end{align}

\subsubsection{Right interval: $1/\sqrt{2} \leq \tilde{r} \leq 1$}
Again, the complete  lower square $0 \leq \theta_1 \leq \pi/2$, $0 \leq \theta_2  \leq \pi/2$   is a  subdomain of $\mathcal{D}(r)$. In the upper square $0 \leq \theta_1 \leq \pi/2$, $\pi/2  \leq \theta_2  \leq \pi$,
we have  the subdomain  consisting of $\theta^{<}_2(\theta_1) \leq \theta_2 \leq \pi - \theta_1$ for  $\theta^{(1)}(\tilde{r}) \leq  \theta_1 \leq \pi/2$ and   its miror image  at the diagonal $\theta_2= \pi - \theta_1$.   This yields
 
 \begin{align} \label{eq:area_3-infinite}
 |\mathcal{D}(r;\epsilon)| =& \pi^2 + 4  \int_{\theta^{(1)}(\tilde{r})}^{\pi/2} d\theta_1 \arcsin(2\tilde{r} \cos \theta_1)   \nonumber\\
 &  - \big(\pi- 2\theta^{(1)}(\tilde{r})\big)^2     .
 \end{align}  

Substitution of  Eqs.~\eqref{eq:area_1-infinite}-\eqref{eq:area_3-infinite} into  Eq.~\eqref{eq:def_v2eff-1} and calculating the integrals numerically leads to the graph shown in Fig.~\ref{fig:soft_2_body_potential} for $X_0 \to \infty$.

\begin{widetext}
\section{Calculation of the leading corrections to the free energy}\label{Sec:Appendix_B}

The goal of this Appendix is to provide formal expressions for the leading contributions for certain terms arising in the thermodynamic perturbation series. We will show that to leading order, the cluster integrals in Eqs.~\eqref{eq:free_energy_expan_1} and \eqref{eq:free_energy_expan_2} of the main text can be expressed in terms of the $m$-particle densities of the reference fluid evaluated at closest contact and angular averages of products of a dimensionless contact function.  

To calculate the contributions $F_k(T,V,N)$ to the excess free energy, we have to calculate the averages in Eqs.~\eqref{eq:free_energy_expan_1} and \eqref{eq:free_energy_expan_2} of the main text.
These averages are of the form $\sum_{i_1 < i_2< \cdots < i_m} \langle \langle h(i_1,i_2,\dots,i_m) \rangle_o \rangle_t $, where $h(\ldots)$ is a product of cluster functions. Here, the orientational average 
\begin{align} \label{eq:o-average} 
\langle (\ldots) \rangle_o =   \int \Big[\prod^m_{\mu=1}  \frac{\diff \mathbf{u}_{\mu} }{\Omega_{D}} \Big] \ (\ldots) 
\end{align}
is performed  over all orientational degrees of freedom, and
$\Omega_{D}$ is the surface area of a D-dimensional unit sphere. 

In contrast, the translational average is with respect to the reference fluid of hard spheres of diameter $\sigma_{\text{min}}$. Formally, it can be obtained using the $m$-particle density $\rho^{(m)}(\mathbf{r}_1, \ldots, \mathbf{r}_m) \equiv n^m g^{(m)}(\mathbf{r}_1,\ldots, \mathbf{r}_m)$ \cite{Hansen:Theory_of_Simple_Liquids:2013}. Using  the invariance of the interaction energy under particle permutations, it follows with the decomposition into translational and orientational d.o.f.\@, $i =(i_t,i_o) =(\mathbf{r}_i,\mathbf{u}_i)$,
\begin{align} \label{eq:average-1} 
 \sum_{i_1 < i_2< \cdots < i_m} \langle \langle h(i_1,i_2,\dots,i_m) \rangle_o \rangle_t \  =
{N \choose m} \langle \langle h(1,\ldots, m) \rangle_o \rangle_t =
 \frac{n^m}{m!}  \int \Big[\prod^m_{\mu=1} \diff \mathbf{r}_{\mu}\Big] \ \langle h(\mathbf{r}_1\mathbf{u}_1,\dots,\mathbf{r}_m\mathbf{u}_m) \rangle_o \ g^{(m)}(\mathbf{r}_1,\dots,\mathbf{r}_m) , 
\end{align}
where in the second identity we appoximated ${N \choose m} \simeq N^m/m!$, anticipating the thermodynamic limit.
By translational invariance, $g^{(m)}$ and $h$ depend only on the relative distances of the particles such that Eq.~\eqref{eq:average-1} is proportional to the volume $V$. 

The main observation is now that each bond $f(ij)$ constrains the relative distance $r_{ij} = |\mathbf{r}_i-\mathbf{r}_j|$ to a small interval, allowing for further analytical progress. Let us exemplify the calculation for a bond $f(1,2)$ introduced in Sec. \ref{Sec:Model_and_Framework} [cf. Eq.~\eqref{eq:cluster-1}]:
\begin{align} \label{eq:cluster-2} 
f(1,2) & = \Theta \Big(r_{12} - d_c(\mathbf{e}_{12},\mathbf{u}_{1},\mathbf{u}_{2};\epsilon)\Big)   -  \Theta\Big(r_{12} - \sigma_\text{min}\Big) .
\end{align}
The preceding identity reveals that the cluster function evaluates to $-1$ for $r_{12} \in [\sigma_\text{min}, d_c(\mathbf{e}_{12}, \mathbf{u}_1,\mathbf{u}_2; \epsilon)]$ and is zero otherwise. 
To make further progress, we introduce the dimensionless contact function $\tilde{d}_c = \tilde{d}_c(\mathbf{e}_{12}, \mathbf{u}_1, \mathbf{u}_2; \epsilon)$ by
\begin{align}\label{eq:scaling} 
d_c(\mathbf{e}_{12},\mathbf{u}_{1},\mathbf{u}_{2};\epsilon ) &= \sigma_\text{min} \big[1 + \epsilon \  \tilde{d}_c(\mathbf{e}_{12},\mathbf{u}_{1},\mathbf{u}_{2};\epsilon)  \big] .
\end{align}  
Note that $\tilde{d}_c \in [0,1]$. Its perturbative calculation of  $\tilde{d}_c(\mathbf{e}_{12},\mathbf{u}_{1},\mathbf{u}_{2};\epsilon)$ is presented in Appendix \ref{Sec:Appendix_C}.

The integral over  the relative coordinate $\mathbf{r}_{12} = \mathbf{r}_1- \mathbf{r}_2$ can be decomposed into an integral over the magnitude $r_{12} = |\mathbf{r}_{12}|$ and the direction $\mathbf{e}_{12} = \mathbf{r}_{12} / r_{12}$. Due to the presence of the bond, the integral over the magnitude can be performed to leading order
\begin{align} \label{eq:integral_decomp} 
 \int \diff \mathbf{r}_{12} \langle f(1,2) \ldots \rangle_o g^{(m)}(\mathbf{r}_1, \ldots, \mathbf{r}_m)  =  \int_{\Omega_D} \diff \mathbf{e}_{12} \Big\langle\int_{\sigma_{\text{min}} }^{d_c(\mathbf{e}_{12},\mathbf{u}_1,\mathbf{u}_2;\epsilon)} r_{12}^{D-1} \diff r_{12} 
 [  f(1,2) \ldots ]  \Big\rangle_o g^{(m)}(\mathbf{r}_1, \mathbf{r}_2, \ldots, \mathbf{r}_m) \nonumber \\
= - \Omega_D \sigma_{\text{min}}^D \epsilon  
\int_{\Omega_D} \frac{\diff \mathbf{e}_{12}}{\Omega_D} \langle \tilde{d}_c(\mathbf{e}_{12}, \mathbf{u}_1, \mathbf{u}_2; 0)  \ldots \rangle_o g^{(m)}(\mathbf{r}_1, \mathbf{r}_2, \ldots, \mathbf{r}_m)  + O(\epsilon^2),
\end{align}
where in the last line $\mathbf{r}_2$ is fixed to $\mathbf{r}_1 - \sigma_{\text{min}} \mathbf{e}_{12}$. 

For the general case, the following rule emerges to obtain the leading order in $\epsilon$. After performing the integral over the relative position $\mathbf{r}_{ij} = \mathbf{r}_i-\mathbf{r}_j$, each bond $f(ij)$ yields a factor $-\Omega_D \sigma_{\text{min}}^D \epsilon$. Furthermore, the distance of the pair $(ij)$ is fixed to $ |\mathbf{r}_i - \mathbf{r}_j|= \sigma_{\text{min}}$. Lastly, an average over the direction over the direction $\mathbf{e}_{ij} = (\mathbf{r}_i - \mathbf{r}_j )/\sigma_{\text{min}}$ is performed.   The average over the directions of all  bonds will be indicated by 
\begin{align}
\overline{(\ldots) }= \int  \Big[ \prod_{ \text{bonds } (ij) }  \frac{\diff \mathbf{e}_{ij} }{\Omega_D } \Big] ( \ldots ) .
\end{align}

 \subsection{First correction to the free energy} 
 The leading correction to the free energy, $F_1(T,V,N;\epsilon)$, is obtained from 
Eq.~\eqref{eq:free_energy_expan_1}, where we need  to calculate $\sum_{i < j} \langle \langle f(i,j) \rangle_o \rangle_t $. Application of the rules introduced above leads to 
\begin{align} \label{eq:F1} 
\sum_{i < j} \langle \langle f(i,j) \rangle_o \rangle_t = {N \choose 2} \langle \langle f(1,2) \rangle_o \rangle_t
&= - \frac{1}{2} N [ n^* \Omega_D \   g^{(2)}(\sigma_\text{min}) ] \  \overline{\langle  \tilde{d}_c(\mathbf{e}_{12},\mathbf{u}_{1},\mathbf{u}_{2};0)  \rangle_o } \  \epsilon   + O(\epsilon^2) .
\end{align}
Here we introduced the dimensionless density $n^* = n \sigma_\text{min}^D$ and used translational and rotational invariance to simplify the pair-distribution function $g^{(2)}(|\mathbf{r}_1-\mathbf{r}_2|) \equiv g^{(2)}(\mathbf{r}_1, \mathbf{r}_2)$. Note that due to isotropy, $\langle  \tilde{d}_c(\mathbf{e}_{12},\mathbf{u}_{1},\mathbf{u}_{2};0)  \rangle_o$
is already independent of $\mathbf{e}_{12}$ such that the average $\overline{(\ldots)}$ does not need to be performed.
Substituting this result  into Eq.~(\ref{eq:free_energy_expan_1}) of the main text yields $F_1(T,V,N;\epsilon)$ in Eq.~(\ref{eq:free_energy_expan_1a}).

We can improve Eq.~\eqref{eq:F1} to order $O(\epsilon^2)$ by going back to the first line in Eq.~\eqref{eq:integral_decomp}: 
\begin{align}
\int \diff \mathbf{r}_{12} \langle f(1,2)  \rangle_o g^{(2)}(r_{12}) & 
= - \Omega_D \overline{\left\langle  \int_{\sigma_{\text{min}} }^{d_c(\mathbf{e}_{12}, \mathbf{u}_1,\mathbf{u}_2; \epsilon)} r_{12}^{D-1} \diff r_{12}  g^{(2)}(r_{12})
 \right\rangle_o} .
\end{align}
Upon series expansion of the integrand for distances  $r_{12}$ close to $\sigma_{\text{min}}$,
\begin{align}
r_{12}^{D-1} g^{(2)}(r_{12}) = \sigma_{\text{min}}^{D-1} g^{(2)}(\sigma_{\text{min}}) + (r_{12} - \sigma_{\text{min}} ) [ (D-1) \sigma_{\text{min}}^{D-2} g^{(2)}(\sigma_{\text{min}}) +  \sigma_{\text{min}}^{D-1} g^{(2)}{}'(\sigma_{\text{min}}) ] + O(r_{12}-\sigma_{\text{min}})^2,
\end{align}
we find
\begin{align}
\int \diff \mathbf{r}_{12} \langle f(1,2)  \rangle_o g^{(2)}(r_{12})   
=&- \epsilon \ \Omega_D \sigma_{\text{min}}^{D} g^{(2)}(\sigma_{\text{min}}) \overline{\langle \tilde{d}_c(\mathbf{e}_{12}, \mathbf{u}_1, \mathbf{u}_2;\epsilon)  \rangle_o }  \nonumber \\
&-\frac{\epsilon^2}{2} \ \Omega_D \sigma_{\text{min}}^{D} \overline{\langle \tilde{d}_c(\mathbf{e}_{12}, \mathbf{u}_1, \mathbf{u}_2;\epsilon)^2 \rangle_o }  [ (D-1)  g^{(2)}(\sigma_{\text{min}}) +  \sigma_{\text{min}} g^{(2)}{}'(\sigma_{\text{min}}) ] + O(\epsilon^3).
\end{align}
Substitution of the  series expansion $\tilde{d}_c(\mathbf{e}_{12}, \mathbf{u}_1,\mathbf{u}_2; \epsilon) = \tilde{d}_c(\mathbf{e}_{12}, \mathbf{u}_1,\mathbf{u}_2; 0) + \epsilon \tilde{d}_c'(\mathbf{e}_{12}, \mathbf{u}_1,\mathbf{u}_2; 0)+ O(\epsilon^2)$ allows one to determine $F_1(T,V,N;\epsilon)$ up to $O(\epsilon^2)$. This demonstrates that the higher-order corrections of, e.g., $F_1(T,V,N;\epsilon)$, involve derivatives of the contact function with respect to $\epsilon$.

\subsection{Second correction to the free energy} 
The next-leading contribution contribution, $F_2(T,V,N)$, already involves  two bonds and consists of two parts. 
The first line of Eq.~(\ref{eq:free_energy_expan_2}), denoted by $F^3_2(T,V,N)$, involves two bonds sharing one label. For the first term, we find upon applying the rules from above
\begin{align} \label{eq:F32-1} 
\lefteqn{ \sum_{i < j < k} \langle \langle f(i,j)  f(j,k)\rangle_o \rangle_t    =
{N \choose 3} \langle \langle f(1,2)  f(2,3)\rangle_o \rangle_t = 
 } \nonumber \\
=   \frac{1}{6} N (n^* \Omega_D)^2  &   \overline{\langle  \tilde{d}_c(\mathbf{e}_{12},\mathbf{u}_{1},\mathbf{u}_{2};0)   \tilde{d}_c(\mathbf{e}_{23},\mathbf{u}_{2},\mathbf{u}_{3};0) \rangle_o  g^{(3)}(\sigma_\text{min},\sigma_\text{min}; \mathbf{e}_{12} \cdot \mathbf{e}_{23})} \epsilon^2    + O(\epsilon^3)  ,
\end{align}
where we used that translational invariance and isotropy implies that  $ g^{(3)}(\mathbf{r}_1, \mathbf{r}_2, \mathbf{r}_3) \equiv g^{(3)}(r_{12}, r_{23}, \mathbf{e}_{12} \cdot \mathbf{e}_{23} ) $ depends only on the magnitudes of the relative distances $r_{12} = |\mathbf{r}_{1}-\mathbf{r}_2|, r_{23} = |\mathbf{r}_2-\mathbf{r}_3|$ and the cosine of their relative angle, 
$\cos \angle (\mathbf{r}_{12}, \mathbf{r}_{23}) = \mathbf{e}_{12} \cdot \mathbf{e}_{23}$. 

The second term follows readily using Eq.~(\ref{eq:F1}) with the result
\begin{align} \label{eq:F32-2} 
& \sum_{i < j < k} \langle \langle f(i,j) \rangle_o \rangle_t   \langle \langle f(j,k)\rangle_o \rangle_t \ 
  = {N \choose 3}\langle \langle f(1,2) \rangle_o \rangle_t   \langle \langle f(2,3)\rangle_o \rangle_t   \  \nonumber\\
& =  \frac{1}{6} N [ n^* \Omega_D g^{(2)}(\sigma_\text{min})] ^2  \overline{\langle  \tilde{d}_c(\mathbf{e}_{12},\mathbf{u}_{1},\mathbf{u}_{2};0)  \rangle_o}  
\ \ \overline{\langle  \tilde{d}_c(\mathbf{e}_{23},\mathbf{u}_{2},\mathbf{u}_{3};0) \rangle_o} \ \epsilon^2 \   + O(\epsilon^3)     .
\end{align}
Collecting results, we arrive at
\begin{align} \label{eq:F32} 
& \sum_{i < j < k} \Big[\langle \langle f(i,j)  f(j,k)\rangle_o \rangle_t \ - \ \langle \langle f(i,j) \rangle_o \rangle_t   \langle \langle f(j,k)\rangle_o \rangle_t \Big]  =  \frac{1}{6} N [ n^* \Omega_D \  g^{(2)}(\sigma_\text{min})]^2   \nonumber\\
& \times \Big[\overline{\langle  \tilde{d}_c(\mathbf{e}_{12},\mathbf{u}_{1},\mathbf{u}_{2};0)   \tilde{d}_c(\mathbf{e}_{23},\mathbf{u}_{2},\mathbf{u}_{3};0) \rangle_o \  g^{(3)}(\sigma_\text{min},\sigma_\text{min}; \mathbf{e}_{12} \cdot \mathbf{e}_{23})}/ [g^{(2)}(\sigma_\text{min})]^2   - 
   \overline{\langle\tilde{d}_c(\mathbf{e}_{12},\mathbf{u}_{1},\mathbf{u}_{2};0) \rangle_o   }  
\ \ \overline{\langle \tilde{d}_c(\mathbf{e}_{23},\mathbf{u}_{2},\mathbf{u}_{3};0) \rangle_o  } \Big] \  \epsilon^2 \nonumber\\
& + O(\epsilon^3)   .
\end{align}
Substituting into the first line of Eq.~(\ref{eq:free_energy_expan_2})  yields $F^{3}_2(T,V,N)$  from  Eq.~\eqref{eq:free_energy_expan_2b}.

The second line of Eq.~(\ref{eq:free_energy_expan_2}), denoted by $F_2^2(T,V,N)$,  consists also of two terms.  The calculation of the first term is performed in close analogy to the first term of Eq.~(\ref{eq:free_energy_expan_2}), yielding
\begin{align} \label{eq:F22-1} 
\lefteqn{ \sum_{i < j < k< l} \langle \langle f(i,j)  f(k,l)\rangle_o \rangle_t = 
{N \choose 4} \langle \langle f(1,2)  f(3,4)\rangle_o \rangle_t
} \nonumber \\
 & =  \frac{1}{24} N  n (n^* \Omega_D)^2  \ \overline{\langle  \tilde{d}_c(\mathbf{e}_{12},\mathbf{u}_{1},\mathbf{u}_{2};0) \rangle_o} \ \  \overline{\langle \tilde{d}_c(\mathbf{e}_{34},\mathbf{u}_{3},\mathbf{u}_{4};0) \rangle_o  }      \int \diff \mathbf{r}_{13}
   \overline{  g^{(4)}(\mathbf{r}_1,\mathbf{r}_2, \mathbf{r}_3, \mathbf{r}_4)} \ \epsilon^2   + O(\epsilon^3)    .
\end{align}
Here only the relative separations $\mathbf{r}_{12}$ and $\mathbf{r}_{34}$ are constrained by the bonds, enforcing $r_{12}= r_{34} = \sigma_{\text{min}}$, 
 whereas an integral over the relative distance $\mathbf{r}_{13}$ needs to be performed. We also observed that the orientational averages $\langle \ldots \rangle_o$ over the two contact functions factorize since they do not a share a label. Then by isotropy, these two averages are already independent of $\mathbf{e}_{12}$, respectively, $\mathbf{e}_{34}$ such that the orientational averages over the bonds $(12), (34)$, $\overline{(\ldots)}$ can be performed solely on the four-point function $g^{(4)}$. From a formal point of view, this expression is of order $O(N^2)$ since the integral over the relative distance $\mathbf{r}_{13}$ yields a contribution of the order of the volume. This superextensive behavior is corrected by the second term in the second line of Eq.~\eqref{eq:free_energy_expan_2},
\begin{align} \label{eq:F22-2} 
 &\sum_{i < j < k < l} \langle \langle f(i,j) \rangle_o \rangle_t   \langle \langle f(k,l)\rangle_o \rangle_t 
  = {N \choose 4} \langle \langle f(1,2) \rangle_o \rangle_t   \langle \langle f(3,4)\rangle_o \rangle_t   \  = \nonumber\\
 & \ \equiv  \frac{1}{24}  n N  [ n^* \Omega_D \   g^{(2)}(\sigma_\text{min}) ]^2   \overline{\langle  \tilde{d}_c(\mathbf{e}_{12},\mathbf{u}_{1},\mathbf{u}_{2};0)  \rangle_o } 
  \overline{\langle  \tilde{d}_c(\mathbf{e}_{34},\mathbf{u}_{3},\mathbf{u}_{4};0)  \rangle_o }
\  \epsilon^2  \int \diff \mathbf{r}_{13}+ O(\epsilon^3)  ,
\end{align}
where we wrote a factor $N = n V = n \int \diff \mathbf{r}_{13}$. The difference of the terms in Eqs.~\eqref{eq:F22-1} and \eqref{eq:F22-2} is then expressed as 
\begin{align} \label{eq:F22} 
& \sum_{i < j < k < l} \left[\langle \langle f(i,j)  f(k,l)\rangle_o \rangle_t \ -  \langle \langle f(i,j) \rangle_o \rangle_t   \langle \langle f(k,l)\rangle_o \rangle_t \right]  =  \frac{1}{24} N \left[n^* \Omega_D \  g^{(2)}(\sigma_\text{min})  \overline{\langle  \tilde{d}_c(\mathbf{e}_{12},\mathbf{u}_{1},\mathbf{u}_{2};0) \rangle_o}  
\right]^2 \ (n \Omega_D)  \nonumber\\
& \times  \int^{\infty}_0 \diff r_{13} (r_{13})^{D-1} \ \Big[\overline{ g^{(4)}(\sigma_\text{min},\sigma_\text{min}, r_{13}; \mathbf{e}_{12},\mathbf{e}_{34},\mathbf{e}_{13})}/ \big(g^{(2)}(\sigma_\text{min})\big)^2   - \ 1 \Big] \ \epsilon^2  \ + O(\epsilon^3)   \  .
\end{align}
Here, we used translational and rotational invariance to express the four-point function as
$ g^{(4)}(\mathbf{r}_1, \ldots, \mathbf{r}_4) \equiv g^{(4)}(r_{12},r_{34},r_{13};\mathbf{e}_{12},\mathbf{e}_{34},  \mathbf{e}_{13}) $.   The latter quantity depends only on the relative distances $r_{12}, r_{34}, r_{13}$ after averaging over the bond directions $(12), (34)$ which allows evaluating the integral over the relative separation $\mathbf{r}_{13}$ in polar coordinates. Note that the integral is finite due to the subtraction term.  The square-bracket term of its integrand is the connected four-particle distribution function.  Introducing  the correlation length $\xi(\mathbf{e}_{12},\mathbf{e}_{34},\mathbf{e}_{13})$ defined by
\begin{align}\label{eq:corr-length} 
\xi(\mathbf{e}_{12},\mathbf{e}_{34},\mathbf{e}_{13})^D =& \big(g^{(2)}(\sigma_\text{min})\big)^{-2} \ \int^{\infty}_0 \diff r_{13} \ r_{13}^{D-1} 
 \Big[ g^{(4)}(\sigma_\text{min},\sigma_\text{min},r_{13};\mathbf{e}_{12},\mathbf{e}_{34},\mathbf{e}_{13}) \ - \ \big(g^{(2)}(\sigma_\text{min})\big)^2 \Big]
\end{align} 
and  substituting the result \eqref{eq:F22}    into the second line of Eq.~(\ref{eq:free_energy_expan_2})  yields $F^{2-2}_2(T,V,N)$  from  Eq.~(\ref{eq:free_energy_expan_2a}). 


\end{widetext}

\section{Perturbative calculation of the contact function}\label{Sec:Appendix_C}

In this Appendix we elaborate on a perturbative calculation of the contact function for small shape anisotropy
$\epsilon$.  This will be conducted for two three-dimensional convex  hard nonspherical bodies of revolution with center-to-center separation vector $\mathbf{d} = d \mathbf{e}$ and orientations $\mathbf{u}_{i}, \ i=1,2$.
In contrast to the main text where the dependence on  $\epsilon$ has been partly suppressed,  it will be made explicit. To keep this Appendix self-contained, we will also include the equations already presented in Sec.~\ref{Sec:contact-f} of the main text.

In the space-fixed frame, the surface of such a  body can be parametrized by [see Sec.~\ref{Sec:Model_and_Framework}]
\begin{align} \label{eq:surface-C}
S(\mathbf{u};\epsilon)  = \{ \mathbf{s}|\  \mathbf{s} = & \mathbf{s}(\omega,\mathbf{u};\epsilon) \equiv  s(\vartheta;\epsilon)  R(\mathbf{u}) \mathbf{e}_s(\omega) , 
\nonumber\\
& 0 \leq \vartheta \leq \pi \ , \ 0 \leq \varphi < 2\pi \ , |\mathbf{e}_s|=1  \}  ,
\end{align}
where $\mathbf{e}_s(\vartheta,\varphi) = 
 ( \sin\vartheta  \cos\varphi, \sin\vartheta \sin\varphi, \cos\vartheta)^T$  is  the unit vector  pointing in the body-fixed frame to the surface point in the direction $\omega=(\vartheta, \varphi)$. The matrix  $R(\mathbf{u})$ rotates the body's symmetry axis from the $z$ direction into the direction of the unit vector $\mathbf{u}$. The  shape function $s(\vartheta;\epsilon)$ is positive and does not depend on $\varphi$. Note:  Since  hard particles are included with shapes completely determined by $\sigma_\text{min}$ and the aspect ratio  $X_0=1+\epsilon$,  $\tilde{s}$ also depends on $\epsilon$ (see discussion in Sec.~\ref{Sec:model}).

Besides $\mathbf{e}_s(\omega)$, we introduce $\mathbf{e}_{\vartheta}(\omega) := \partial \mathbf{e}_s/ \partial  \vartheta =  (\cos\vartheta  \cos\varphi, \cos\vartheta \sin\varphi, -\sin\vartheta)^T$ and $\mathbf{e}_{\varphi}(\omega) := (\partial \mathbf{e}_s/ \partial  \varphi)/\sin\vartheta =  (-\sin\varphi,\cos\varphi,0)^T$, which form an orthonormal basis. In the following  their  derivatives are also needed:
\begin{align} \label{eq:derivative}
\partial \mathbf{e}_{\vartheta}/ \partial  \vartheta & = - \mathbf{e}_{s}    \ , \   \  \partial \mathbf{e}_{\vartheta} / \partial  \varphi =   \mathbf{e}_{\varphi}  \  \cos\vartheta ,   \nonumber\\
\partial \mathbf{e}_{\varphi}/ \partial  \vartheta & = 0   \ , \   \  \partial \mathbf{e}_{\varphi}/ \partial  \varphi =   - \mathbf{e}_{s}  \  \sin\vartheta  - \mathbf{e}_{\vartheta} \cos \vartheta .
\end{align}
To find the conditions for an exterior tangential contact, we also need the normal vector 
$\mathbf{n} = \mathbf{t}_{\vartheta} \times \mathbf{t}_{\varphi}$,    which is the cross product of  
the orthonormal  tangential vectors
$\mathbf{t}_{\vartheta}= (\partial  \mathbf{s}/ \partial \vartheta)/| \partial  \mathbf{s}/ \partial \vartheta|$ and 
$\mathbf{t}_{\varphi} = (\partial  \mathbf{s}/ \partial \varphi )/| \partial  \mathbf{s}/ \partial \varphi|$.
Using $ \mathbf{s}(\omega,\mathbf{u};\epsilon)$ in Eqs.~(\ref{eq:surface-C}) and  ~(\ref{eq:derivative}),
  we obtain for the tangential vectors 
\begin{align}\label{eq:tangent-vector} 
\mathbf{t}_{\vartheta}(\omega,\mathbf{u};\epsilon) & = R(\mathbf{u}) \frac{ s'(\vartheta;\epsilon)\mathbf{e}_s(\omega) + s(\vartheta;\epsilon)\mathbf{e}_{\vartheta}(\omega) }{\sqrt{( s'(\vartheta;\epsilon))^2 + ( s(\vartheta;\epsilon))^2}}  ,
\nonumber\\
 \mathbf{t}_{\varphi}(\omega,\mathbf{u};\epsilon) & = R(\mathbf{u}) \mathbf{e}_{\varphi}(\omega)  ,
\end{align}  
where $s' = \partial  s/ \partial \vartheta$ .   Then,  with 
$\mathbf{e}_{s} \times \mathbf{e}_{\varphi} = - \ \mathbf{e}_{\vartheta}$ and $\mathbf{e}_{\vartheta} \times \mathbf{e}_{\varphi} = \mathbf{e}_{s}$. we find for the normal vector
\begin{align}\label{eq:normal-vector} 
\mathbf{n}(\omega,\mathbf{u};\epsilon) & = R(\mathbf{u}) \frac{ -  s'(\vartheta;\epsilon)\mathbf{e}_{\vartheta}(\omega) + s(\vartheta;\epsilon)\mathbf{e}_{s}(\omega) }{\sqrt{( s'(\vartheta;\epsilon))^2 + ( s(\vartheta;\epsilon))^2}} .
\end{align}  
Making use of these relations, the conditions for an exterior tangential contact read
\begin{subequations}
\begin{align}\label{eq:contact-condition} 
\mathbf{s}(\omega_1,\mathbf{u}_1;\epsilon) & = d_c( \mathbf{e}_{12},\mathbf{u}_1,\mathbf{u}_2;\epsilon) \ \mathbf{e}_{12} +  \mathbf{s}(\omega_2,\mathbf{u}_2;\epsilon)  , 
\\
\mathbf{n}(\omega_1,\mathbf{u}_1;\epsilon) & = - \  \mathbf{n}(\omega_2,\mathbf{u}_2;\epsilon)  \ .
\end{align}  
\end{subequations}
The solution of Eq.~\eqref{eq:contact-condition} yields $\omega_i( \mathbf{e}_{12},\mathbf{u}_1,\mathbf{u}_2;\epsilon), \ i=1,2$ for the point of contact and  $d_c( \mathbf{e}_{12},\mathbf{u}_1,\mathbf{u}_2;\epsilon)$, the contact 
function  which is the most important quantity in our analysis., Even for ellipses, one of the simplest hard bodies,
these equations cannot be solved analytically. Therefore, we elaborate on a perturbative approach using  $\epsilon$ as smallness parameter.   Accordingly,  suppressing  the dependence  on $(\mathbf{e}_{12},\mathbf{u}_1,\mathbf{u}_2)$,  we use the series expansion 
\begin{align}\label{eq:omega-series} 
 \omega_i( \epsilon)  = \sum^{\infty}_{\nu=0}  \omega^{(\nu)}_i   \  \epsilon^{\nu}      \ ,  
 \  \  \omega^{(\nu)}_i=( \vartheta^{(\nu)}_i, \varphi^{(\nu)}_i)         \  , 
\end{align}  
and rewrite the shape and contact function
\begin{subequations}
\begin{align}\label{eq:paramet} 
s(\vartheta;\epsilon) & =   \frac{\sigma_\text{min}}{2} \big[1 \ + \  \epsilon \ \tilde{s}(\vartheta;\epsilon) \big]       ,  \\ 
\label{eq:paramet2} d_c(\epsilon) & = \sigma_\text{min}  \big[1 \ + \  \epsilon \ \tilde{d}_c(\epsilon) \big]     ,   
\end{align} 
\end{subequations}
and use  the  pertubative series
\begin{align}\label{eq:s-series} 
 \tilde{s}(\vartheta;\epsilon) = \sum^{\infty}_{\nu=1}  \tilde{s}_{\nu} (\vartheta)  \  \epsilon^{\nu-1}     ,
\end{align}  
and
\begin{align}\label{eq:d-expan_C} 
 \tilde{d}_c(\epsilon) =  \sum^{\infty}_{\nu=1} \tilde{d}_{\nu}   \epsilon^{\nu -1} .
\end{align} 
First,  we replace $ \omega_i$ in $s( \omega_i,\dots)$ and $\mathbf{n}(\omega_i,\dots)$ by 
$ \omega_i( \epsilon)$ from Eq.~\eqref{eq:omega-series} and expand with respect to $\epsilon$. Use of Eqs.~\eqref{eq:derivative} leads to 
\begin{align}\label{eq:s-expan-1_C} 
& \mathbf{s}(\omega_i( \epsilon) ;\epsilon) =  R(\mathbf{u}_i) \Big\{s(\vartheta^{(0)}_i;\epsilon) \ \mathbf{e}_{s}(\omega^{(0)}_i ) \ +  \ \nonumber\\
& + \Big[\vartheta^{(1)}_i    \Big(s'(\vartheta^{(0)}_i;\epsilon) \mathbf{e}_{s}(\omega^{(0)}_i) \ + \
   s(\vartheta^{(0)}_i;\epsilon)\mathbf{e}_{\vartheta}(\omega^{(0)}_i) \Big)    \nonumber\\  
 & \ +  \ \varphi^{(1)}_i  \sin\vartheta^{(0)}_i s(\vartheta^{(0)}_i;\epsilon) \mathbf{e}_{\varphi}(\omega^{(0)}_i)  \Big] \ \epsilon  \ +O(\epsilon^2)  \Big\}  \  .
\end{align} 
Taking into account $s'(\vartheta_i;\epsilon) = O(\epsilon) $ as well as $s(\vartheta_i;\epsilon) > 0$,  we obtain for the normal vector
\begin{align}\label{eq:n-expan-1_C} 
& \mathbf{n}(\omega_i( \epsilon) ;\epsilon) =  R(\mathbf{u}_i) \Big\{ \Big[\mathbf{e}_{s}(\omega^{(0)}_i )  - 
\frac{s'(\vartheta^{(0)}_i;\epsilon)}{s(\vartheta^{(0)}_i;\epsilon)} \ \mathbf{e}_{\vartheta}(\omega^{(0)}_i ) \Big]  \nonumber\\
&  + \Big[\vartheta^{(1)}_i  \mathbf{e}_{\vartheta}(\omega^{(0)}_i) 
   +  \ \varphi^{(1)}_i  \sin\vartheta^{(0)}_i  \mathbf{e}_{\varphi}(\omega^{(0)}_i)  \Big] \epsilon  +O(\epsilon^2)  \Big\} .
\end{align} 
Second, using  Eq.~\eqref{eq:paramet} and the series, Eq.~\eqref{eq:s-series}, we  expand  the r.h.s.\@ of Eqs.~\eqref{eq:s-expan-1_C} and \eqref{eq:n-expan-1_C} with respect to~$\epsilon$. This yields
\begin{align}\label{eq:s-expan-2_C} 
& \mathbf{s}(\omega_i( \epsilon) ;\epsilon) =  \frac{\sigma_\text{min}}{2}  R(\mathbf{u}_i) \Big\{\ \mathbf{e}_{s}(\omega^{(0)}_i ) \ + \ \Big[\tilde{s}_1(\vartheta^{(0)}_i) \ \mathbf{e}_{s}(\omega^{(0)}_i )    \ +            \nonumber\\
& \ + \ \vartheta^{(1)}_i  \mathbf{e}_{\vartheta}(\omega^{(0)}_i )    
\ +  \ \varphi^{(1)}_i  \sin\vartheta^{(0)}_i \mathbf{e}_{\varphi}(\omega^{(0)}_i)  \Big]  \  \epsilon \ +O(\epsilon^2)  \Big\} 
\end{align} 
and
\begin{align}\label{eq:n-expan-2_C} 
\lefteqn{  \mathbf{n}(\omega_i( \epsilon) ;\epsilon) =} \nonumber \\
=&  R(\mathbf{u}_i) \Big\{ \mathbf{e}_{s}(\omega^{(0)}_i )   +  
 \Big[\Big(- \tilde{s}_1 '(\vartheta^{(0)}_i) +  \vartheta^{(1)}_i \Big)  \mathbf{e}_{\vartheta}(\omega^{(0)}_i)
\nonumber\\
&  \ + \
 \varphi^{(1)}_i  \sin\vartheta^{(0)}_i  \mathbf{e}_{\varphi}(\omega^{(0)}_i)  \Big] \epsilon  +O(\epsilon^2)  \Big\} .
\end{align} 
In the final step, we substitute Eqs.~\eqref{eq:s-expan-2_C}, \eqref{eq:n-expan-2_C}, and \eqref{eq:paramet2} with $\tilde{d_c}(\epsilon )$ from  Eq.~\eqref{eq:d-expan_C}  into Eq.~\eqref{eq:contact-condition}. Comparison of the coefficients of order 
$\epsilon^{\nu}$  yields for the zeroth order
\begin{align}\label{eq:contact-condition-0} 
R(\mathbf{u}_1) \mathbf{e}_s(\omega^{(0)}_1 )  &=  2\mathbf{e}_{12} + R(\mathbf{u}_2) \mathbf{e}_s(\omega^{(0)}_2 )  , \nonumber \\
R(\mathbf{u}_1) \mathbf{e}_s(\omega^{(0)}_1 )  &=  - R(\mathbf{u}_2) \mathbf{e}_s(\omega^{(0)}_2 )  ,
\end{align}
and for the  first order
\begin{align}\label{eq:contact-condition-1-a} 
& R(\mathbf{u}_1) \Big[ \tilde{s}_1(\vartheta^{(0)}_1) \ \mathbf{e}_{s}(\omega^{(0)}_1)  +  \vartheta^{(1)}_1 \mathbf{e}_{\vartheta}(\omega^{(0)}_1 )  +  \varphi^{(1)}_1 \sin\vartheta^{(0)}_1 \mathbf{e}_{\varphi}(\omega^{(0)}_1)\Big]   \nonumber\\          
 & = 2\tilde{d_1}\mathbf{e}_{12} +  R(\mathbf{u}_2) \Big[ \tilde{s}_1(\vartheta^{(0)}_2) \ \mathbf{e}_{s}(\omega^{(0)}_2)  +  \vartheta^{(1)}_2 \mathbf{e}_{\vartheta}(\omega^{(0)}_2 )  \nonumber\\
 & \ + \  \varphi^{(1)}_2 \sin\vartheta^{(0)}_2 \mathbf{e}_{\varphi}(\omega^{(0)}_2)\Big]   ,
\end{align} 
as well as 
\begin{align}\label{eq:contact-condition-1-b} 
& R(\mathbf{u}_1) \Big[ \big(- \tilde{s}_1'(\vartheta^{(0)}_1) + \vartheta^{(1)}_1 \big)  \mathbf{e}_{\vartheta}(\omega^{(0)}_1) + \varphi^{(1)}_1 \sin\vartheta^{(0)}_1 \mathbf{e}_{\varphi}(\omega^{(0)}_1) \Big]  \nonumber\\
& = - R(\mathbf{u}_2) \Big[ \big(- \tilde{s}_1'(\vartheta^{(0)}_2) + \vartheta^{(1)}_2 \big)  \mathbf{e}_{\vartheta}(\omega^{(0)}_2)   \nonumber\\
&  \ + \ \varphi^{(1)}_2 \sin\vartheta^{(0)}_2 \mathbf{e}_{\varphi}(\omega^{(0)}_2) \Big] \ .
\end{align} 
Equations~\eqref{eq:contact-condition-0} imply
\begin{align}\label{eq:contact-condition-0-d} 
R(\mathbf{u}_1) \mathbf{e}_s(\omega^{(0)}_1 )  = \mathbf{e}_{12} = - \  R(\mathbf{u}_2) \mathbf{e}_s(\omega^{(0)}_2 ) .  
\end{align} 
Multiplying Eq.~\eqref{eq:contact-condition-1-a} by  $\mathbf{e}_{12}$ and taking into account that 
$R(\mathbf{u}_i)\mathbf{e}_{\vartheta}(\omega^{(0)}_i )$ and  $R(\mathbf{u}_i)\mathbf{e}_{\varphi}(\omega^{(0)}_i )$ are orthogonal to $\mathbf{e}_{12}$ [which follows from  Eq.~\eqref{eq:contact-condition-0-d} and the orthonormality of $\big(\mathbf{e}_s(\omega^{(0)}_i ),\mathbf{e}_{\vartheta}(\omega^{(0)}_i ),\mathbf{e}_{\varphi}(\omega^{(0)}_i )  \big)$]  yields  
\begin{align}\label{eq:d-1-order-1} 
\lefteqn{\tilde{d}_1(\mathbf{e}_{12},\mathbf{u}_1,\mathbf{u}_2) = } \nonumber \\
&= \frac{1}{2}  \big[\tilde{s}_1(\vartheta^{(0)}_1(\mathbf{e}_{12},\mathbf{u}_1,\mathbf{u}_2)) + 
\tilde{s}_1(\vartheta^{(0)}_2(\mathbf{e}_{12},\mathbf{u}_1,\mathbf{u}_2)) \big]  ,
\end{align}
the relation between the contact function and the shape function in first order. Here we reintroduced the dependence on $(\mathbf{e}_{12},\mathbf{u}_1,\mathbf{u}_2)$.

For a complete solution of the first-order conditions, Eqs.~\eqref{eq:contact-condition-1-a} and  \eqref{eq:contact-condition-1-b}, one has to proceed as follows. Due to Eq.~\eqref{eq:contact-condition-0-d} 
$R(\mathbf{u}_1) \mathbf{e}_s(\omega^{(0)}_1 )$ and $R(\mathbf{u}_2) \mathbf{e}_s(\omega^{(0)}_2 )$ correspond
to two antipodal points on the unit sphere. Accordingly, the two remaining orthonormal basis vectors at these two points are related by 
\begin{align}\label{eq:contact-condition-0-c} 
R(\mathbf{u}_1) \mathbf{e}_\varphi(\omega^{(0)}_1 ) & =\  -\ R(\mathbf{u}_2) \mathbf{e}_\varphi(\omega^{(0)}_2 ) ,  \nonumber\\
R(\mathbf{u}_1) \mathbf{e}_\vartheta(\omega^{(0)}_1 ) & =\  \ R(\mathbf{u}_2) \mathbf{e}_\vartheta(\omega^{(0)}_2) .
\end{align} 
Replacing $\mathbf{e}_{12}$ in Eq.~\eqref{eq:contact-condition-1-a}   by $R(\mathbf{u}_1) \mathbf{e}_s(\omega^{(0)}_1 )$  and making use of Eq.~\eqref{eq:contact-condition-0-c} leads to a linear combination  of the orthonomal basis vectors   $R(\mathbf{u}_1) \mathbf{e}_m(\omega^{(0)}_1 )$, $m=s, \vartheta, \varphi$ which equals zero. The condition that its coefficients vanish, yields the result, Eq.~\eqref{eq:contact-condition-0-d},  and conditions between
$(\vartheta^{(0)}_1, \varphi^{(0)}_1)$  and  $(\vartheta^{(0)}_2, \varphi^{(0)}_2)$ which, however, are not needed for our present purpose.

What remains to be determined is the relation between $\omega^{(0)}_i=(\vartheta^{(0)}_i,\varphi^{(0)}_i)$ and $(\mathbf{e}_{12},\mathbf{u}_1,\mathbf{u}_2)$. This relation follows from Eq.~\eqref{eq:contact-condition-0-d} by using
$\mathbf{e}_{s}(\omega^{(0)}_i) = \mathbf{e}_{z} \cos\vartheta^{(0)}_i + \mathbf{e}_\varphi(\omega^{(0)}_i)\times \mathbf{e}_z \sin\vartheta^{(0)}_i$ 
 and $R(\mathbf{u}_i)\mathbf{e}_{z} = \mathbf{u}_i$. 
 Then,  we find after multiplication of both sides of  Eq.~\eqref{eq:contact-condition-0-d} with $\mathbf{u}_i, i = 1,2$,
\begin{align}\label{eq:theta_0} 
\vartheta^{(0)}_1  (\mathbf{e}_{12},\mathbf{u}_1,\mathbf{u}_2) & =  \arccos(\mathbf{e}_{12} \cdot \mathbf{u}_1) ,   \nonumber\\
\vartheta^{(0)}_2(\mathbf{e}_{12},\mathbf{u}_1,\mathbf{u}_2) & =  \  \arccos(- \mathbf{e}_{12} \cdot \mathbf{u}_2)   .  
\end{align}
 Equations.~\eqref{eq:d-1-order-1} and \eqref{eq:theta_0} lead to the final  result
\begin{align}\label{eq:1-order-3} 
\lefteqn{
\tilde{d}_1(\mathbf{e}_{12},\mathbf{u}_1,\mathbf{u}_2) = } \nonumber \\
&= \frac{1}{2}  \big[\tilde{s}_1\big( \arccos(\mathbf{e}_{12} \cdot \mathbf{u}_1) \big) + 
\tilde{s}_1\big(\arccos( - \  \mathbf{e}_{12} \cdot \mathbf{u}_2)\big) \big]   ,
\end{align}
which is Eq.~\eqref{eq:d-expan-0}  of the main text.  

From Eq.~\eqref{eq:contact-condition-0-d} we also obtain 
$\mathbf{e}_s(\omega^{(0)}_{1/2} ) = \pm  \ R^{-1}(\mathbf{u}_{1/2}) \mathbf{e}_{12}$. Multiplication of both sides with $ \mathbf{e}_x $  yields with $(\mathbf{e}_x \cdot \mathbf{e}_s(\omega^{(0)}_{1/2} )) = \sin\vartheta^{(0)}_i \cos\varphi^{(0)}_i $ and $\sin\vartheta^{(0)}_i = \sqrt{1 - (\mathbf{e}_{12} \cdot \mathbf{u}_i)^2}$ [which follows from Eq.~\eqref{eq:theta_0}]:
\begin{align}\label{eq:1-order-4} 
& \sqrt{1 - (\mathbf{e}_{12} \cdot \mathbf{u}_{1/2})^2} \ \cos\varphi^{(0)}_{1/2}(\mathbf{e}_{12},\mathbf{u}_1,\mathbf{u}_2)  = 
 \nonumber \\ 
=& \pm \  \mathbf{e}_x \cdot  R^{-1}(\mathbf{u}_{1/2}) \mathbf{e}_{12}   .
\end{align}
For $|\mathbf{e}_{12} \cdot \mathbf{u}_{i}| \neq 1$, this allows one to calculate $\varphi^{(0)}_{i}(\mathbf{e}_{12},\mathbf{u}_1,\mathbf{u}_2)$. For $|\mathbf{e}_{12} \cdot \mathbf{u}_{i}| = 1$, Equations~\eqref{eq:theta_0} imply $\vartheta^{(0)}_1 =0$  and  $\vartheta^{(0)}_2 = \pi$. Due to the singularity of the polar coordinates at the poles , $\varphi^{(0)}_{i}$ is not defined for $\vartheta^{(0)}_{1}, \vartheta^{(0)}_{2}$ equal to $0$ or $\pi$.

Accordingly, we succeeded in expressing the first-order contribution of the  contact function by the
first-order term of an arbitrary shape function $s(\vartheta;\epsilon)$ and  determining the polar coordinates of the contact point in zeroth order.

Since the thermodynamic quantities in the main text are calculated only up to the first order in $\epsilon$,
the analytical knowledge of $\tilde{d}_1(\mathbf{e}_{12},\mathbf{u}_1,\mathbf{u}_2)$ is  sufficient.  Of course, 
from a purely theoretical point of view it would interesting to continue the perturbative procedure systematically
to higher orders. 
We expect that already the next-to-leading order, $\tilde{d}_2(\mathbf{e}_{12},\mathbf{u}_1,\mathbf{u}_2)$, will also depend on $(\mathbf{u}_1 \cdot \mathbf{u}_2)$, in contrast to $\tilde{d}_1(\mathbf{e}_{12},\mathbf{u}_1,\mathbf{u}_2)$.


\begin{thebibliography}{76}%
\makeatletter
\providecommand \@ifxundefined [1]{%
 \@ifx{#1\undefined}
}%
\providecommand \@ifnum [1]{%
 \ifnum #1\expandafter \@firstoftwo
 \else \expandafter \@secondoftwo
 \fi
}%
\providecommand \@ifx [1]{%
 \ifx #1\expandafter \@firstoftwo
 \else \expandafter \@secondoftwo
 \fi
}%
\providecommand \natexlab [1]{#1}%
\providecommand \enquote  [1]{``#1''}%
\providecommand \bibnamefont  [1]{#1}%
\providecommand \bibfnamefont [1]{#1}%
\providecommand \citenamefont [1]{#1}%
\providecommand \href@noop [0]{\@secondoftwo}%
\providecommand \href[0]{\begingroup \@sanitize@url \@href}%
\providecommand \@href[1]{\@@startlink{#1}\@@href}%
\providecommand \@@href[1]{\endgroup#1\@@endlink}%
\providecommand \@sanitize@url [0]{\catcode `\\12\catcode `\$12\catcode
  `\&12\catcode `\#12\catcode `\^12\catcode `\_12\catcode `\%12\relax}%
\providecommand \@@startlink[1]{}%
\providecommand \@@endlink[0]{}%
\providecommand \url  [0]{\begingroup\@sanitize@url \@url }%
\providecommand \@url [1]{\endgroup\@href {#1}{\urlprefix }}%
\providecommand \urlprefix  [0]{URL }%
\providecommand \Eprint [0]{\href }%
\providecommand \doibase [0]{http://dx.doi.org/}%
\providecommand \selectlanguage [0]{\@gobble}%
\providecommand \bibinfo  [0]{\@secondoftwo}%
\providecommand \bibfield  [0]{\@secondoftwo}%
\providecommand \translation [1]{[#1]}%
\providecommand \BibitemOpen [0]{}%
\providecommand \bibitemStop [0]{}%
\providecommand \bibitemNoStop [0]{.\EOS\space}%
\providecommand \EOS [0]{\spacefactor3000\relax}%
\providecommand \BibitemShut  [1]{\csname bibitem#1\endcsname}%
\let\auto@bib@innerbib\@empty
\bibitem [{\citenamefont {Onsager}(1949)}]{Onsager:AnnalsAcademy_51:1949}%
  \BibitemOpen
  \bibfield  {author} {\bibinfo {author} {\bibfnamefont {L.}~\bibnamefont
  {Onsager}},\ }\bibfield  {title} {\enquote {\bibinfo {title} {The effects of
  shape on the interaction of colloidal particles},}\
  }\href{https://nyaspubs.onlinelibrary.wiley.com/doi/abs/10.1111/j.1749-6632.1949.tb27296.x}
  {\bibfield  {journal} {\bibinfo  {journal} {Annals of the New York Academy of
  Sciences}\ }\textbf {\bibinfo {volume} {51}},\ \bibinfo {pages} {627}
  (\bibinfo {year} {1949})}\BibitemShut {NoStop}%
\bibitem [{\citenamefont {Hansen}\ and\ \citenamefont
  {McDonald}(2013)}]{Hansen:Theory_of_Simple_Liquids:2013}%
  \BibitemOpen
  \bibfield  {author} {\bibinfo {author} {\bibfnamefont {J.~P.}\ \bibnamefont
  {Hansen}}\ and\ \bibinfo {author} {\bibfnamefont {I.~R.}\ \bibnamefont
  {McDonald}},\ }\href{\doibase 10.1016/C2010-0-66723-X} {\emph {\bibinfo
  {title} {Theory of {Simple} {Liquids}}}},\ \bibinfo {edition} {4th}\ ed.\
  (\bibinfo  {publisher} {Academic Press},\ \bibinfo {address} {London},\
  \bibinfo {year} {2013})\BibitemShut {NoStop}%
\bibitem [{\citenamefont {Priestly}(2012)}]{Priestly:Liquid_Crystals:2012}%
  \BibitemOpen
  \bibfield  {author} {\bibinfo {author} {\bibfnamefont {E.}~\bibnamefont
  {Priestly}},\
  }\href{https://books.google.at/books?hl=de&lr=&id=5uviBwAAQBAJ&oi=fnd&pg=PA1&ots=ZKrBUKqWpP&sig=ayXaiK7G8Pzp1Xwk9LCmwYZ40Zs&redir_esc=y#v=onepage&q&f=false}
  {\emph {\bibinfo {title} {Introduction to liquid crystals}}}\ (\bibinfo
  {publisher} {Springer Science \& Business Media},\ \bibinfo {year}
  {2012})\BibitemShut {NoStop}%
\bibitem [{\citenamefont {Stephen}\ and\ \citenamefont
  {Straley}(1974)}]{Stephen:RMP_46:1974}%
  \BibitemOpen
  \bibfield  {author} {\bibinfo {author} {\bibfnamefont {M.~J.}\ \bibnamefont
  {Stephen}}\ and\ \bibinfo {author} {\bibfnamefont {J.~P.}\ \bibnamefont
  {Straley}},\ }\bibfield  {title} {\enquote {\bibinfo {title} {Physics of
  liquid crystals},}\ }\href{\doibase 10.1103/RevModPhys.46.617} {\bibfield
  {journal} {\bibinfo  {journal} {Rev. Mod. Phys.}\ }\textbf {\bibinfo {volume}
  {46}},\ \bibinfo {pages} {617} (\bibinfo {year} {1974})}\BibitemShut
  {NoStop}%
\bibitem [{\citenamefont {Glotzer}\ and\ \citenamefont
  {Solomon}(2007)}]{Glotzer:NatureMaterials_6:2007}%
  \BibitemOpen
  \bibfield  {author} {\bibinfo {author} {\bibfnamefont {S.~C.}\ \bibnamefont
  {Glotzer}}\ and\ \bibinfo {author} {\bibfnamefont {M.~J.}\ \bibnamefont
  {Solomon}},\ }\bibfield  {title} {\enquote {\bibinfo {title} {Anisotropy of
  building blocks and their assembly into complex structures},}\
  }\href{https://www.nature.com/articles/nmat1949} {\bibfield  {journal}
  {\bibinfo  {journal} {Nature materials}\ }\textbf {\bibinfo {volume} {6}},\
  \bibinfo {pages} {557} (\bibinfo {year} {2007})}\BibitemShut {NoStop}%
\bibitem [{\citenamefont
  {Vieillard‐Baron}(1972)}]{Vieillard-Baron:JCP_56:1972}%
  \BibitemOpen
  \bibfield  {author} {\bibinfo {author} {\bibfnamefont {J.}~\bibnamefont
  {Vieillard‐Baron}},\ }\bibfield  {title} {\enquote {\bibinfo {title} {Phase
  transitions of the classical hard‐ellipse system},}\ }\href{\doibase
  https://doi.org/10.1063/1.1676946} {\bibfield  {journal} {\bibinfo  {journal}
  {The Journal of Chemical Physics}\ }\textbf {\bibinfo {volume} {56}},\
  \bibinfo {pages} {4729} (\bibinfo {year} {1972})}\BibitemShut {NoStop}%
\bibitem [{\citenamefont {Frenkel}\ \emph {et~al.}(1984)\citenamefont
  {Frenkel}, \citenamefont {Mulder},\ and\ \citenamefont
  {McTague}}]{Frenkel:PRL_52:1984}%
  \BibitemOpen
  \bibfield  {author} {\bibinfo {author} {\bibfnamefont {D.}~\bibnamefont
  {Frenkel}}, \bibinfo {author} {\bibfnamefont {B.~M.}\ \bibnamefont {Mulder}},
  \ and\ \bibinfo {author} {\bibfnamefont {J.~P.}\ \bibnamefont {McTague}},\
  }\bibfield  {title} {\enquote {\bibinfo {title} {Phase diagram of a system of
  hard ellipsoids},}\ }\href{\doibase 10.1103/PhysRevLett.52.287} {\bibfield
  {journal} {\bibinfo  {journal} {Phys. Rev. Lett.}\ }\textbf {\bibinfo
  {volume} {52}},\ \bibinfo {pages} {287} (\bibinfo {year} {1984})}\BibitemShut
  {NoStop}%
\bibitem [{\citenamefont {Odriozola}(2012)}]{Odriozola:JCP_136:2012}%
  \BibitemOpen
  \bibfield  {author} {\bibinfo {author} {\bibfnamefont {G.}~\bibnamefont
  {Odriozola}},\ }\bibfield  {title} {\enquote {\bibinfo {title} {{Revisiting
  the phase diagram of hard ellipsoids}},}\ }\href{\doibase 10.1063/1.3699331}
  {\bibfield  {journal} {\bibinfo  {journal} {The Journal of Chemical Physics}\
  }\textbf {\bibinfo {volume} {136}},\ \bibinfo {pages} {134505} (\bibinfo
  {year} {2012})}\BibitemShut {NoStop}%
\bibitem [{\citenamefont {Foulaadvand}\ and\ \citenamefont
  {Yarifard}(2013)}]{Foulaadvand:PRE_88:2013}%
  \BibitemOpen
  \bibfield  {author} {\bibinfo {author} {\bibfnamefont {M.~E.}\ \bibnamefont
  {Foulaadvand}}\ and\ \bibinfo {author} {\bibfnamefont {M.}~\bibnamefont
  {Yarifard}},\ }\bibfield  {title} {\enquote {\bibinfo {title}
  {Two-dimensional system of hard ellipses: A molecular dynamics study},}\
  }\href{\doibase 10.1103/PhysRevE.88.052504} {\bibfield  {journal} {\bibinfo
  {journal} {Phys. Rev. E}\ }\textbf {\bibinfo {volume} {88}},\ \bibinfo
  {pages} {052504} (\bibinfo {year} {2013})}\BibitemShut {NoStop}%
\bibitem [{\citenamefont {Xu}\ \emph {et~al.}(2013)\citenamefont {Xu},
  \citenamefont {Li}, \citenamefont {Sun},\ and\ \citenamefont
  {An}}]{Xu:JCP_139:2013}%
  \BibitemOpen
  \bibfield  {author} {\bibinfo {author} {\bibfnamefont {W.-S.}\ \bibnamefont
  {Xu}}, \bibinfo {author} {\bibfnamefont {Y.-W.}\ \bibnamefont {Li}}, \bibinfo
  {author} {\bibfnamefont {Z.-Y.}\ \bibnamefont {Sun}}, \ and\ \bibinfo
  {author} {\bibfnamefont {L.-J.}\ \bibnamefont {An}},\ }\bibfield  {title}
  {\enquote {\bibinfo {title} {{Hard ellipses: Equation of state, structure,
  and self-diffusion}},}\ }\href{\doibase 10.1063/1.4812361} {\bibfield
  {journal} {\bibinfo  {journal} {The Journal of Chemical Physics}\ }\textbf
  {\bibinfo {volume} {139}},\ \bibinfo {pages} {024501} (\bibinfo {year}
  {2013})}\BibitemShut {NoStop}%
\bibitem [{\citenamefont {Bautista-Carbajal}\ and\ \citenamefont
  {Odriozola}(2014)}]{Bautista-Carbajal:JCP_140:2014}%
  \BibitemOpen
  \bibfield  {author} {\bibinfo {author} {\bibfnamefont {G.}~\bibnamefont
  {Bautista-Carbajal}}\ and\ \bibinfo {author} {\bibfnamefont {G.}~\bibnamefont
  {Odriozola}},\ }\bibfield  {title} {\enquote {\bibinfo {title} {{Phase
  diagram of two-dimensional hard ellipses}},}\ }\href{\doibase
  https://doi.org/10.1063/1.4878411} {\bibfield  {journal} {\bibinfo  {journal}
  {The Journal of Chemical Physics}\ }\textbf {\bibinfo {volume} {140}},\
  \bibinfo {pages} {204502} (\bibinfo {year} {2014})}\BibitemShut {NoStop}%
\bibitem [{\citenamefont {Torres-D{\'i}az}\ \emph {et~al.}(2022)\citenamefont
  {Torres-D{\'i}az}, \citenamefont {Hendley}, \citenamefont {Mishra},
  \citenamefont {Yeh},\ and\ \citenamefont {Bevan}}]{Torres-Diaz:SM_18:2022}%
  \BibitemOpen
  \bibfield  {author} {\bibinfo {author} {\bibfnamefont {I.}~\bibnamefont
  {Torres-D{\'i}az}}, \bibinfo {author} {\bibfnamefont {R.~S.}\ \bibnamefont
  {Hendley}}, \bibinfo {author} {\bibfnamefont {A.}~\bibnamefont {Mishra}},
  \bibinfo {author} {\bibfnamefont {A.~J.}\ \bibnamefont {Yeh}}, \ and\
  \bibinfo {author} {\bibfnamefont {M.~A.}\ \bibnamefont {Bevan}},\ }\bibfield
  {title} {\enquote {\bibinfo {title} {Hard superellipse phases: particle shape
  anisotropy \& curvature},}\ }\href{\doibase 10.1039/D1SM01523K} {\bibfield
  {journal} {\bibinfo  {journal} {Soft Matter}\ }\textbf {\bibinfo {volume}
  {18}},\ \bibinfo {pages} {1319} (\bibinfo {year} {2022})}\BibitemShut
  {NoStop}%
\bibitem [{\citenamefont {Marienhagen}\ and\ \citenamefont
  {Wagner}(2022)}]{Marienhagen:PRE_105:2022}%
  \BibitemOpen
  \bibfield  {author} {\bibinfo {author} {\bibfnamefont {P.}~\bibnamefont
  {Marienhagen}}\ and\ \bibinfo {author} {\bibfnamefont {J.}~\bibnamefont
  {Wagner}},\ }\bibfield  {title} {\enquote {\bibinfo {title} {Reexamining
  equations of state of oblate hard ellipsoids of revolution: Numerical
  simulation utilizing a cluster {M}onte {C}arlo algorithm and comparison to
  virial theory},}\ }\href{\doibase 10.1103/PhysRevE.105.014125} {\bibfield
  {journal} {\bibinfo  {journal} {Phys. Rev. E}\ }\textbf {\bibinfo {volume}
  {105}},\ \bibinfo {pages} {014125} (\bibinfo {year} {2022})}\BibitemShut
  {NoStop}%
\bibitem [{\citenamefont {De~Michele}\ \emph {et~al.}(2007)\citenamefont
  {De~Michele}, \citenamefont {Schilling},\ and\ \citenamefont
  {Sciortino}}]{DeMichele:PRL_98:2007}%
  \BibitemOpen
  \bibfield  {author} {\bibinfo {author} {\bibfnamefont {C.}~\bibnamefont
  {De~Michele}}, \bibinfo {author} {\bibfnamefont {R.}~\bibnamefont
  {Schilling}}, \ and\ \bibinfo {author} {\bibfnamefont {F.}~\bibnamefont
  {Sciortino}},\ }\bibfield  {title} {\enquote {\bibinfo {title} {Dynamics of
  uniaxial hard ellipsoids},}\ }\href{\doibase 10.1103/PhysRevLett.98.265702}
  {\bibfield  {journal} {\bibinfo  {journal} {Phys. Rev. Lett.}\ }\textbf
  {\bibinfo {volume} {98}},\ \bibinfo {pages} {265702} (\bibinfo {year}
  {2007})}\BibitemShut {NoStop}%
\bibitem [{\citenamefont {Pfleiderer}\ \emph {et~al.}(2008)\citenamefont
  {Pfleiderer}, \citenamefont {Milinkovic},\ and\ \citenamefont
  {Schilling}}]{Pfleiderer:EPL_84:2008}%
  \BibitemOpen
  \bibfield  {author} {\bibinfo {author} {\bibfnamefont {P.}~\bibnamefont
  {Pfleiderer}}, \bibinfo {author} {\bibfnamefont {K.}~\bibnamefont
  {Milinkovic}}, \ and\ \bibinfo {author} {\bibfnamefont {T.}~\bibnamefont
  {Schilling}},\ }\bibfield  {title} {\enquote {\bibinfo {title} {Glassy
  dynamics in monodisperse hard ellipsoids},}\ }\href{\doibase
  https://doi.org/10.1209/0295-5075/84/16003} {\bibfield  {journal} {\bibinfo
  {journal} {Europhysics Letters}\ }\textbf {\bibinfo {volume} {84}},\ \bibinfo
  {pages} {16003} (\bibinfo {year} {2008})}\BibitemShut {NoStop}%
\bibitem [{\citenamefont {Xu}\ \emph {et~al.}(2015)\citenamefont {Xu},
  \citenamefont {Sun},\ and\ \citenamefont {An}}]{Xu:SM_11:2015}%
  \BibitemOpen
  \bibfield  {author} {\bibinfo {author} {\bibfnamefont {W.-S.}\ \bibnamefont
  {Xu}}, \bibinfo {author} {\bibfnamefont {Z.-Y.}\ \bibnamefont {Sun}}, \ and\
  \bibinfo {author} {\bibfnamefont {L.-J.}\ \bibnamefont {An}},\ }\bibfield
  {title} {\enquote {\bibinfo {title} {Relaxation dynamics in a binary
  hard-ellipse liquid},}\ }\href{\doibase 10.1039/C4SM02290D} {\bibfield
  {journal} {\bibinfo  {journal} {Soft Matter}\ }\textbf {\bibinfo {volume}
  {11}},\ \bibinfo {pages} {627} (\bibinfo {year} {2015})}\BibitemShut
  {NoStop}%
\bibitem [{\citenamefont {Alhissi}\ \emph {et~al.}(2024)\citenamefont
  {Alhissi}, \citenamefont {Zumbusch},\ and\ \citenamefont
  {Fuchs}}]{Alhissi:JCP_160:2024}%
  \BibitemOpen
  \bibfield  {author} {\bibinfo {author} {\bibfnamefont {M.}~\bibnamefont
  {Alhissi}}, \bibinfo {author} {\bibfnamefont {A.}~\bibnamefont {Zumbusch}}, \
  and\ \bibinfo {author} {\bibfnamefont {M.}~\bibnamefont {Fuchs}},\ }\bibfield
   {title} {\enquote {\bibinfo {title} {{Observation of liquid glass in
  molecular dynamics simulations}},}\ }\href{\doibase 10.1063/5.0196599}
  {\bibfield  {journal} {\bibinfo  {journal} {The Journal of Chemical Physics}\
  }\textbf {\bibinfo {volume} {160}},\ \bibinfo {pages} {164502} (\bibinfo
  {year} {2024})}\BibitemShut {NoStop}%
\bibitem [{\citenamefont {Rowlinson}(1985)}]{Rowlinson:PROSL_402:1985}%
  \BibitemOpen
  \bibfield  {author} {\bibinfo {author} {\bibfnamefont {J.~S.}\ \bibnamefont
  {Rowlinson}},\ }\bibfield  {title} {\enquote {\bibinfo {title} {Virial
  expansions in an inhomogeneous system},}\ }\href{\doibase
  https://doi.org/10.1098/rspa.1985.0108} {\bibfield  {journal} {\bibinfo
  {journal} {Proc. Roy. Soc. (London)}\ }\textbf {\bibinfo {volume} {A402}},\
  \bibinfo {pages} {67} (\bibinfo {year} {1985})}\BibitemShut {NoStop}%
\bibitem [{\citenamefont {Isihara}(1950)}]{Isihara:JCP_18:1950}%
  \BibitemOpen
  \bibfield  {author} {\bibinfo {author} {\bibfnamefont {A.}~\bibnamefont
  {Isihara}},\ }\bibfield  {title} {\enquote {\bibinfo {title} {Determination
  of molecular shape by osmotic measurement},}\ }\href{\doibase
  https://doi.org/10.1063/1.1747510} {\bibfield  {journal} {\bibinfo  {journal}
  {The Journal of Chemical Physics}\ }\textbf {\bibinfo {volume} {18}},\
  \bibinfo {pages} {1446} (\bibinfo {year} {1950})}\BibitemShut {NoStop}%
\bibitem [{\citenamefont {Hadwiger}(1951)}]{Hadwiger:Experimentia_7:1951}%
  \BibitemOpen
  \bibfield  {author} {\bibinfo {author} {\bibfnamefont {H.}~\bibnamefont
  {Hadwiger}},\ }\bibfield  {title} {\enquote {\bibinfo {title} {Der kinetische
  {R}adius nichtkugelf\"ormiger {M}olek\"ule},}\
  }\href{https://link.springer.com/article/10.1007/BF02168922#citeas}
  {\bibfield  {journal} {\bibinfo  {journal} {Experimentia}\ }\textbf {\bibinfo
  {volume} {7}},\ \bibinfo {pages} {395} (\bibinfo {year} {1951})}\BibitemShut
  {NoStop}%
\bibitem [{\citenamefont {Kihara}(1953)}]{Kihara:RMP_25:1953}%
  \BibitemOpen
  \bibfield  {author} {\bibinfo {author} {\bibfnamefont {T.}~\bibnamefont
  {Kihara}},\ }\bibfield  {title} {\enquote {\bibinfo {title} {Virial
  coefficients and models of molecules in gases},}\ }\href{\doibase
  https://doi.org/10.1103/RevModPhys.25.831} {\bibfield  {journal} {\bibinfo
  {journal} {Rev. Mod. Phys.}\ }\textbf {\bibinfo {volume} {25}},\ \bibinfo
  {pages} {831} (\bibinfo {year} {1953})}\BibitemShut {NoStop}%
\bibitem [{\citenamefont {Freasier}\ and\ \citenamefont
  {Bearman}(1976)}]{Freasier:MolPhys_32:1976}%
  \BibitemOpen
  \bibfield  {author} {\bibinfo {author} {\bibfnamefont {B.}~\bibnamefont
  {Freasier}}\ and\ \bibinfo {author} {\bibfnamefont {R.}~\bibnamefont
  {Bearman}},\ }\bibfield  {title} {\enquote {\bibinfo {title} {Virial
  expansion for hard ellipsoids of revolution},}\ }\href{\doibase
  https://doi.org/10.1080/00268977600103281} {\bibfield  {journal} {\bibinfo
  {journal} {Molecular Physics}\ }\textbf {\bibinfo {volume} {32}},\ \bibinfo
  {pages} {551} (\bibinfo {year} {1976})}\BibitemShut {NoStop}%
\bibitem [{\citenamefont {Nezbeda}(1976)}]{Nezbeda:ChemPhysLett_41:1976}%
  \BibitemOpen
  \bibfield  {author} {\bibinfo {author} {\bibfnamefont {I.}~\bibnamefont
  {Nezbeda}},\ }\bibfield  {title} {\enquote {\bibinfo {title} {Virial
  expansion and an improved equation of state for the hard convex molecule
  system},}\ }\href{\doibase https://doi.org/10.1016/0009-2614(76)85246-3}
  {\bibfield  {journal} {\bibinfo  {journal} {Chemical Physics Letters}\
  }\textbf {\bibinfo {volume} {41}},\ \bibinfo {pages} {55} (\bibinfo {year}
  {1976})}\BibitemShut {NoStop}%
\bibitem [{\citenamefont {Rigby}(1989)}]{Rigby:MolPhys_66:1989}%
  \BibitemOpen
  \bibfield  {author} {\bibinfo {author} {\bibfnamefont {M.}~\bibnamefont
  {Rigby}},\ }\bibfield  {title} {\enquote {\bibinfo {title} {Hard ellipsoids
  of revolution},}\ }\href{\doibase https://doi.org/10.1080/00268978900100851}
  {\bibfield  {journal} {\bibinfo  {journal} {Molecular Physics}\ }\textbf
  {\bibinfo {volume} {66}},\ \bibinfo {pages} {1261} (\bibinfo {year}
  {1989})}\BibitemShut {NoStop}%
\bibitem [{\citenamefont {G.~Tarjus}\ and\ \citenamefont
  {Talbot}(1991)}]{Tarjus:MolPhys_73:1991}%
  \BibitemOpen
  \bibfield  {author} {\bibinfo {author} {\bibfnamefont {S.~R.}\ \bibnamefont
  {G.~Tarjus}, \bibfnamefont {P.~Viot}}\ and\ \bibinfo {author} {\bibfnamefont
  {J.}~\bibnamefont {Talbot}},\ }\bibfield  {title} {\enquote {\bibinfo {title}
  {New analytical and numerical results on virial coefficients for 2-d hard
  convex bodies},}\ }\href{\doibase 10.1080/00268979100101541} {\bibfield
  {journal} {\bibinfo  {journal} {Molecular Physics}\ }\textbf {\bibinfo
  {volume} {73}},\ \bibinfo {pages} {773} (\bibinfo {year} {1991})}\BibitemShut
  {NoStop}%
\bibitem [{\citenamefont {Rigby}(1993)}]{Rigby:MolPhys_78:1993}%
  \BibitemOpen
  \bibfield  {author} {\bibinfo {author} {\bibfnamefont {M.}~\bibnamefont
  {Rigby}},\ }\bibfield  {title} {\enquote {\bibinfo {title} {Virial
  coefficients of hard convex molecules in two dimensions},}\ }\href{\doibase
  https://doi.org/10.1080/00268979300100031} {\bibfield  {journal} {\bibinfo
  {journal} {Molecular Physics}\ }\textbf {\bibinfo {volume} {78}},\ \bibinfo
  {pages} {21} (\bibinfo {year} {1993})}\BibitemShut {NoStop}%
\bibitem [{\citenamefont {You}\ \emph {et~al.}(2005)\citenamefont {You},
  \citenamefont {Vlasov},\ and\ \citenamefont {Masters}}]{You:JCP_123:2005}%
  \BibitemOpen
  \bibfield  {author} {\bibinfo {author} {\bibfnamefont {X.-M.}\ \bibnamefont
  {You}}, \bibinfo {author} {\bibfnamefont {A.~Y.}\ \bibnamefont {Vlasov}}, \
  and\ \bibinfo {author} {\bibfnamefont {A.~J.}\ \bibnamefont {Masters}},\
  }\bibfield  {title} {\enquote {\bibinfo {title} {{The equation of state of
  isotropic fluids of hard convex bodies from a high-level virial
  expansion}},}\ }\href{\doibase 10.1063/1.1992471} {\bibfield  {journal}
  {\bibinfo  {journal} {The Journal of Chemical Physics}\ }\textbf {\bibinfo
  {volume} {123}},\ \bibinfo {pages} {034510} (\bibinfo {year}
  {2005})}\BibitemShut {NoStop}%
\bibitem [{\citenamefont {Masters}(2008)}]{Masters:JPhys_20:2008}%
  \BibitemOpen
  \bibfield  {author} {\bibinfo {author} {\bibfnamefont {A.~J.}\ \bibnamefont
  {Masters}},\ }\bibfield  {title} {\enquote {\bibinfo {title} {Virial
  expansions},}\ }\href{\doibase 10.1088/0953-8984/20/28/283102} {\bibfield
  {journal} {\bibinfo  {journal} {Journal of Physics: Condensed Matter}\
  }\textbf {\bibinfo {volume} {20}},\ \bibinfo {pages} {283102} (\bibinfo
  {year} {2008})}\BibitemShut {NoStop}%
\bibitem [{\citenamefont {Herold}\ \emph {et~al.}(2017)\citenamefont {Herold},
  \citenamefont {Hellmann},\ and\ \citenamefont
  {Wagner}}]{Herold:JCP_147:2017}%
  \BibitemOpen
  \bibfield  {author} {\bibinfo {author} {\bibfnamefont {E.}~\bibnamefont
  {Herold}}, \bibinfo {author} {\bibfnamefont {R.}~\bibnamefont {Hellmann}}, \
  and\ \bibinfo {author} {\bibfnamefont {J.}~\bibnamefont {Wagner}},\
  }\bibfield  {title} {\enquote {\bibinfo {title} {{Virial coefficients of
  anisotropic hard solids of revolution: The detailed influence of the particle
  geometry}},}\ }\href{\doibase 10.1063/1.5004687} {\bibfield  {journal}
  {\bibinfo  {journal} {The Journal of Chemical Physics}\ }\textbf {\bibinfo
  {volume} {147}},\ \bibinfo {pages} {204102} (\bibinfo {year}
  {2017})}\BibitemShut {NoStop}%
\bibitem [{\citenamefont {Kulossa}\ \emph {et~al.}(2022)\citenamefont
  {Kulossa}, \citenamefont {Marienhagen},\ and\ \citenamefont
  {Wagner}}]{Kulossa:PRE_105:2022}%
  \BibitemOpen
  \bibfield  {author} {\bibinfo {author} {\bibfnamefont {M.}~\bibnamefont
  {Kulossa}}, \bibinfo {author} {\bibfnamefont {P.}~\bibnamefont
  {Marienhagen}}, \ and\ \bibinfo {author} {\bibfnamefont {J.}~\bibnamefont
  {Wagner}},\ }\bibfield  {title} {\enquote {\bibinfo {title} {{Virial
  coefficients of hard hyperspherocylinders in ${\mathbb{R}}^{4}$: Influence of
  the aspect ratio}},}\ }\href{\doibase 10.1103/PhysRevE.105.064121} {\bibfield
   {journal} {\bibinfo  {journal} {Phys. Rev. E}\ }\textbf {\bibinfo {volume}
  {105}},\ \bibinfo {pages} {064121} (\bibinfo {year} {2022})}\BibitemShut
  {NoStop}%
\bibitem [{\citenamefont {Kulossa}\ and\ \citenamefont
  {Wagner}(2023)}]{Kulossa:MolPhys_0:2023}%
  \BibitemOpen
  \bibfield  {author} {\bibinfo {author} {\bibfnamefont {M.}~\bibnamefont
  {Kulossa}}\ and\ \bibinfo {author} {\bibfnamefont {J.}~\bibnamefont
  {Wagner}},\ }\bibfield  {title} {\enquote {\bibinfo {title} {Virial
  coefficients of hard, two-dimensional, convex particles up to the eighth
  order},}\ }\href{\doibase 10.1080/00268976.2023.2289699} {\bibfield
  {journal} {\bibinfo  {journal} {Molecular Physics}\ ,\ \bibinfo {pages}
  {e2289699}} (\bibinfo {year} {2023})}\BibitemShut {NoStop}%
\bibitem [{\citenamefont {Kulossa}\ and\ \citenamefont
  {Wagner}(2025)}]{Kulossa:PRE_111:2025}%
  \BibitemOpen
  \bibfield  {author} {\bibinfo {author} {\bibfnamefont {M.}~\bibnamefont
  {Kulossa}}\ and\ \bibinfo {author} {\bibfnamefont {J.}~\bibnamefont
  {Wagner}},\ }\bibfield  {title} {\enquote {\bibinfo {title} {Geometric
  measures of uniaxial solids of revolution in higher-dimensional euclidean
  spaces and their relation to the second virial coefficient},}\
  }\href{\doibase 10.1103/PhysRevE.111.024112} {\bibfield  {journal} {\bibinfo
  {journal} {Phys. Rev. E}\ }\textbf {\bibinfo {volume} {111}},\ \bibinfo
  {pages} {024112} (\bibinfo {year} {2025})}\BibitemShut {NoStop}%
\bibitem [{\citenamefont {Gray}\ and\ \citenamefont
  {Gubbins}(1984)}]{Gray:Molecular_Fluids:1984}%
  \BibitemOpen
  \bibfield  {author} {\bibinfo {author} {\bibfnamefont {C.~G.}\ \bibnamefont
  {Gray}}\ and\ \bibinfo {author} {\bibfnamefont {K.~E.}\ \bibnamefont
  {Gubbins}},\
  }\href{https://www.google.de/books/edition/Theory_of_Molecular_Fluids/ABE1DQAAQBAJ?hl=de&gbpv=0}
  {\emph {\bibinfo {title} {Theory of Molecular Fluids {I}: Fundamentals}}}\
  (\bibinfo  {publisher} {Clarendon Press},\ \bibinfo {year}
  {1984})\BibitemShut {NoStop}%
\bibitem [{\citenamefont
  {Solana}(2013)}]{Solana:Perturbation_Theory_Fluids:2013}%
  \BibitemOpen
  \bibfield  {author} {\bibinfo {author} {\bibfnamefont {J.~R.}\ \bibnamefont
  {Solana}},\
  }\href{https://www.routledge.com/Perturbation-Theories-for-the-Thermodynamic-Properties-of-Fluids-and-Solids/Solana/p/book/9780367380250}
  {\emph {\bibinfo {title} {Perturbation Theories for the Thermodynamic
  Properties of Fluids and Solids}}}\ (\bibinfo  {publisher} {CRC Press, Boca
  Raton, FL},\ \bibinfo {year} {2013})\BibitemShut {NoStop}%
\bibitem [{\citenamefont {Barboy}\ and\ \citenamefont
  {Gelbart}(1979)}]{Barboy:JCP_71:1979}%
  \BibitemOpen
  \bibfield  {author} {\bibinfo {author} {\bibfnamefont {B.}~\bibnamefont
  {Barboy}}\ and\ \bibinfo {author} {\bibfnamefont {W.~M.}\ \bibnamefont
  {Gelbart}},\ }\bibfield  {title} {\enquote {\bibinfo {title} {Series
  representation of the equation of state for hard particle fluids},}\
  }\href{\doibase 10.1063/1.438711} {\bibfield  {journal} {\bibinfo  {journal}
  {The Journal of Chemical Physics}\ }\textbf {\bibinfo {volume} {71}},\
  \bibinfo {pages} {3053} (\bibinfo {year} {1979})}\BibitemShut {NoStop}%
\bibitem [{\citenamefont {Mulder}\ and\ \citenamefont
  {Frenkel}(1985)}]{Mulder:MolPhys_55:1985}%
  \BibitemOpen
  \bibfield  {author} {\bibinfo {author} {\bibfnamefont {B.}~\bibnamefont
  {Mulder}}\ and\ \bibinfo {author} {\bibfnamefont {D.}~\bibnamefont
  {Frenkel}},\ }\bibfield  {title} {\enquote {\bibinfo {title} {The hard
  ellipsoid-of-revolution fluid {II}. {T}he $y$-expansion equation of state},}\
  }\href{\doibase 10.1080/00268978500101981} {\bibfield  {journal} {\bibinfo
  {journal} {Molecular Physics}\ }\textbf {\bibinfo {volume} {55}},\ \bibinfo
  {pages} {1193} (\bibinfo {year} {1985})}\BibitemShut {NoStop}%
\bibitem [{\citenamefont {Bellemans}(1968)}]{Belleman:PRL_21:1968}%
  \BibitemOpen
  \bibfield  {author} {\bibinfo {author} {\bibfnamefont {A.}~\bibnamefont
  {Bellemans}},\ }\bibfield  {title} {\enquote {\bibinfo {title} {Free energy
  of an assembly of nonspherical molecules with a hard core},}\ }\href{\doibase
  https://doi.org/10.1103/PhysRevLett.21.527} {\bibfield  {journal} {\bibinfo
  {journal} {Phys. Rev. Lett.}\ }\textbf {\bibinfo {volume} {21}},\ \bibinfo
  {pages} {527} (\bibinfo {year} {1968})}\BibitemShut {NoStop}%
\bibitem [{\citenamefont {Nezbeda}\ and\ \citenamefont
  {Thomas}(1979)}]{Nezbeda:JChemSoc_75:1979}%
  \BibitemOpen
  \bibfield  {author} {\bibinfo {author} {\bibfnamefont {I.}~\bibnamefont
  {Nezbeda}}\ and\ \bibinfo {author} {\bibfnamefont {T.~W.~L.}\ \bibnamefont
  {Thomas}},\ }\bibfield  {title} {\enquote {\bibinfo {title} {Conformal theory
  of hard non-spherical molecule fluids},}\ }\href{\doibase
  https://doi.org/10.1039/F29797500193} {\bibfield  {journal} {\bibinfo
  {journal} {J. Chem. Soc., Faraday Trans. 2}\ }\textbf {\bibinfo {volume}
  {75}},\ \bibinfo {pages} {193} (\bibinfo {year} {1979})}\BibitemShut
  {NoStop}%
\bibitem [{\citenamefont {Percus}(1985)}]{Percus:AnnNYAcad_221:1954}%
  \BibitemOpen
  \bibfield  {author} {\bibinfo {author} {\bibfnamefont {J.~K.}\ \bibnamefont
  {Percus}},\ }\bibfield  {title} {\enquote {\bibinfo {title} {Sphericalized
  molecular interactions},}\ }\href{\doibase
  https://doi.org/10.1111/j.1749-6632.1985.tb30001.x} {\bibfield  {journal}
  {\bibinfo  {journal} {Ann. N.Y. Acad. of Sci.}\ }\textbf {\bibinfo {volume}
  {452}},\ \bibinfo {pages} {67} (\bibinfo {year} {1985})}\BibitemShut
  {NoStop}%
\bibitem [{\citenamefont {Reiss}\ \emph {et~al.}(1959)\citenamefont {Reiss},
  \citenamefont {Frisch},\ and\ \citenamefont {Lebowitz}}]{Reiss:JCP_31:1959}%
  \BibitemOpen
  \bibfield  {author} {\bibinfo {author} {\bibfnamefont {H.}~\bibnamefont
  {Reiss}}, \bibinfo {author} {\bibfnamefont {H.~L.}\ \bibnamefont {Frisch}}, \
  and\ \bibinfo {author} {\bibfnamefont {J.~L.}\ \bibnamefont {Lebowitz}},\
  }\bibfield  {title} {\enquote {\bibinfo {title} {Statistical mechanics of
  rigid spheres},}\ }\href{\doibase 10.1063/1.1730361} {\bibfield  {journal}
  {\bibinfo  {journal} {The Journal of Chemical Physics}\ }\textbf {\bibinfo
  {volume} {31}},\ \bibinfo {pages} {369} (\bibinfo {year} {1959})}\BibitemShut
  {NoStop}%
\bibitem [{\citenamefont {Gibbons}(1969)}]{Gibbons:MolPhys_17:1969}%
  \BibitemOpen
  \bibfield  {author} {\bibinfo {author} {\bibfnamefont {R.}~\bibnamefont
  {Gibbons}},\ }\bibfield  {title} {\enquote {\bibinfo {title} {The scaled
  particle theory for particles of arbitrary shape},}\ }\href{\doibase
  https://doi.org/10.1080/00268976900100811} {\bibfield  {journal} {\bibinfo
  {journal} {Molecular Physics}\ }\textbf {\bibinfo {volume} {17}},\ \bibinfo
  {pages} {81} (\bibinfo {year} {1969})}\BibitemShut {NoStop}%
\bibitem [{\citenamefont {Boubl{\'i}k}(1974)}]{Boublik:MolPhys_27:1974}%
  \BibitemOpen
  \bibfield  {author} {\bibinfo {author} {\bibfnamefont {T.}~\bibnamefont
  {Boubl{\'i}k}},\ }\bibfield  {title} {\enquote {\bibinfo {title} {Statistical
  thermodynamics of convex molecule fluids},}\ }\href{\doibase
  https://doi.org/10.1080/00268977400101191} {\bibfield  {journal} {\bibinfo
  {journal} {Molecular Physics}\ }\textbf {\bibinfo {volume} {27}},\ \bibinfo
  {pages} {1415} (\bibinfo {year} {1974})}\BibitemShut {NoStop}%
\bibitem [{\citenamefont {Boubl{\'i}k}(1981)}]{Boublik:MolPhys_42:1981}%
  \BibitemOpen
  \bibfield  {author} {\bibinfo {author} {\bibfnamefont {T.}~\bibnamefont
  {Boubl{\'i}k}},\ }\bibfield  {title} {\enquote {\bibinfo {title} {Equation of
  state of hard convex body fluids},}\ }\href{\doibase
  10.1080/00268978100100161} {\bibfield  {journal} {\bibinfo  {journal}
  {Molecular Physics}\ }\textbf {\bibinfo {volume} {42}},\ \bibinfo {pages}
  {209} (\bibinfo {year} {1981})}\BibitemShut {NoStop}%
\bibitem [{\citenamefont {Rosenfeld}(1994)}]{Rosenfeld:PRE_50:1994}%
  \BibitemOpen
  \bibfield  {author} {\bibinfo {author} {\bibfnamefont {Y.}~\bibnamefont
  {Rosenfeld}},\ }\bibfield  {title} {\enquote {\bibinfo {title} {Density
  functional theory of molecular fluids: Free-energy model for the
  inhomogeneous hard-body fluid},}\ }\href{\doibase 10.1103/PhysRevE.50.R3318}
  {\bibfield  {journal} {\bibinfo  {journal} {Phys. Rev. E}\ }\textbf {\bibinfo
  {volume} {50}},\ \bibinfo {pages} {R3318} (\bibinfo {year}
  {1994})}\BibitemShut {NoStop}%
\bibitem [{\citenamefont {Rosenfeld}(1996)}]{Rosenfeld:ACS_Symposium:1996}%
  \BibitemOpen
  \bibfield  {author} {\bibinfo {author} {\bibfnamefont {Y.}~\bibnamefont
  {Rosenfeld}},\ }\enquote {\bibinfo {title} {Chemical {A}pplications of
  {D}ensity-{F}unctional {T}heory},}\ \ (\bibinfo  {publisher} {ACS
  Publications},\ \bibinfo {year} {1996})\ p.\ \bibinfo {pages}
  {198}\BibitemShut {NoStop}%
\bibitem [{\citenamefont {Cinacchi}\ and\ \citenamefont
  {Schmid}(2002)}]{Cinacchi:JPhys_14:2002}%
  \BibitemOpen
  \bibfield  {author} {\bibinfo {author} {\bibfnamefont {G.}~\bibnamefont
  {Cinacchi}}\ and\ \bibinfo {author} {\bibfnamefont {F.}~\bibnamefont
  {Schmid}},\ }\bibfield  {title} {\enquote {\bibinfo {title} {Density
  functional for anisotropic fluids},}\ }\href{\doibase
  https://doi.org/10.1088/0953-8984/14/46/323} {\bibfield  {journal} {\bibinfo
  {journal} {Journal of Physics: Condensed Matter}\ }\textbf {\bibinfo {volume}
  {14}},\ \bibinfo {pages} {12223} (\bibinfo {year} {2002})}\BibitemShut
  {NoStop}%
\bibitem [{\citenamefont {Wittmann}\ \emph {et~al.}(2016)\citenamefont
  {Wittmann}, \citenamefont {Marechal},\ and\ \citenamefont
  {Mecke}}]{Wittmann:JPhys_28:2016}%
  \BibitemOpen
  \bibfield  {author} {\bibinfo {author} {\bibfnamefont {R.}~\bibnamefont
  {Wittmann}}, \bibinfo {author} {\bibfnamefont {M.}~\bibnamefont {Marechal}},
  \ and\ \bibinfo {author} {\bibfnamefont {K.}~\bibnamefont {Mecke}},\
  }\bibfield  {title} {\enquote {\bibinfo {title} {Fundamental measure theory
  for non-spherical hard particles: {P}redicting liquid crystal properties from
  the particle shape},}\ }\href{\doibase 10.1088/0953-8984/28/24/244003}
  {\bibfield  {journal} {\bibinfo  {journal} {Journal of Physics: Condensed
  Matter}\ }\textbf {\bibinfo {volume} {28}},\ \bibinfo {pages} {244003}
  (\bibinfo {year} {2016})}\BibitemShut {NoStop}%
\bibitem [{\citenamefont {Parsons}(1979)}]{Parsons:PRA_19:1979}%
  \BibitemOpen
  \bibfield  {author} {\bibinfo {author} {\bibfnamefont {J.~D.}\ \bibnamefont
  {Parsons}},\ }\bibfield  {title} {\enquote {\bibinfo {title} {Nematic
  ordering in a system of rods},}\ }\href{\doibase 10.1103/PhysRevA.19.1225}
  {\bibfield  {journal} {\bibinfo  {journal} {Phys. Rev. A}\ }\textbf {\bibinfo
  {volume} {19}},\ \bibinfo {pages} {1225} (\bibinfo {year}
  {1979})}\BibitemShut {NoStop}%
\bibitem [{\citenamefont {Lee}(1987)}]{Lee:JCP_87:1987}%
  \BibitemOpen
  \bibfield  {author} {\bibinfo {author} {\bibfnamefont {S.-D.}\ \bibnamefont
  {Lee}},\ }\bibfield  {title} {\enquote {\bibinfo {title} {{A numerical
  investigation of nematic ordering based on a simple hard‐rod mode}},}\
  }\href{\doibase 10.1063/1.452811} {\bibfield  {journal} {\bibinfo  {journal}
  {The Journal of Chemical Physics}\ }\textbf {\bibinfo {volume} {87}},\
  \bibinfo {pages} {4972} (\bibinfo {year} {1987})}\BibitemShut {NoStop}%
\bibitem [{\citenamefont {Song}\ and\ \citenamefont
  {Mason}(1990)}]{Song:PRA_41:1990}%
  \BibitemOpen
  \bibfield  {author} {\bibinfo {author} {\bibfnamefont {Y.}~\bibnamefont
  {Song}}\ and\ \bibinfo {author} {\bibfnamefont {E.~A.}\ \bibnamefont
  {Mason}},\ }\bibfield  {title} {\enquote {\bibinfo {title} {Equation of state
  for a fluid of hard convex bodies in any number of dimensions},}\
  }\href{\doibase 10.1103/PhysRevA.41.3121} {\bibfield  {journal} {\bibinfo
  {journal} {Phys. Rev. A}\ }\textbf {\bibinfo {volume} {41}},\ \bibinfo
  {pages} {3121} (\bibinfo {year} {1990})}\BibitemShut {NoStop}%
\bibitem [{\citenamefont {Vega}(1997)}]{Vega:MolPhys_92:1997}%
  \BibitemOpen
  \bibfield  {author} {\bibinfo {author} {\bibfnamefont {C.}~\bibnamefont
  {Vega}},\ }\bibfield  {title} {\enquote {\bibinfo {title} {Virial
  coefficients and equation of state of hard ellipsoids},}\ }\href{\doibase
  https://doi.org/10.1080/002689797169934} {\bibfield  {journal} {\bibinfo
  {journal} {Molecular Physics}\ }\textbf {\bibinfo {volume} {92}},\ \bibinfo
  {pages} {651} (\bibinfo {year} {1997})}\BibitemShut {NoStop}%
\bibitem [{\citenamefont {Solana}(2015)}]{Solana:MolPhys_113:2015}%
  \BibitemOpen
  \bibfield  {author} {\bibinfo {author} {\bibfnamefont {J.~R.}\ \bibnamefont
  {Solana}},\ }\bibfield  {title} {\enquote {\bibinfo {title} {Equations of
  state of hard-body fluids: a new proposal},}\ }\href{\doibase
  https://doi.org/10.1080/00268976.2014.975764} {\bibfield  {journal} {\bibinfo
   {journal} {Molecular Physics}\ }\textbf {\bibinfo {volume} {113}},\ \bibinfo
  {pages} {1003} (\bibinfo {year} {2015})}\BibitemShut {NoStop}%
\bibitem [{\citenamefont {Allen}\ \emph {et~al.}(1993)\citenamefont {Allen},
  \citenamefont {Evans}, \citenamefont {Frenkel},\ and\ \citenamefont
  {Mulder}}]{Allen:AdChemPhys_86:1993}%
  \BibitemOpen
  \bibfield  {author} {\bibinfo {author} {\bibfnamefont {M.~P.}\ \bibnamefont
  {Allen}}, \bibinfo {author} {\bibfnamefont {G.~T.}\ \bibnamefont {Evans}},
  \bibinfo {author} {\bibfnamefont {D.}~\bibnamefont {Frenkel}}, \ and\
  \bibinfo {author} {\bibfnamefont {B.}~\bibnamefont {Mulder}},\ }\bibfield
  {title} {\enquote {\bibinfo {title} {Hard convex body fluids},}\
  }\href{https://onlinelibrary.wiley.com/doi/book/10.1002/9780470141458#page=9}
  {\bibfield  {journal} {\bibinfo  {journal} {Advances in Chemical Physics}\
  }\textbf {\bibinfo {volume} {86}},\ \bibinfo {pages} {1} (\bibinfo {year}
  {1993})}\BibitemShut {NoStop}%
\bibitem [{\citenamefont {Dijkstra}(2014)}]{Dijkstra:AdvChemPhys_156:2014}%
  \BibitemOpen
  \bibfield  {author} {\bibinfo {author} {\bibfnamefont {M.}~\bibnamefont
  {Dijkstra}},\ }\bibfield  {title} {\enquote {\bibinfo {title} {Entropy-driven
  phase transitions in colloids: From spheres to anisotropic particles},}\
  }\href{https://doi.org/10.1002/9781118949702.ch2} {\bibfield  {journal}
  {\bibinfo  {journal} {Adv. Chem. Phys.}\ }\textbf {\bibinfo {volume} {156}},\
  \bibinfo {pages} {35} (\bibinfo {year} {2014})}\BibitemShut {NoStop}%
\bibitem [{\citenamefont {Mecke}(2000)}]{Mecke:Minkowski_functionals:2000}%
  \BibitemOpen
  \bibfield  {author} {\bibinfo {author} {\bibfnamefont {K.~R.}\ \bibnamefont
  {Mecke}},\ }\href{\doibase 10.1016/C2010-0-66723-X} {\emph {\bibinfo {title}
  {Additivity, convexity, and beyond: Applications of Minkowski functionals in
  statistical physics Statistical Physics and Spatial Statistics}}}\ (\bibinfo
  {publisher} {Springer},\ \bibinfo {year} {2000})\ pp.\ \bibinfo {pages}
  {111--184}\BibitemShut {NoStop}%
\bibitem [{\citenamefont {Torquato}\ and\ \citenamefont
  {Jiao}(2022)}]{Torquato:JStatMech:2022}%
  \BibitemOpen
  \bibfield  {author} {\bibinfo {author} {\bibfnamefont {S.}~\bibnamefont
  {Torquato}}\ and\ \bibinfo {author} {\bibfnamefont {Y.}~\bibnamefont
  {Jiao}},\ }\bibfield  {title} {\enquote {\bibinfo {title} {Exclusion volumes
  of convex bodies in high space dimensions: applications to virial
  coefficients and continuum percolation},}\ }\href{\doibase
  10.1088/1742-5468/ac8c8b} {\bibfield  {journal} {\bibinfo  {journal} {J.
  Stat. Mech.}\ ,\ \bibinfo {pages} {093404}} (\bibinfo {year}
  {2022})}\BibitemShut {NoStop}%
\bibitem [{\citenamefont {Boubl{\'i}k}(1975)}]{Boublik:MolPhys_29:1975}%
  \BibitemOpen
  \bibfield  {author} {\bibinfo {author} {\bibfnamefont {T.}~\bibnamefont
  {Boubl{\'i}k}},\ }\bibfield  {title} {\enquote {\bibinfo {title}
  {Two-dimensional convex particle liquid},}\ }\href{\doibase
  https://doi.org/10.1080/00268977500100361} {\bibfield  {journal} {\bibinfo
  {journal} {Molecular Physics}\ }\textbf {\bibinfo {volume} {29}},\ \bibinfo
  {pages} {421} (\bibinfo {year} {1975})}\BibitemShut {NoStop}%
\bibitem [{\citenamefont {Strandburg}(1988)}]{Strandburg:RMP_60:1988}%
  \BibitemOpen
  \bibfield  {author} {\bibinfo {author} {\bibfnamefont {K.~J.}\ \bibnamefont
  {Strandburg}},\ }\bibfield  {title} {\enquote {\bibinfo {title}
  {Two-dimensional melting},}\ }\href{\doibase 10.1103/RevModPhys.60.161}
  {\bibfield  {journal} {\bibinfo  {journal} {Rev. Mod. Phys.}\ }\textbf
  {\bibinfo {volume} {60}},\ \bibinfo {pages} {161} (\bibinfo {year}
  {1988})}\BibitemShut {NoStop}%
\bibitem [{\citenamefont {Perram}\ and\ \citenamefont
  {Wertheim}(1985)}]{Perram:JCompPhys_58:1985}%
  \BibitemOpen
  \bibfield  {author} {\bibinfo {author} {\bibfnamefont {J.~W.}\ \bibnamefont
  {Perram}}\ and\ \bibinfo {author} {\bibfnamefont {M.}~\bibnamefont
  {Wertheim}},\ }\bibfield  {title} {\enquote {\bibinfo {title} {Statistical
  mechanics of hard ellipsoids. {I}. {O}verlap algorithm and the contact
  function},}\ }\href{\doibase https://doi.org/10.1016/0021-9991(85)90171-8}
  {\bibfield  {journal} {\bibinfo  {journal} {Journal of Computational
  Physics}\ }\textbf {\bibinfo {volume} {58}},\ \bibinfo {pages} {409}
  (\bibinfo {year} {1985})}\BibitemShut {NoStop}%
\bibitem [{\citenamefont {Asakura}\ and\ \citenamefont
  {Oosawa}(1954)}]{Asakura:JCP_22:1954}%
  \BibitemOpen
  \bibfield  {author} {\bibinfo {author} {\bibfnamefont {S.}~\bibnamefont
  {Asakura}}\ and\ \bibinfo {author} {\bibfnamefont {F.}~\bibnamefont
  {Oosawa}},\ }\bibfield  {title} {\enquote {\bibinfo {title} {On interaction
  between two bodies immersed in a solution of macromolecules},}\
  }\href{\doibase 10.1063/1.1740347} {\bibfield  {journal} {\bibinfo  {journal}
  {The Journal of Chemical Physics}\ }\textbf {\bibinfo {volume} {22}},\
  \bibinfo {pages} {1255} (\bibinfo {year} {1954})}\BibitemShut {NoStop}%
\bibitem [{\citenamefont {Asakura}\ and\ \citenamefont
  {Oosawa}(1958)}]{Asakura:JPS_33:1958}%
  \BibitemOpen
  \bibfield  {author} {\bibinfo {author} {\bibfnamefont {S.}~\bibnamefont
  {Asakura}}\ and\ \bibinfo {author} {\bibfnamefont {F.}~\bibnamefont
  {Oosawa}},\ }\bibfield  {title} {\enquote {\bibinfo {title} {Interaction
  between particles suspended in solutions of macromolecules},}\
  }\href{\doibase 10.1002/pol.1958.1203312618} {\bibfield  {journal} {\bibinfo
  {journal} {Journal of Polymer Science}\ }\textbf {\bibinfo {volume} {3}},\
  \bibinfo {pages} {183} (\bibinfo {year} {1958})}\BibitemShut {NoStop}%
\bibitem [{\citenamefont {Franosch}\ \emph {et~al.}(2012)\citenamefont
  {Franosch}, \citenamefont {Lang},\ and\ \citenamefont
  {Schilling}}]{Franosch:PRL_109:2012}%
  \BibitemOpen
  \bibfield  {author} {\bibinfo {author} {\bibfnamefont {T.}~\bibnamefont
  {Franosch}}, \bibinfo {author} {\bibfnamefont {S.}~\bibnamefont {Lang}}, \
  and\ \bibinfo {author} {\bibfnamefont {R.}~\bibnamefont {Schilling}},\
  }\bibfield  {title} {\enquote {\bibinfo {title} {{Fluids in Extreme
  Confinement}},}\ }\href{\doibase 10.1103/PhysRevLett.109.240601} {\bibfield
  {journal} {\bibinfo  {journal} {Physical Review Letters}\ }\textbf {\bibinfo
  {volume} {109}},\ \bibinfo {pages} {240601} (\bibinfo {year}
  {2012})}\BibitemShut {NoStop}%
\bibitem [{\citenamefont {Franosch}\ \emph {et~al.}(2013)\citenamefont
  {Franosch}, \citenamefont {Lang},\ and\ \citenamefont
  {Schilling}}]{Franosch:PRL_110:2013}%
  \BibitemOpen
  \bibfield  {author} {\bibinfo {author} {\bibfnamefont {T.}~\bibnamefont
  {Franosch}}, \bibinfo {author} {\bibfnamefont {S.}~\bibnamefont {Lang}}, \
  and\ \bibinfo {author} {\bibfnamefont {R.}~\bibnamefont {Schilling}},\
  }\bibfield  {title} {\enquote {\bibinfo {title} {Erratum: Fluids in extreme
  confinement [{P}hys. {R}ev. {L}ett. \textbf{109}, 240601 (2012)]},}\
  }\href{\doibase 10.1103/PhysRevLett.110.059901} {\bibfield  {journal}
  {\bibinfo  {journal} {Phys. Rev. Lett.}\ }\textbf {\bibinfo {volume} {110}},\
  \bibinfo {pages} {059901} (\bibinfo {year} {2013})}\BibitemShut {NoStop}%
\bibitem [{\citenamefont {Franosch}\ and\ \citenamefont
  {Schilling}(2022)}]{Franosch:PRL_128:2022}%
  \BibitemOpen
  \bibfield  {author} {\bibinfo {author} {\bibfnamefont {T.}~\bibnamefont
  {Franosch}}\ and\ \bibinfo {author} {\bibfnamefont {R.}~\bibnamefont
  {Schilling}},\ }\bibfield  {title} {\enquote {\bibinfo {title} {Erratum:
  Fluids in extreme confinement [{P}hys. {R}ev. {L}ett. 109, 240601 (2012)]},}\
  }\href{\doibase 10.1103/PhysRevLett.128.209902} {\bibfield  {journal}
  {\bibinfo  {journal} {Phys. Rev. Lett.}\ }\textbf {\bibinfo {volume} {128}},\
  \bibinfo {pages} {209902} (\bibinfo {year} {2022})}\BibitemShut {NoStop}%
\bibitem [{\citenamefont {Zheng}\ and\ \citenamefont
  {Palffy-Muhoray}(2007)}]{Zheng:PRE_75:2007}%
  \BibitemOpen
  \bibfield  {author} {\bibinfo {author} {\bibfnamefont {X.}~\bibnamefont
  {Zheng}}\ and\ \bibinfo {author} {\bibfnamefont {P.}~\bibnamefont
  {Palffy-Muhoray}},\ }\bibfield  {title} {\enquote {\bibinfo {title} {Distance
  of closest approach of two arbitrary hard ellipses in two dimensions},}\
  }\href{\doibase https://doi.org/10.1103/PhysRevE.75.061709} {\bibfield
  {journal} {\bibinfo  {journal} {Phys. Rev. E}\ }\textbf {\bibinfo {volume}
  {75}},\ \bibinfo {pages} {061709} (\bibinfo {year} {2007})}\BibitemShut
  {NoStop}%
\bibitem [{\citenamefont {Zheng}\ \emph {et~al.}(2009)\citenamefont {Zheng},
  \citenamefont {Iglesias},\ and\ \citenamefont
  {Palffy-Muhoray}}]{Zheng:PRE_79:2009}%
  \BibitemOpen
  \bibfield  {author} {\bibinfo {author} {\bibfnamefont {X.}~\bibnamefont
  {Zheng}}, \bibinfo {author} {\bibfnamefont {W.}~\bibnamefont {Iglesias}}, \
  and\ \bibinfo {author} {\bibfnamefont {P.}~\bibnamefont {Palffy-Muhoray}},\
  }\bibfield  {title} {\enquote {\bibinfo {title} {Distance of closest approach
  of two arbitrary hard ellipsoids},}\ }\href{\doibase
  https://doi.org/10.1103/PhysRevE.79.057702} {\bibfield  {journal} {\bibinfo
  {journal} {Phys. Rev. E}\ }\textbf {\bibinfo {volume} {79}},\ \bibinfo
  {pages} {057702} (\bibinfo {year} {2009})}\BibitemShut {NoStop}%
\bibitem [{\citenamefont {Lebowitz}\ \emph {et~al.}(1987)\citenamefont
  {Lebowitz}, \citenamefont {Percus},\ and\ \citenamefont
  {Talbot}}]{Lebowitz:JStat_49:1987}%
  \BibitemOpen
  \bibfield  {author} {\bibinfo {author} {\bibfnamefont {J.}~\bibnamefont
  {Lebowitz}}, \bibinfo {author} {\bibfnamefont {J.}~\bibnamefont {Percus}}, \
  and\ \bibinfo {author} {\bibfnamefont {J.}~\bibnamefont {Talbot}},\
  }\bibfield  {title} {\enquote {\bibinfo {title} {On the orientational
  properties of some one-dimensional model systems},}\ }\href{\doibase
  https://doi.org/10.1007/BF01017568} {\bibfield  {journal} {\bibinfo
  {journal} {Journal of Statistical Physics}\ }\textbf {\bibinfo {volume}
  {49}},\ \bibinfo {pages} {1221} (\bibinfo {year} {1987})}\BibitemShut
  {NoStop}%
\bibitem [{\citenamefont {Tonks}(1936)}]{Tonks:PhysRev_50:1936}%
  \BibitemOpen
  \bibfield  {author} {\bibinfo {author} {\bibfnamefont {L.}~\bibnamefont
  {Tonks}},\ }\bibfield  {title} {\enquote {\bibinfo {title} {The complete
  equation of state of one, two and three-dimensional gases of hard elastic
  spheres},}\ }\href{\doibase 10.1103/PhysRev.50.955} {\bibfield  {journal}
  {\bibinfo  {journal} {Phys. Rev.}\ }\textbf {\bibinfo {volume} {50}},\
  \bibinfo {pages} {955} (\bibinfo {year} {1936})}\BibitemShut {NoStop}%
\bibitem [{\citenamefont {Salsburg}\ \emph {et~al.}(1953)\citenamefont
  {Salsburg}, \citenamefont {Zwanzig},\ and\ \citenamefont
  {Kirkwood}}]{Salsburg:JCP_21:1953}%
  \BibitemOpen
  \bibfield  {author} {\bibinfo {author} {\bibfnamefont {Z.~W.}\ \bibnamefont
  {Salsburg}}, \bibinfo {author} {\bibfnamefont {R.~W.}\ \bibnamefont
  {Zwanzig}}, \ and\ \bibinfo {author} {\bibfnamefont {J.~G.}\ \bibnamefont
  {Kirkwood}},\ }\bibfield  {title} {\enquote {\bibinfo {title} {Molecular
  distribution functions in a one‐dimensional fluid},}\
  }\href{https://doi.org/10.1063/1.1699116} {\bibfield  {journal} {\bibinfo
  {journal} {The Journal of Chemical Physics}\ }\textbf {\bibinfo {volume}
  {21}},\ \bibinfo {pages} {1098} (\bibinfo {year} {1953})}\BibitemShut
  {NoStop}%
\bibitem [{\citenamefont {Kolafa}\ and\ \citenamefont
  {Rottner}(2006)}]{Kolafa:MolPhys_104:2006}%
  \BibitemOpen
  \bibfield  {author} {\bibinfo {author} {\bibfnamefont {J.}~\bibnamefont
  {Kolafa}}\ and\ \bibinfo {author} {\bibfnamefont {M.}~\bibnamefont
  {Rottner}},\ }\bibfield  {title} {\enquote {\bibinfo {title}
  {Simulation-based equation of state of the hard disk fluid and prediction of
  higher-order virial coefficients},}\ }\href{\doibase
  https://doi.org/10.1080/00268970600967963} {\bibfield  {journal} {\bibinfo
  {journal} {Molecular Physics}\ }\textbf {\bibinfo {volume} {104}},\ \bibinfo
  {pages} {3435} (\bibinfo {year} {2006})}\BibitemShut {NoStop}%
\bibitem [{\citenamefont {Helfand}\ \emph {et~al.}(1961)\citenamefont
  {Helfand}, \citenamefont {Frisch},\ and\ \citenamefont
  {Lebowitz}}]{Helfand:JCP_34:1961}%
  \BibitemOpen
  \bibfield  {author} {\bibinfo {author} {\bibfnamefont {E.}~\bibnamefont
  {Helfand}}, \bibinfo {author} {\bibfnamefont {H.~L.}\ \bibnamefont {Frisch}},
  \ and\ \bibinfo {author} {\bibfnamefont {J.~L.}\ \bibnamefont {Lebowitz}},\
  }\bibfield  {title} {\enquote {\bibinfo {title} {Theory of the two‐ and
  one‐dimensional rigid sphere fluids},}\ }\href{\doibase 10.1063/1.1731629}
  {\bibfield  {journal} {\bibinfo  {journal} {The Journal of Chemical Physics}\
  }\textbf {\bibinfo {volume} {34}},\ \bibinfo {pages} {1037} (\bibinfo {year}
  {1961})}\BibitemShut {NoStop}%
\bibitem [{\citenamefont {NIST}()}]{bworld}%
  \BibitemOpen
  \bibfield  {author} {\bibinfo {author} {\bibnamefont {NIST}},\
  }\href{https://dlmf.nist.gov/19.2} {\enquote {\bibinfo {title} {Elliptic
  integrals},}\ }\BibitemShut {NoStop}%
\bibitem [{\citenamefont {T{\'o}th}(1950)}]{Toth:ActaSci_12:1950}%
  \BibitemOpen
  \bibfield  {author} {\bibinfo {author} {\bibfnamefont {L.~F.}\ \bibnamefont
  {T{\'o}th}},\ }\bibfield  {title} {\enquote {\bibinfo {title} {Some packing
  and covering theorems},}\ }\href@noop {} {\bibfield  {journal} {\bibinfo
  {journal} {Acta Sci. Math. Szeged}\ }\textbf {\bibinfo {volume} {12}},\
  \bibinfo {pages} {62} (\bibinfo {year} {1950})}\BibitemShut {NoStop}%
\bibitem [{\citenamefont {Bernard}\ and\ \citenamefont
  {Krauth}(2011)}]{Bernard:PRL_107:2011}%
  \BibitemOpen
  \bibfield  {author} {\bibinfo {author} {\bibfnamefont {E.~P.}\ \bibnamefont
  {Bernard}}\ and\ \bibinfo {author} {\bibfnamefont {W.}~\bibnamefont
  {Krauth}},\ }\bibfield  {title} {\enquote {\bibinfo {title} {Two-step melting
  in two dimensions: First-order liquid-hexatic transition},}\ }\href{\doibase
  https://doi.org/10.1103/PhysRevLett.107.155704} {\bibfield  {journal}
  {\bibinfo  {journal} {Phys. Rev. Lett.}\ }\textbf {\bibinfo {volume} {107}},\
  \bibinfo {pages} {155704} (\bibinfo {year} {2011})}\BibitemShut {NoStop}%
\bibitem [{\citenamefont {Kapfer}\ and\ \citenamefont
  {Krauth}(2015)}]{Kapfer:PRL_114:2015}%
  \BibitemOpen
  \bibfield  {author} {\bibinfo {author} {\bibfnamefont {S.~C.}\ \bibnamefont
  {Kapfer}}\ and\ \bibinfo {author} {\bibfnamefont {W.}~\bibnamefont
  {Krauth}},\ }\bibfield  {title} {\enquote {\bibinfo {title} {Two-dimensional
  melting: From liquid-hexatic coexistence to continuous transitions},}\
  }\href{\doibase 10.1103/PhysRevLett.114.035702} {\bibfield  {journal}
  {\bibinfo  {journal} {Phys. Rev. Lett.}\ }\textbf {\bibinfo {volume} {114}},\
  \bibinfo {pages} {035702} (\bibinfo {year} {2015})}\BibitemShut {NoStop}%
\bibitem [{\citenamefont {Talbot}\ \emph {et~al.}(1990)\citenamefont {Talbot},
  \citenamefont {Kivelson}, \citenamefont {Allen}, \citenamefont {Evans},\ and\
  \citenamefont {Frenkel}}]{Talbot:JCP_92:1990}%
  \BibitemOpen
  \bibfield  {author} {\bibinfo {author} {\bibfnamefont {J.}~\bibnamefont
  {Talbot}}, \bibinfo {author} {\bibfnamefont {D.}~\bibnamefont {Kivelson}},
  \bibinfo {author} {\bibfnamefont {M.~P.}\ \bibnamefont {Allen}}, \bibinfo
  {author} {\bibfnamefont {G.~T.}\ \bibnamefont {Evans}}, \ and\ \bibinfo
  {author} {\bibfnamefont {D.}~\bibnamefont {Frenkel}},\ }\bibfield  {title}
  {\enquote {\bibinfo {title} {Structure of the hard ellipsoid fluid},}\
  }\href{\doibase 10.1063/1.457902} {\bibfield  {journal} {\bibinfo  {journal}
  {The Journal of Chemical Physics}\ }\textbf {\bibinfo {volume} {92}},\
  \bibinfo {pages} {3048} (\bibinfo {year} {1990})}\BibitemShut {NoStop}%
\end{thebibliography}
%

\end{document}